\documentclass[12pt]{article}

\pdfoutput=1
\usepackage[dvipdfm]{graphicx}
\usepackage{epsfig}
\usepackage{color}
\usepackage{atlasphysics}
\usepackage{subfigure}
\usepackage{mathrsfs}
\usepackage{multirow}
\usepackage{url}
\usepackage{hyperref}
\usepackage{amssymb}
\usepackage{epstopdf}
\usepackage{feynmp}
\usepackage{tabularx}
\graphicspath{{figures/}}

\newcommand{\eqref}{Eq.}

\usepackage{authblk}

\begin{document}
\title{Studies of Vector Boson Scattering And Triboson Production with {\sc delphes}  Parametrized Fast Simulation for Snowmass 2013}

\author[a]{C.~Degrande}
\author[b]{J.~L.~Holzbauer}
\author[c]{S.-C.~Hsu}
\author[d]{A.~V.~Kotwal}
\author[d]{S.~Li}
\author[c]{M.~Marx}
\author[a]{O.~Mattelaer}
\author[e]{J.~Metcalfe}
\author[e]{M.-A.~Pleier}
\author[d]{C.~Pollard}
\author[f]{M.~Rominsky}
\author[g]{D.~Wackeroth}

\affil[a]{University of Illinois at Urbana-Champaign}
\affil[b]{University of Mississippi}
\affil[c]{University of Washington}
\affil[d]{Duke University}
\affil[e]{Brookhaven National Laboratory}
\affil[f]{Fermi National Accelerator Laboratory}
\affil[g]{University at Buffalo, the State University of New York}

\maketitle

\begin{abstract}
Multiboson production provides a unique way to probe Electroweak Symmetry Breaking (EWSB) and physics beyond the Standard Model (SM). With the discovery of the Higgs boson, the default model is that EWSB occurs according to the Higgs mechanism. Deviations from the SM in Higgs and gauge boson properties due to new physics at a higher energy scale can be parameterized by higher-dimension operators in an Effective Field Theory (EFT). We present sensitivity studies for dimension-6 and dimension-8 operators in an EFT by looking for anomalous vector boson scattering and triboson production, at proton-proton colliders with center-of-mass energies of 14 TeV, 33 TeV and 100 TeV, respectively.
\end{abstract}


\section{Introduction}
\label{sec:intro}

Multiboson production provides a unique way to probe Electroweak Symmetry Breaking (EWSB) and physics beyond the Standard Model (SM). With the discovery of the Higgs boson, the default model is that EWSB occurs according to the Higgs mechanism. In this case, all the couplings are completely specified. The gauge and Higgs sectors of the SM can be probed by studying vector boson scattering (VBS) and triboson production. These processes provide unique probes of quartic couplings which to date have not been studied extensively, since past experiments were
 mostly sensitive to diboson production. Deviations from SM predictions will give clues to physics beyond the SM.  

Assuming that the energy scale associated with this new physics is sufficiently high compared to the masses
 of the gauge and Higgs bosons, we are motivated to use Effective Field Theory~\cite{Degrande:2012wf} to
 parameterize the new physics in channels involving these particles.  The truncated EFT Lagrangian employed in our study is
\begin{equation}
\mathcal{L} = \mathcal{L}^{SM} + \displaystyle\sum\limits_i \frac{c_{i}}{\Lambda^{2}} \mathcal{O}_{i} + \displaystyle\sum\limits_j \frac{f_{j}}{\Lambda^{4}} \mathcal{O}_{j} 
\label{eq:eft}
\end{equation}
where $\mathcal{O}_{i}$ and  $\mathcal{O}_{j}$ are the dimension-6 and dimension-8 operators~\cite{Eboli:2006wa} 
 respectively, and 
$c_{i}$ and $f_{j}$ represent the numerical coefficients 
 associated with these operators. These operators are induced by integrating out the new degrees of freedom,
 and the numerical coeffients are meant to be derivable from a more complete high-energy theory. $\Lambda$
 is a mass-dimension parameter associated with the energy scale of the new degrees of freedom which have 
 been integrated out. 

 We choose
 operators that are Lorentz-invariant and gauge-invariant under the SM electroweak group $SU(2)_L \times U(1)_Y$, and contain the electroweak gauge and Higgs fields. Thus,
 EWSB is induced only by the spontaneous symmetry breaking as in the SM, due to the Higgs vacuum expectation
 value, and there is no explicit breaking of electroweak symmetry. This latter point is one of the differences between this EFT expansion and the traditional method of anomalous couplings where some operators 
 explicitly break gauge invariance. 

 For this study,  we tested both dimension-6 operators and dimension-8 operators at $pp$ colliders with
 $\sqrt {s} = 14, \; 33 $ and 100~TeV. 
This Snowmass exercise was performed using {\sc madgraph}~\cite{Alwall:2011uj} for event generation with {\sc CTEQ6L1}~\cite{CTEQ6L1} PDF set, {\sc pythia} for showering and a special version of {\sc delphes}~\cite{Ovyn:2009tx} for the detector simulation. The effect of
 multiple interactions at high instantaneous luminosity was simulated using 
  specially designed pileup files~\cite{delphes1, delphes2, delphes3}.

For the vector boson scattering process, we have chosen to study the fully-leptonic decay channels of the 
$ZZ$, $WZ$ and same-sign $WW$ final states. These final states do not suffer from large $t \bar{t}$ 
 backgrounds. For the triboson final states involving heavy gauge bosons, we have chosen the fully-leptonic
 $WWW$ final state because it has the largest production cross section $\times$ branching ratio compared
 to all other combinations of heavy gauge bosons, and is separable from diboson backgrounds. We have also studied the
 photonic channels using the $Z \gamma \gamma$ final state. 

\section{VBS $ZZ \to 4 \ell $}

In this vector boson scattering channel, we explore anomalous production by studying the invariant mass distribution of the $ZZ$ diboson pair. New physics is parameterized by the following dimension-6 operator
\begin{equation} 
{\cal L}_{\phi W} = \frac{c_{\phi W}}{\Lambda^2} {\rm Tr} (W^{\mu \nu} W_{\mu \nu}) \phi^\dagger \phi
\end{equation} 
or by one of the  dimension-8 operators
\begin{eqnarray}
{\cal L}_{T,8} & = & \frac{f_{T8}}{\Lambda^4} B_{\mu \nu}B^{\mu \nu} B_{\alpha \beta} B^{\alpha \beta} \nonumber \\ 
{\cal L}_{T,9} & = & \frac{f_{T9}}{\Lambda^4} B_{\alpha \mu}B^{\mu \beta} B_{\beta \nu}B^{\nu \alpha}
\label{eq:lllljjdim8}
\end{eqnarray}
where $\phi$ is the SM Higgs field, and $W^{\mu \nu}$ ($B^{\mu \nu}$) are the field strength tensors derived from the $SU(2)_L$ ($U(1)_Y$) gauge fields. 
The fully-leptonic VBS $ZZ \rightarrow 4 \ell$ channel provides a fully reconstructible $ZZ$ final state with small mis-identification 
 backgrounds~\cite{atlas:ZZconf} which can be neglected in this sensitivity study.  The contribution
 from non-VBS diboson production accompanied by QCD jets  
 is reduced by requiring the 
forward jet-jet mass to be greater than 1 TeV, and the surviving background is included in this study. 

The operator ${\cal L}_{\phi W}$ was chosen for this study based on comparing cross section enhancements due to individual dimension-6 operators with their coefficients set to a non-zero, nominal value.
 This operator was found to give the largest enhancement. Similar comparisons of cross section enhancements due to individual dimension-8 operators were also performed and are shown in
  Table~\ref{tab:ZZ_xsec_ILC}. The ${\cal L}_{T,0}$ and ${\cal L}_{T,1}$ operators give the largest enhancement, followed by ${\cal L}_{T,2}$, ${\cal L}_{T,8}$ and ${\cal L}_{T,9}$. The first three
 of these operators contain the same field strength tensors, and the study of ${\cal L}_{T,1}$ using the VBS $WZ$ final state is presented in the next section. For the VBS $ZZ$ final state, we
 chose the dimension-8 operators in Eqn.~\ref{eq:lllljjdim8} because they involve only the electrically-neutral gauge fields and can be probed by the $ZZ$ final state but not the $WZ$ or $WW$ final states. 

{\sc madgraph} 5.1.5.10 is used for the generation of VBS $ZZ$ (SM) and non-VBS (SM $ZZ$ QCD) processes as well as the non-SM processes mentioned above. $Z$ bosons were required to decay to electron or muon pairs.

\subsection{Event Selection}
After {\sc pythia} 6.4~\cite{pythia6} parton showering, additional detector effects are applied using {\sc delphes} 3.0.9~\cite{deFavereau:2013fsa} with the Snowmass parameterization~\cite{delphes1, delphes2, 
 delphes3}. 
Candidate VBS $ZZ$ events are selected according to the following criteria:

\begin{itemize}
    \item Exactly four selected leptons (each with $p_T > 25$~GeV) which can be
        separated into two opposite sign, same flavor pairs (No $Z$ mass window requirement)
    \item At least two selected jets, each with $p_T > 50$~GeV
    \item $m_{jj} > 1$ TeV, where $m_{jj}$ is the invariant mass of the
        two highest-$p_T$ selected jets
\end{itemize}

\subsection{Statistical Analysis}
\label{zzjjStats}

To determine the expected sensitivity to beyond-SM (BSM)  $ZZ$ contribution,
the background-only $p_0$-value expected for signal+background is calculated using the $m_{4\ell}$ spectrum.  In order to show the improvement possible with the increased luminosity and center-of-mass energy, the $5 \sigma$ discovery potential and 95\% CL limits are studied.
Since the $4-$lepton mass is the process $\sqrt{\hat{s}}$, the study of its distribution directly probes the energy-dependence of the new physics.

At sufficiently high energy, the amplitude predicted by higher-dimension operators will eventually violate unitarity. In this regime, the new physics that presumably restores unitarity is expected to be probed directly, such as the 
 production of on-shell resonances. This is a very interesting regime because the masses and couplings of new resonances can be measured independently, which is a much more powerful probe as compared to the low-energy regime where
 only the appropriate ratio of coupling and mass can be probed. Furthermore, in the high energy regime it is also possible to study new decay modes of the resonances, whereas in the low-energy regime of EFT applicability we can only
 study the anomalous production of SM particles. The regime above the unitarity bound is probed more strongly by the higher energy colliders. 

We present the sensitivity to the higher-dimension operators in two ways. In the first case, we assume that new physics is only probed ``virtually'' by higher-dimension operators involving SM fields, and we require the
 generated events to lie below the unitarity bound in the diboson mass. In the second case, we allow the collider to probe the sensitivity to new physics above the unitarity bound through direct production of new resonances and 
 measuring their masses, couplings and decay branching ratios. Since an ultraviolet-complete theory of strongly-interacting electroweak sector is not available  for the additional physics that would be accessible in this high-energy regime, as a proxy we also quote the sensitivity to the higher-dimension operators without making the unitarity bound requirement on the diboson mass.

\begin{figure}[h]
  \centering
  \includegraphics[width=0.49\textwidth]{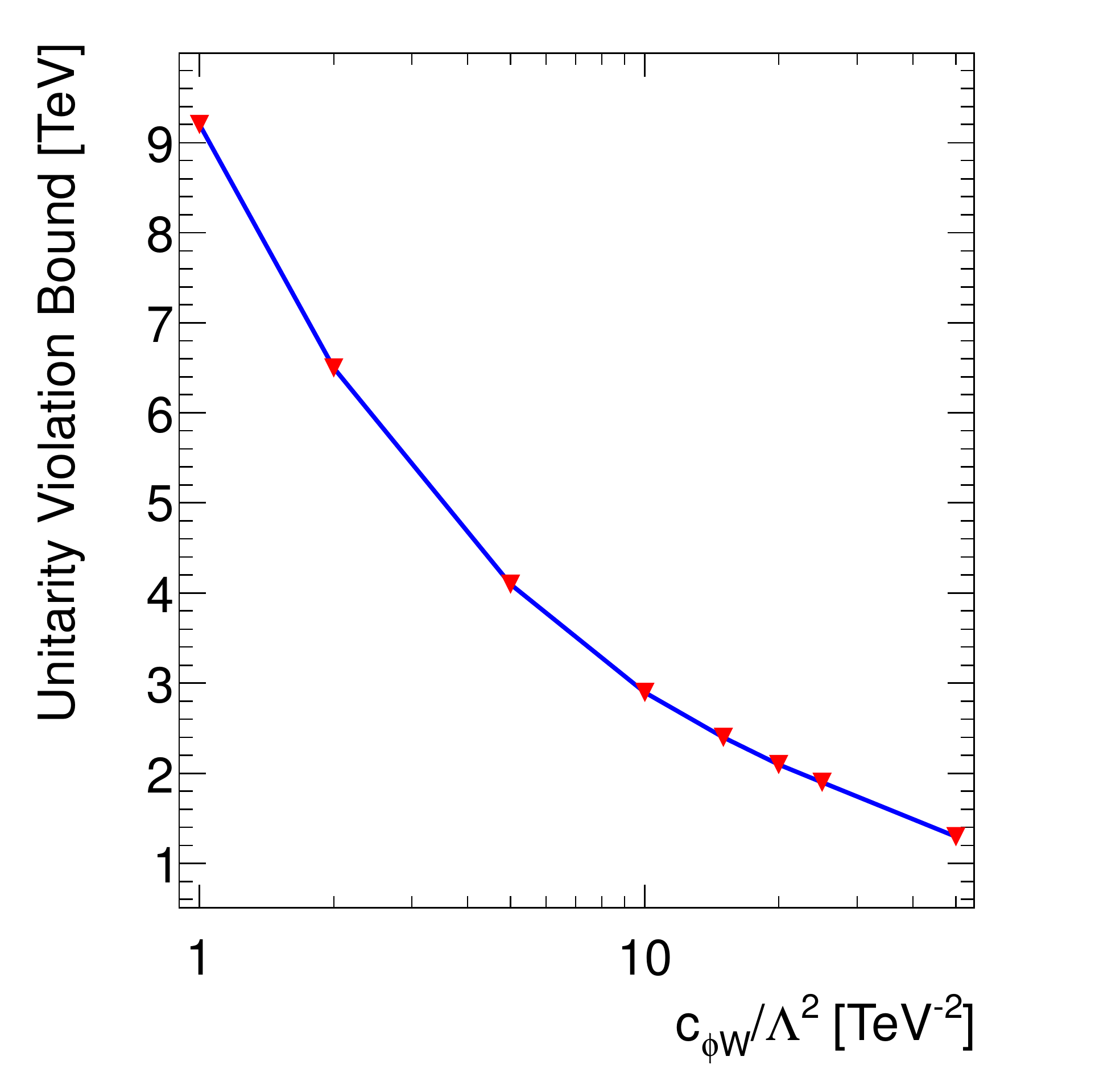}
  \includegraphics[width=0.49\textwidth]{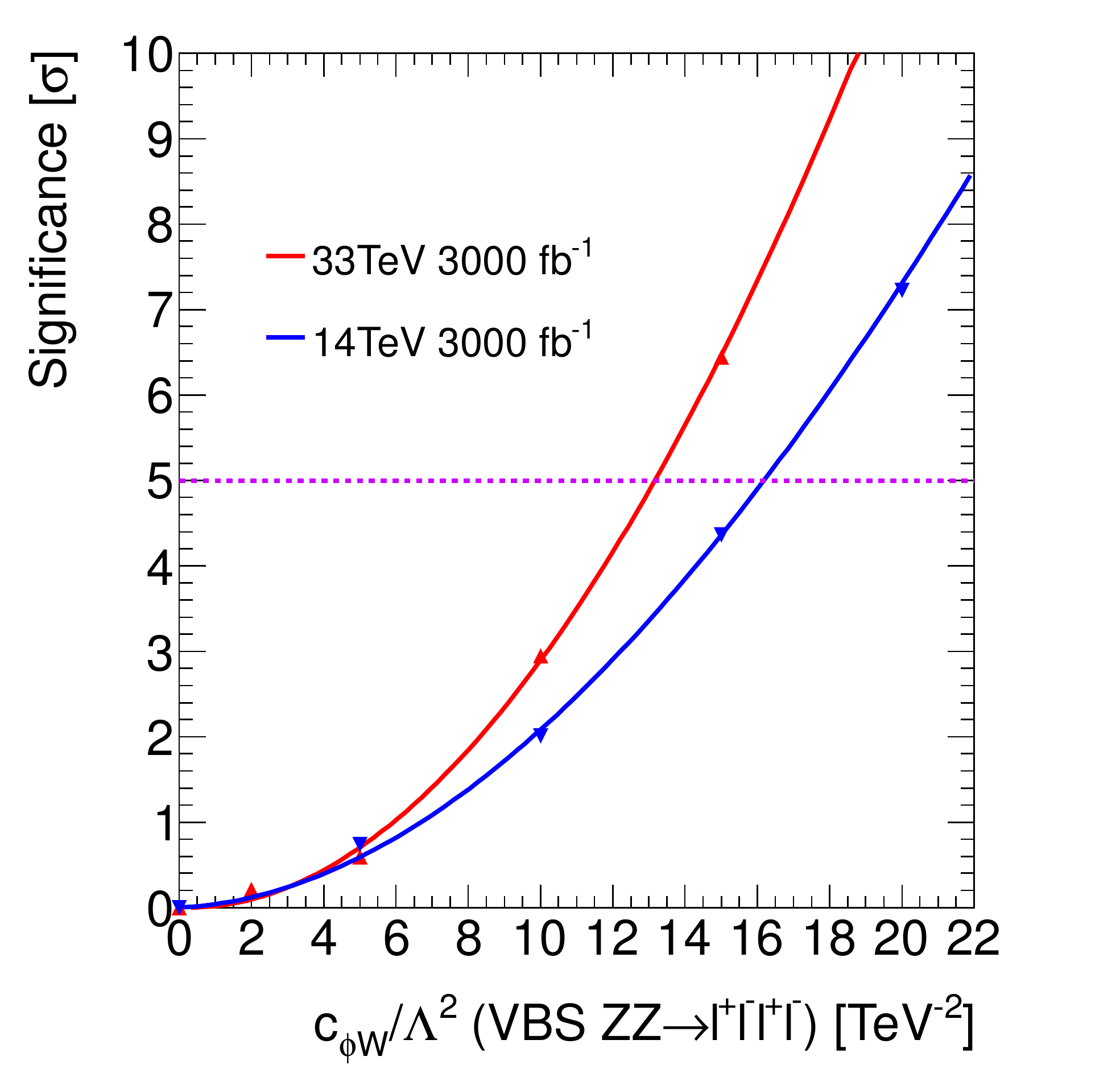}
  \caption{
The unitarity violation bounds (left) and signal significances (right) are shown as a function of dimension-6 operator $c_{\phi W}/\Lambda^2$
  coefficient values in $pp \to ZZ + 2j \to 4 \ell + 2j$ processes. The UV bounds are not applied in the significance plot. }
\label{fig:zz_cphiw_curve}
\end{figure}

\begin{figure}[h]
  \centering
  \includegraphics[width=0.49\textwidth]{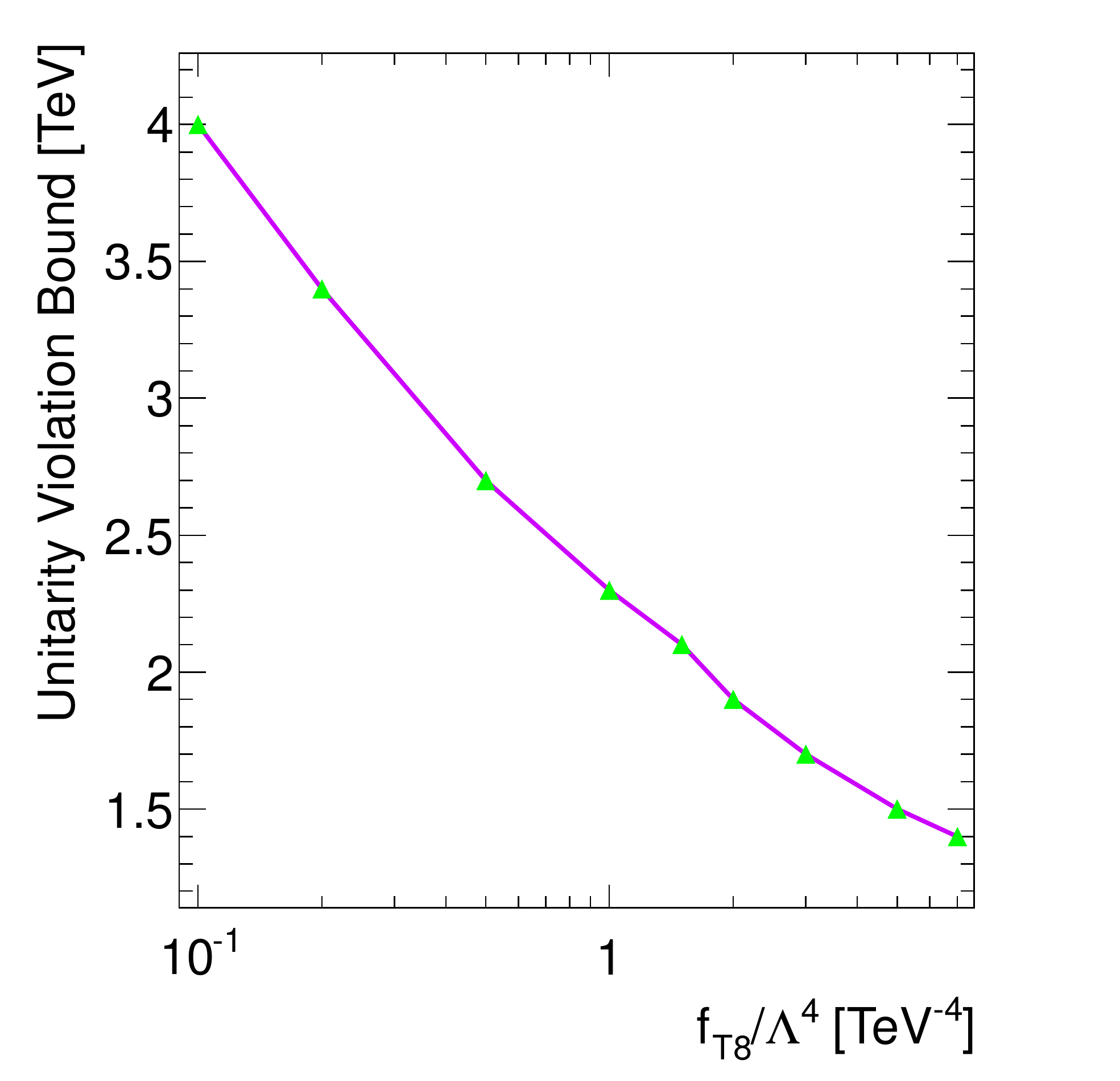}
  \includegraphics[width=0.49\textwidth]{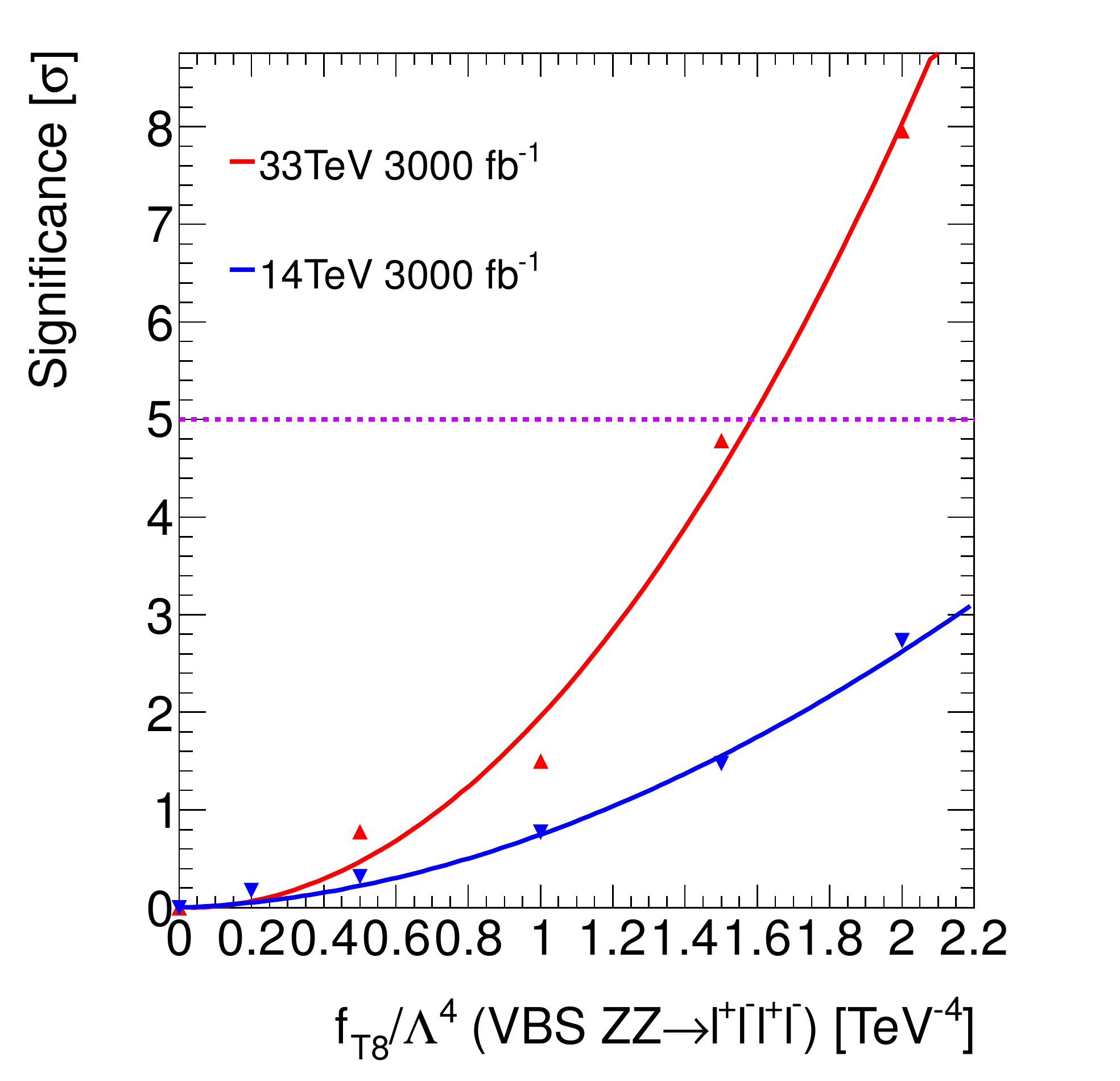} \\
  \includegraphics[width=0.49\textwidth]{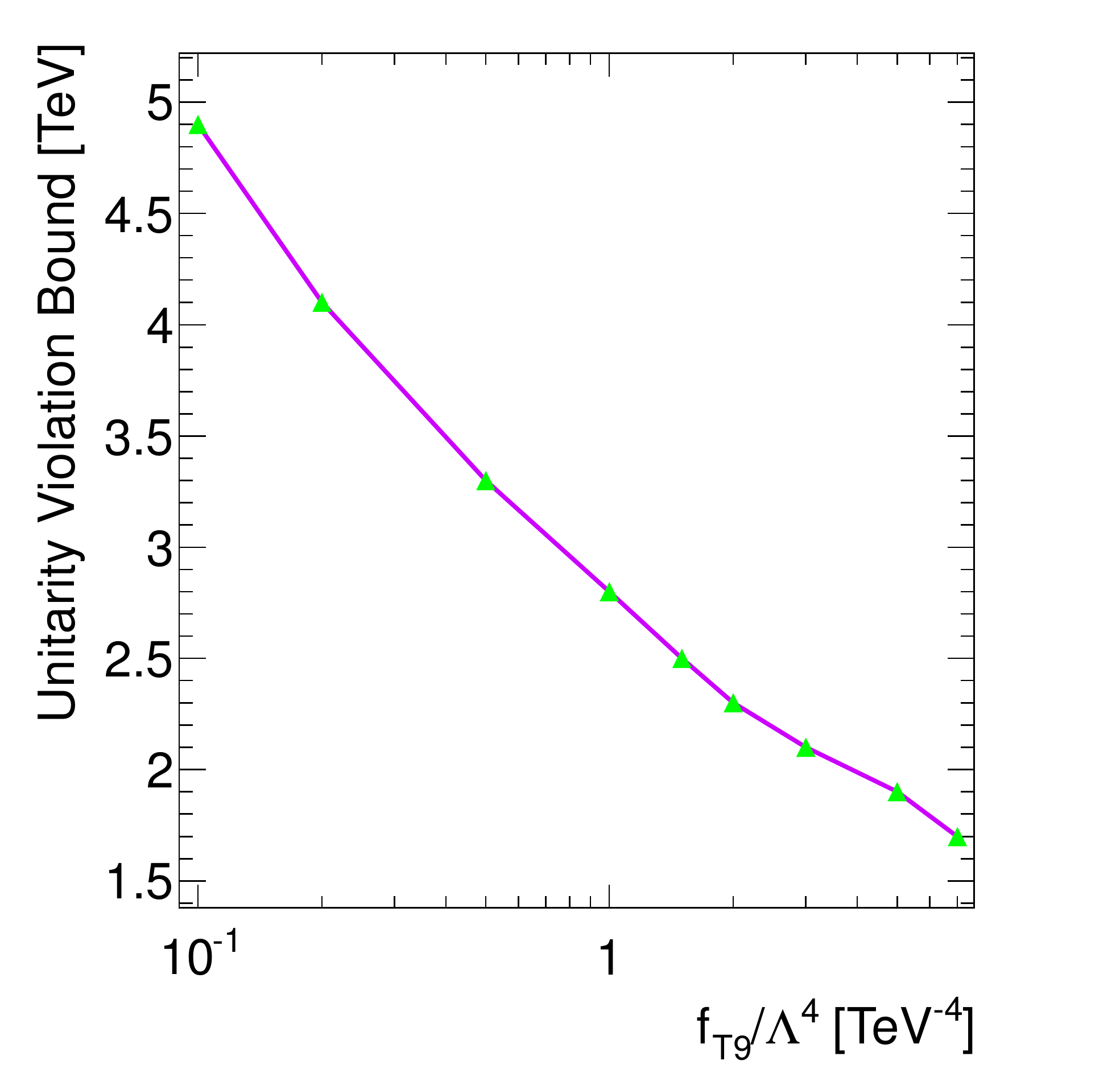}
  \includegraphics[width=0.49\textwidth]{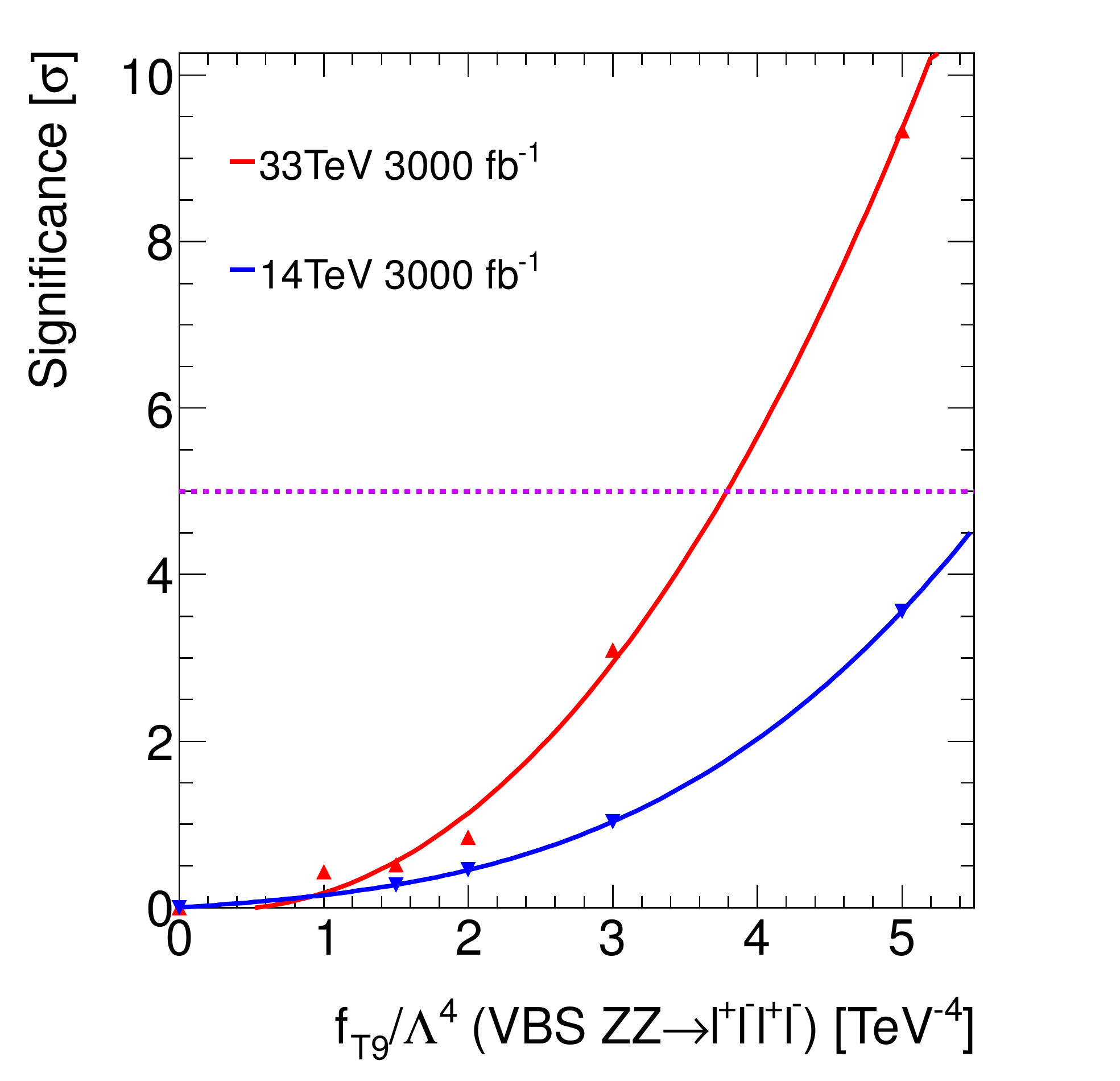}
  \caption{
The unitarity violation bounds (left) and signal significances (right) are shown as a function of dimension-8 operator ${\cal L}_{T,8}$ (top) and ${\cal L}_{T,9}$ (bottom)
  coefficient values in $pp \to ZZ + 2j \to 4 \ell + 2j$ processes. The UV bounds are not applied in the significance plots.}
\label{fig:zz_ft89_curve}
\end{figure}

\begin{figure}[h]
  \centering
  \includegraphics[width=0.49\textwidth]{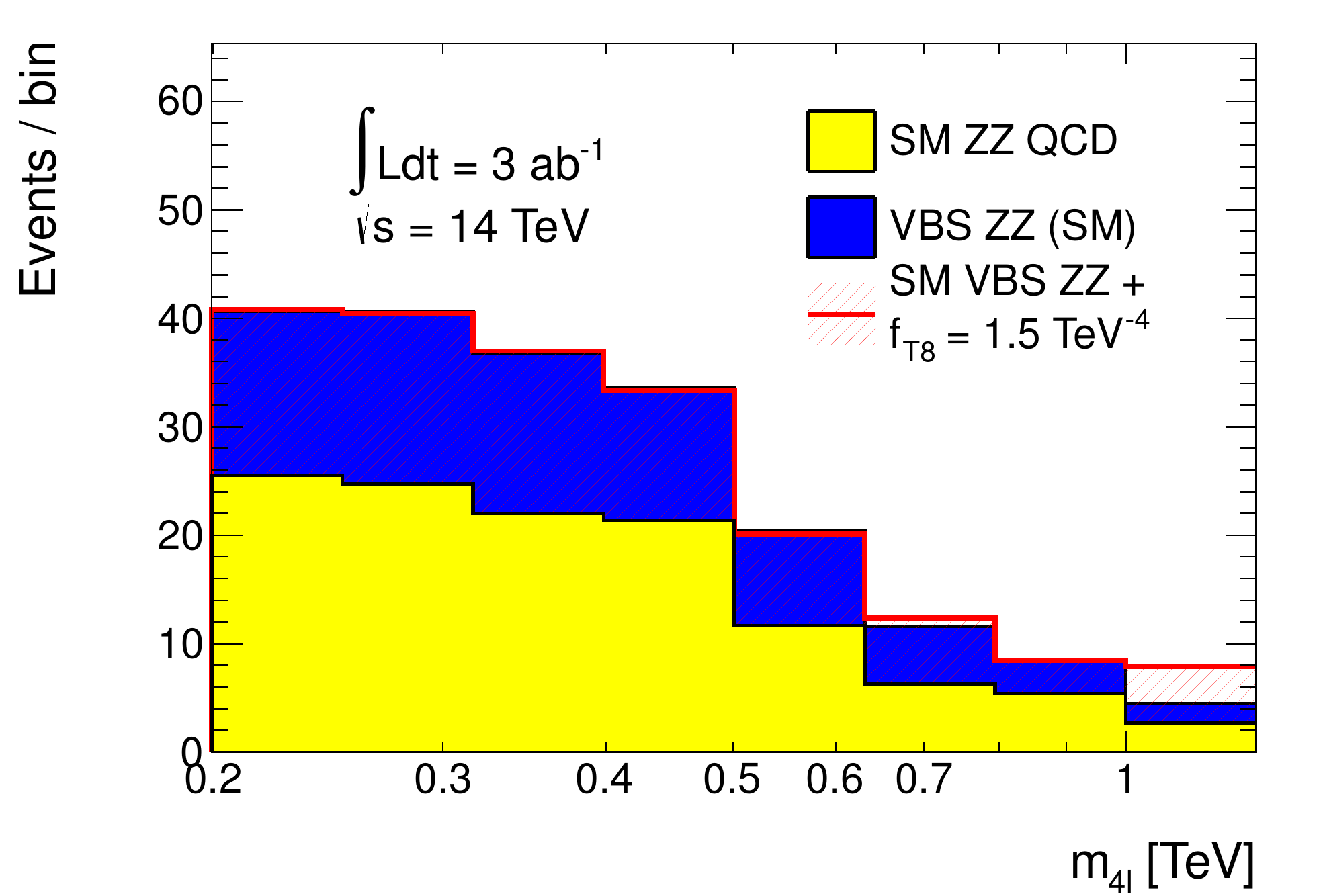} 
  \includegraphics[width=0.49\textwidth]{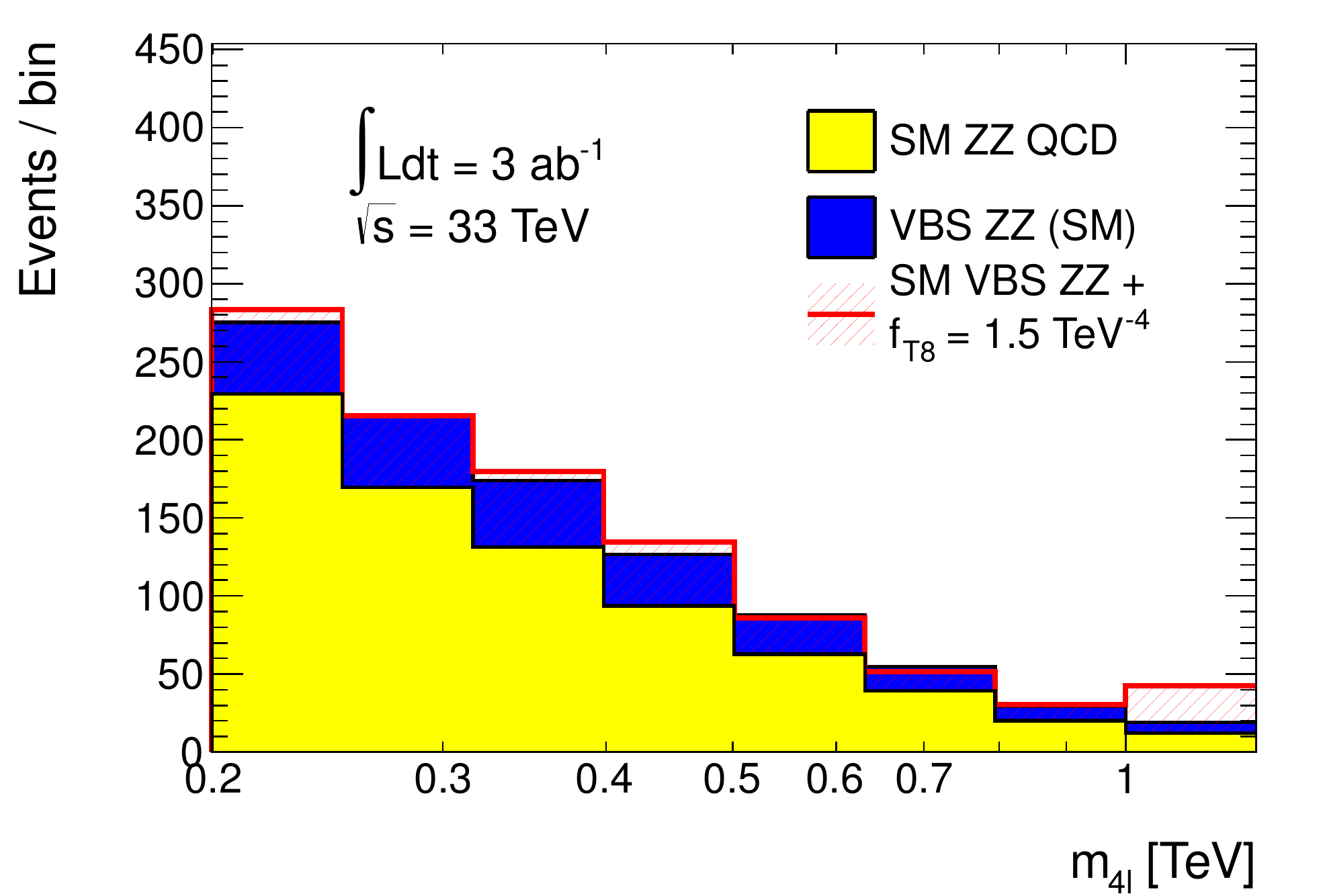} \\
  \includegraphics[width=0.49\textwidth]{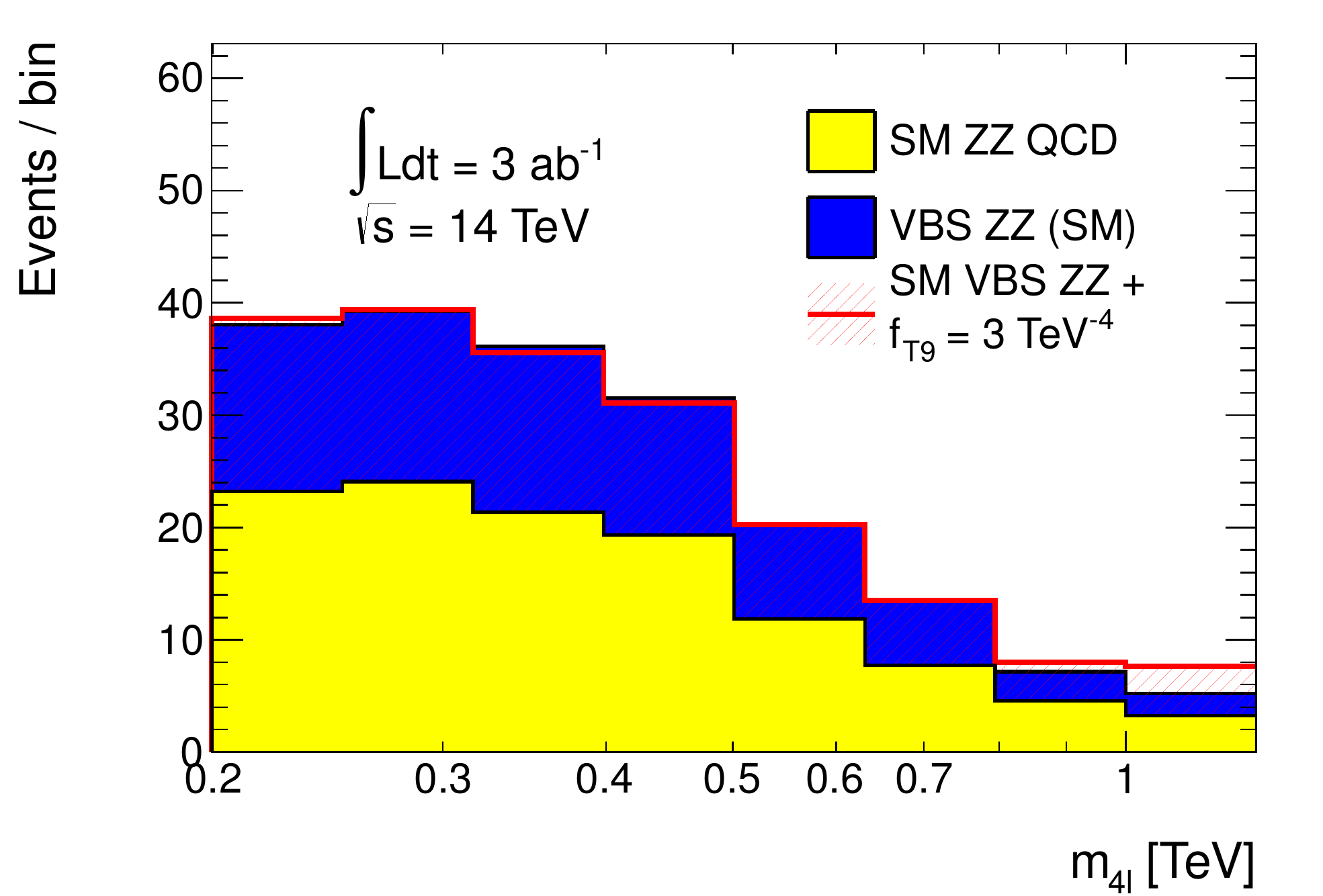}
  \includegraphics[width=0.49\textwidth]{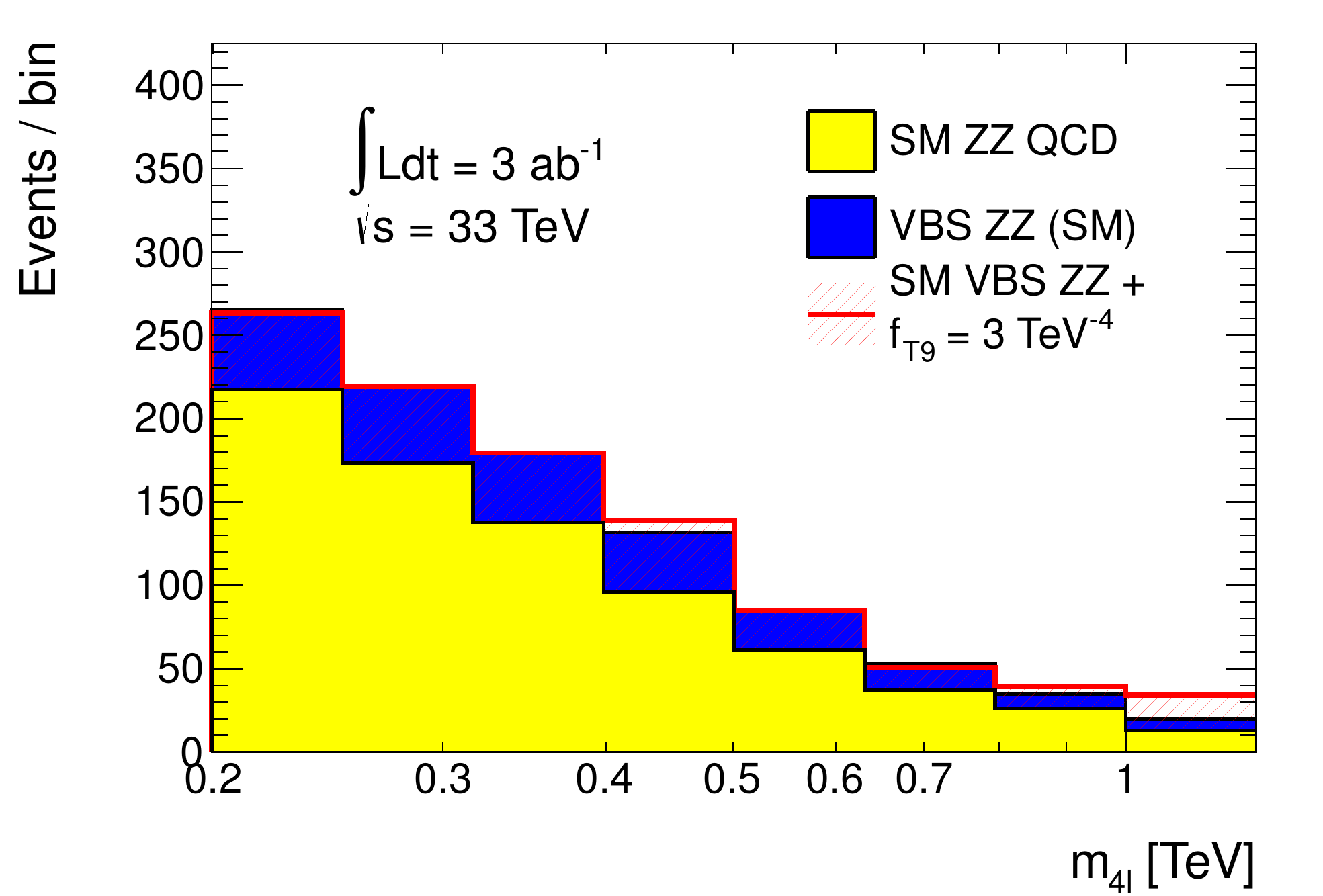} \\
  \includegraphics[width=0.49\textwidth]{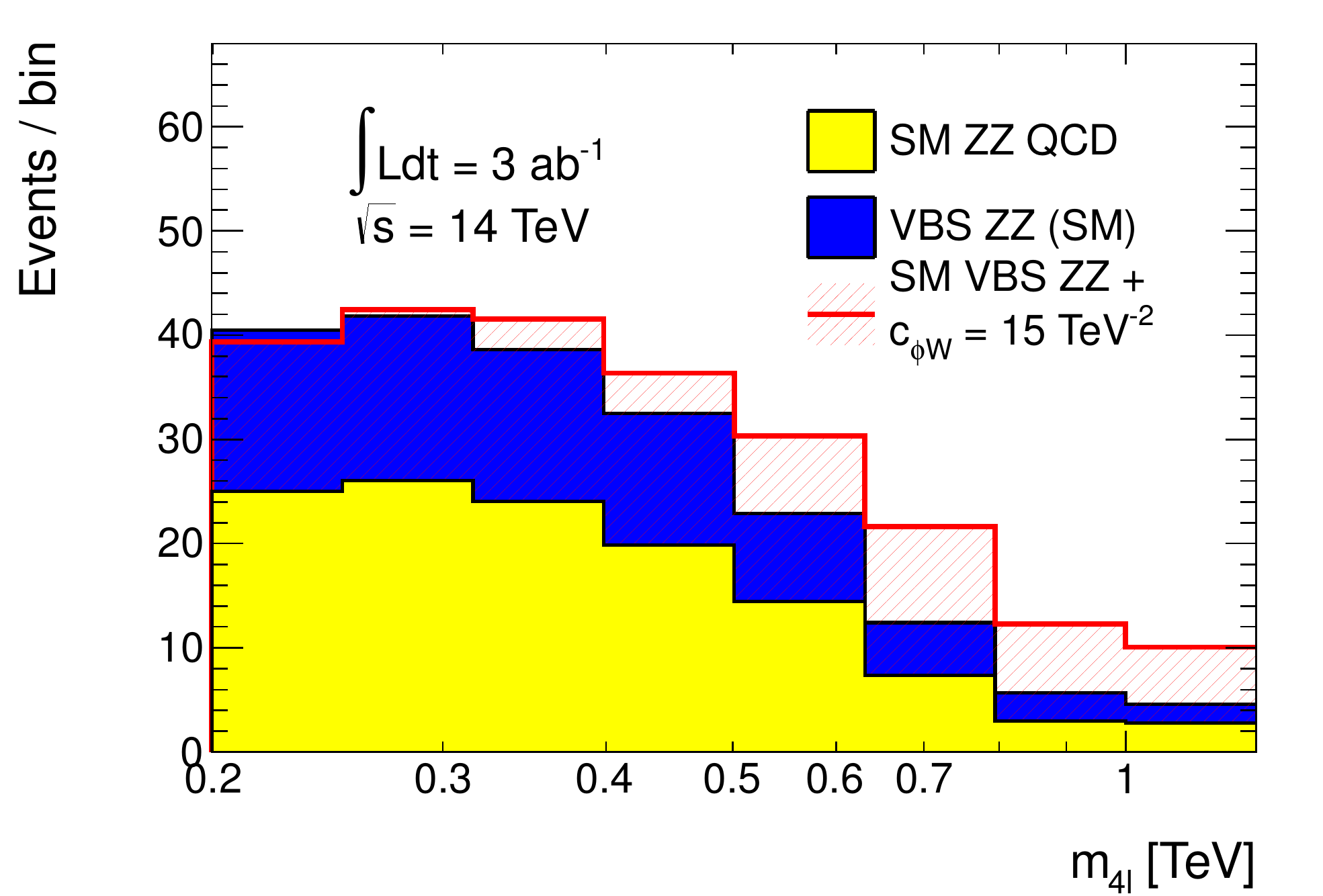} 
  \includegraphics[width=0.49\textwidth]{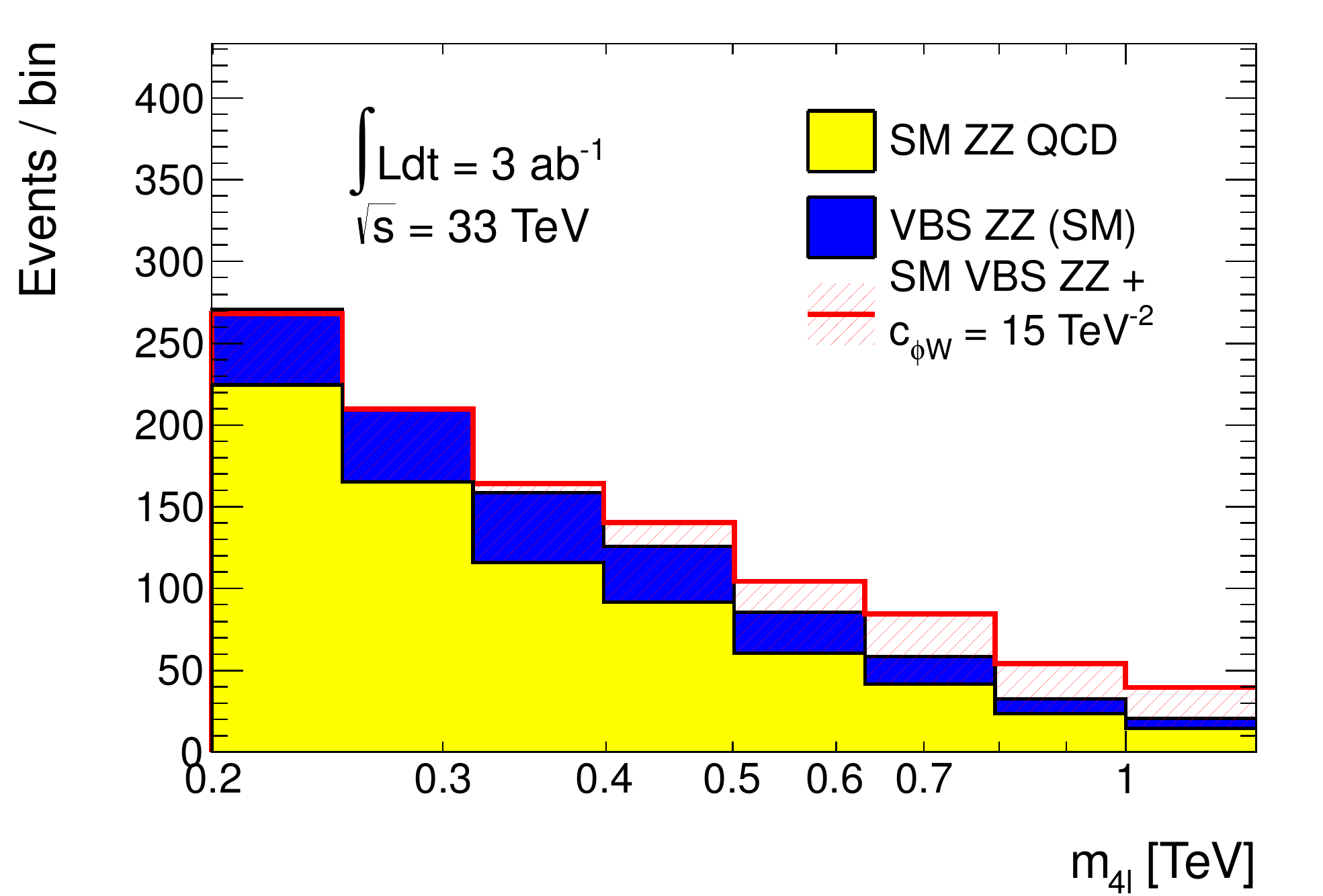}
  \caption{\label{fig:zz_ft8_ft9_cphiw_noUVCutOff} 
 In the $pp \to ZZ + 2j \to 4 \ell + 2j$ process, 
 the reconstructed 4-lepton mass ($m_{4 \ell}$)
 spectrum comparisons between Standard Model and dimension-8 operator coefficient $f_{T8}/\Lambda^{4} = 1.5$~TeV$^{-4}$ (top), $f_{T9}/\Lambda^{4} = 3$~TeV$^{-4}$ (middle) and dimension-6 operator coefficient $c_{\phi W}/\Lambda^2 = 15$~TeV$^{-2}$ (bottom) are shown after requiring $m_{jj} > 1$~TeV at $\sqrt{s}=14$ TeV (left) and 33 TeV (right). The overflow and underflow bins are included in the  plots. The UV bound is not applied. }
\end{figure}

\begin{figure}[h]
  \centering
  \includegraphics[width=0.49\textwidth]{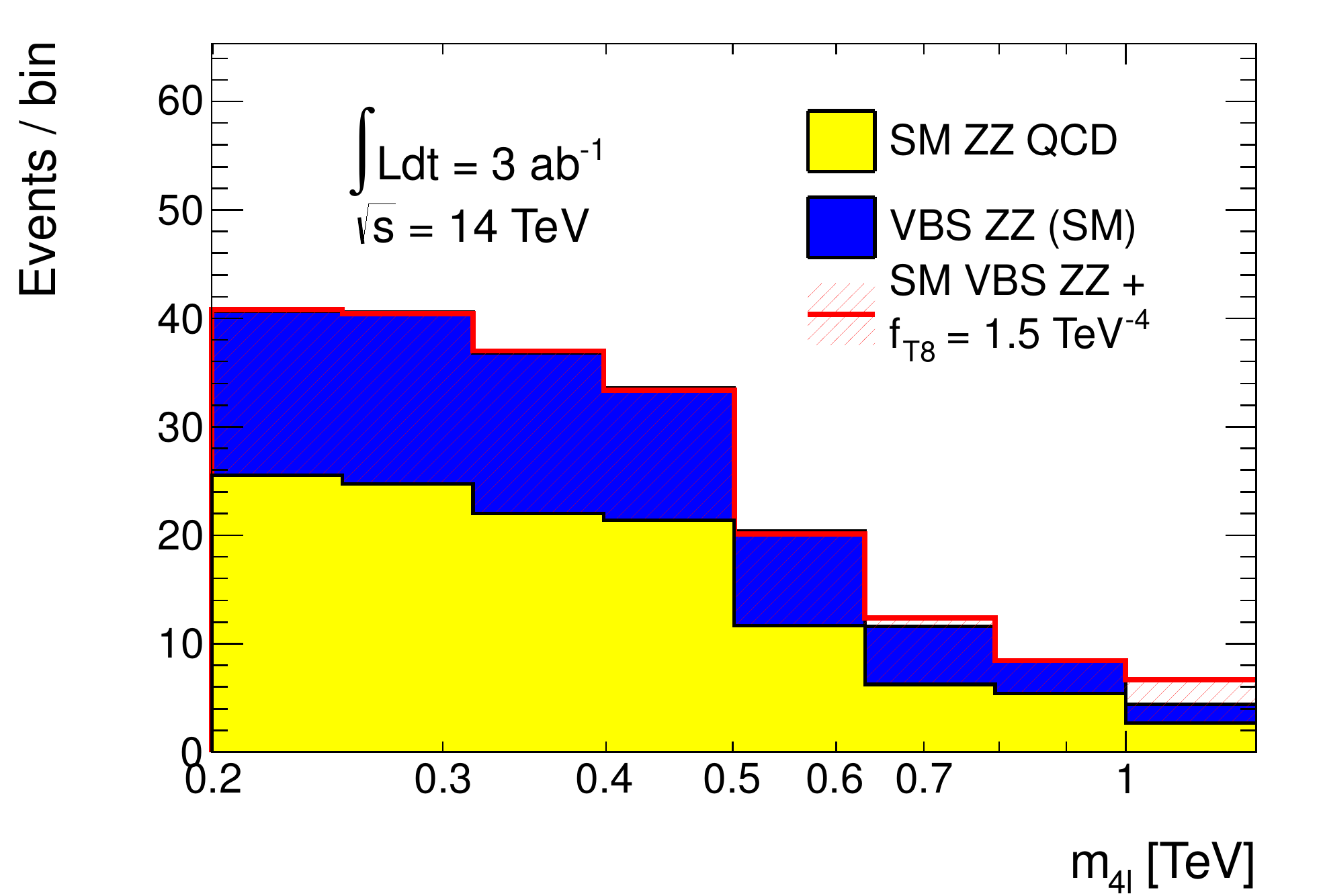} 
  \includegraphics[width=0.49\textwidth]{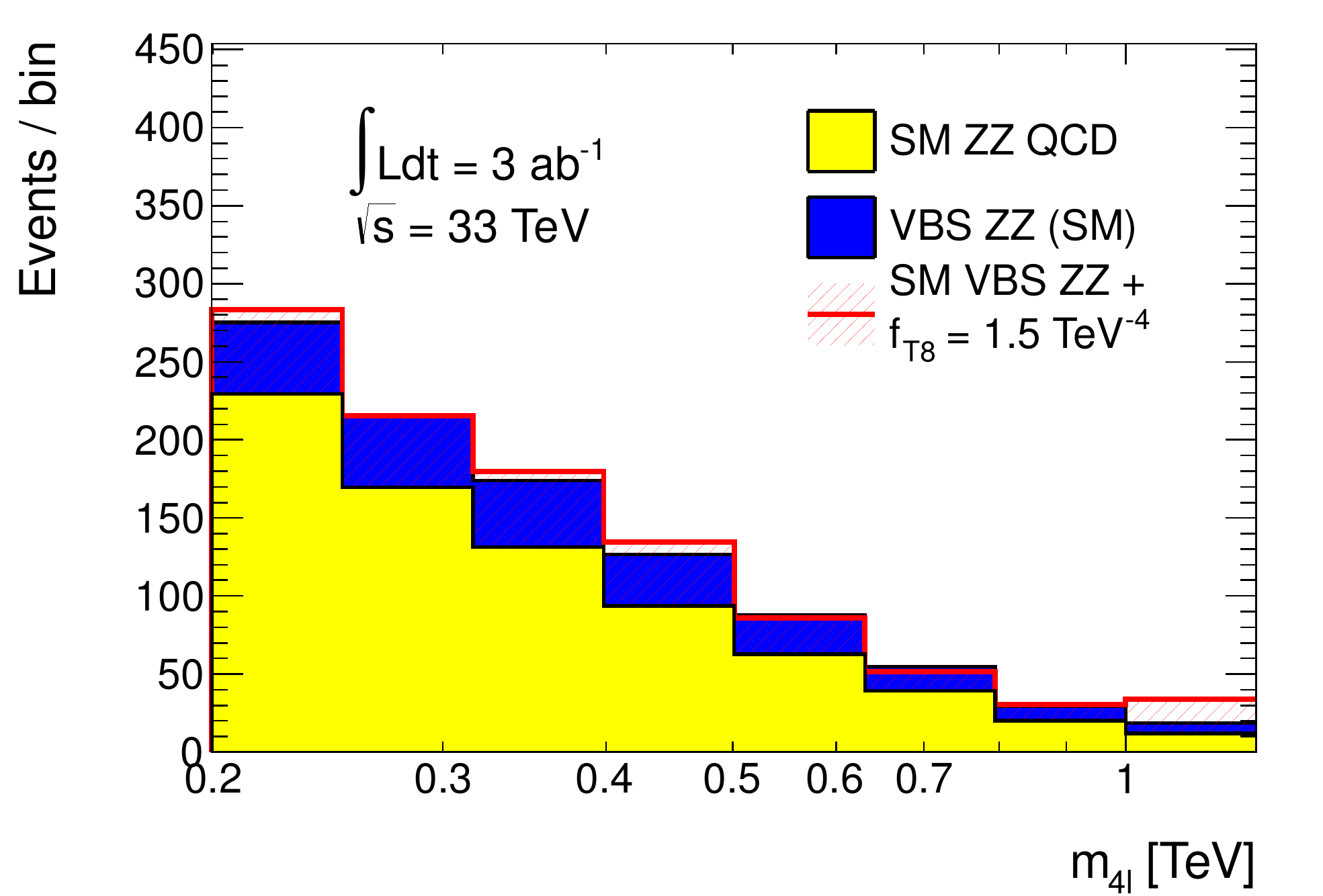} \\
  \includegraphics[width=0.49\textwidth]{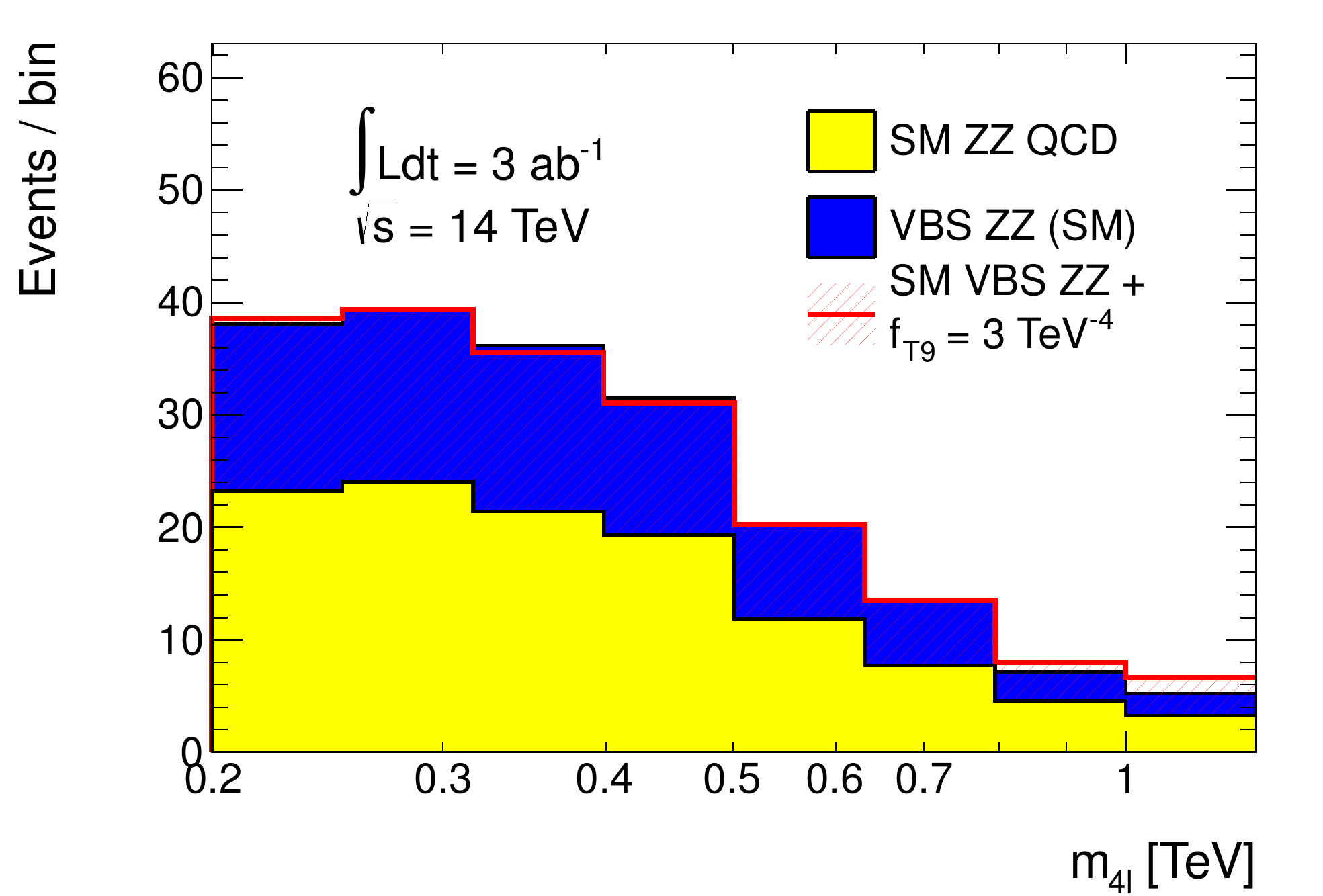}
  \includegraphics[width=0.49\textwidth]{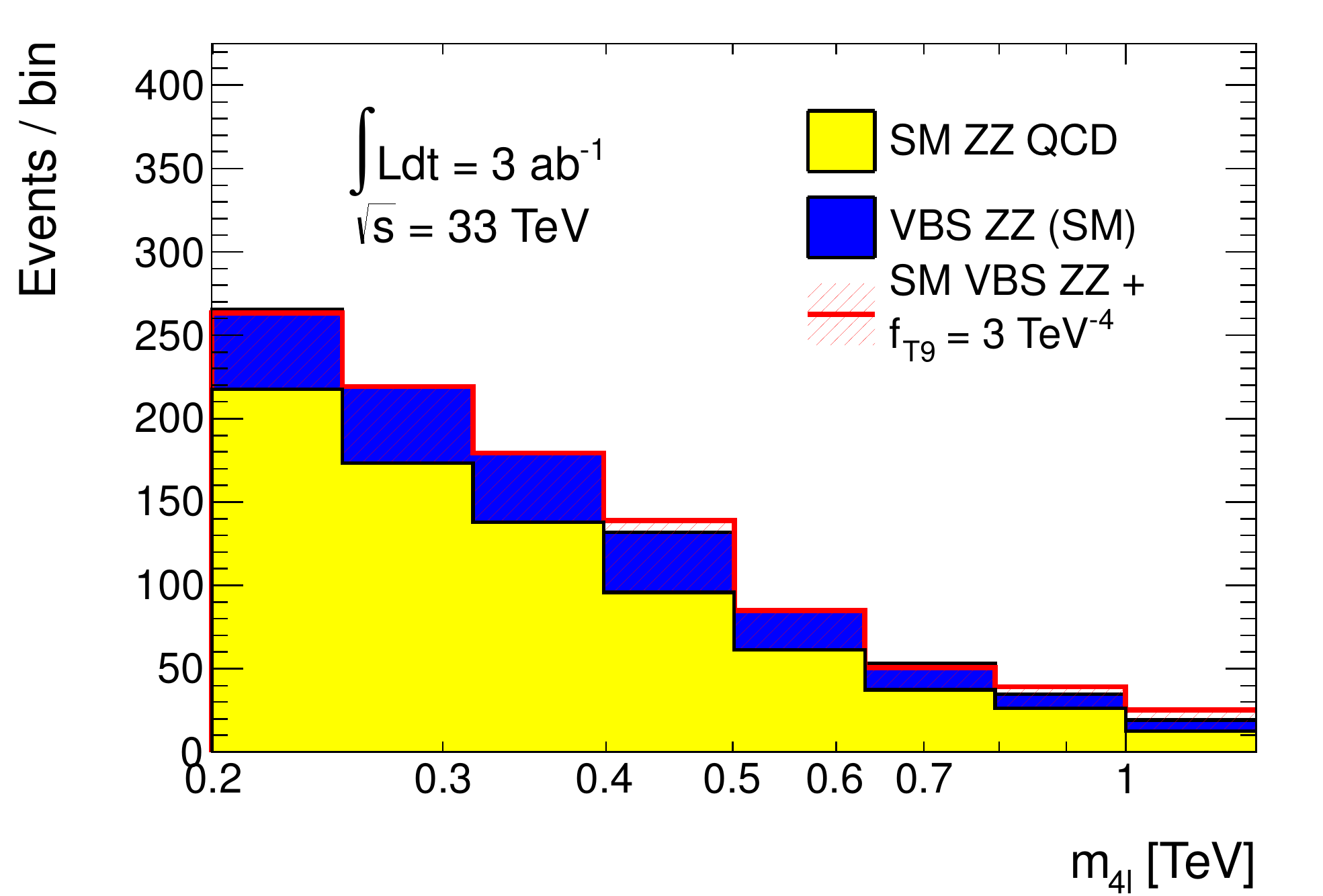} \\
  \includegraphics[width=0.49\textwidth]{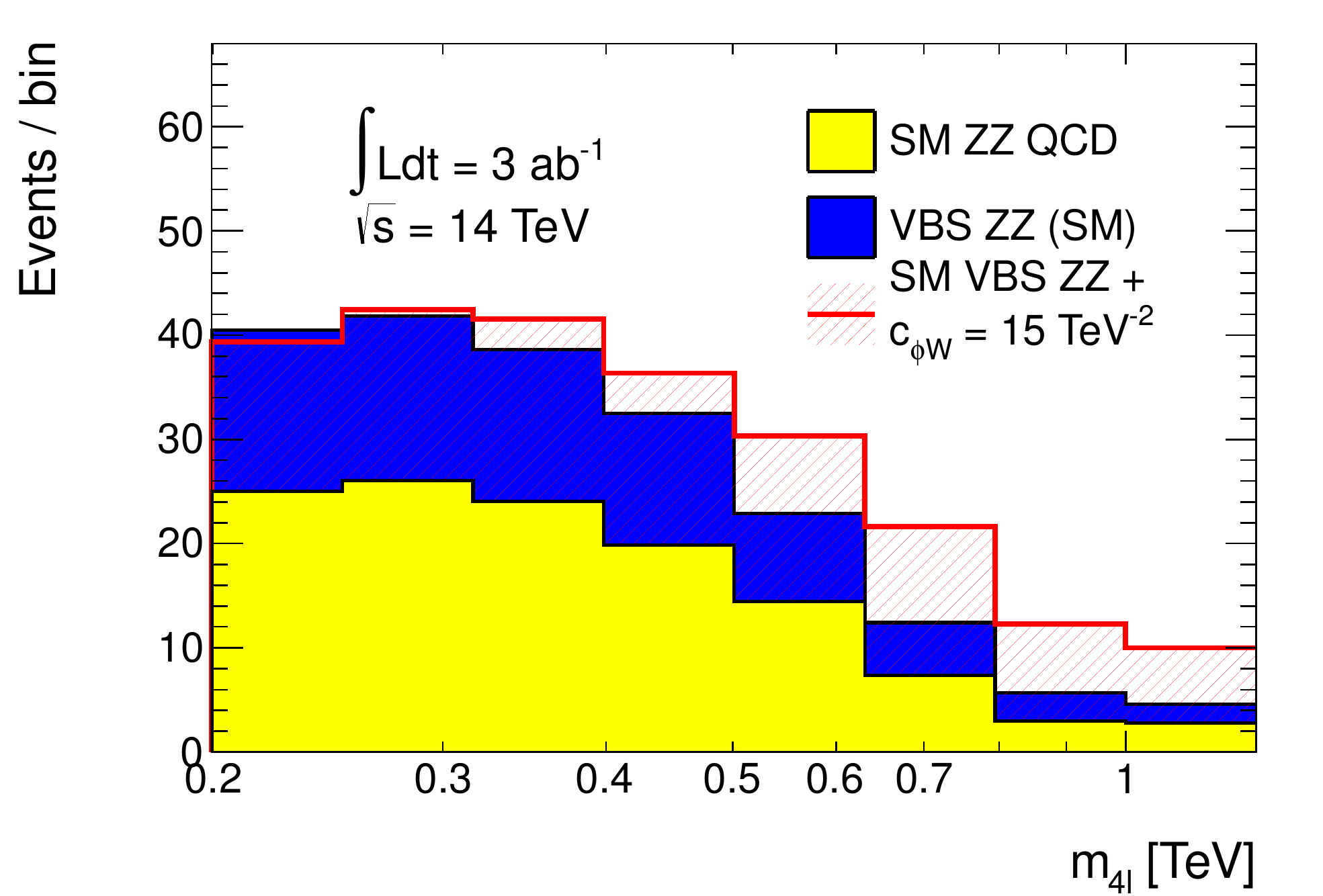}
  \includegraphics[width=0.49\textwidth]{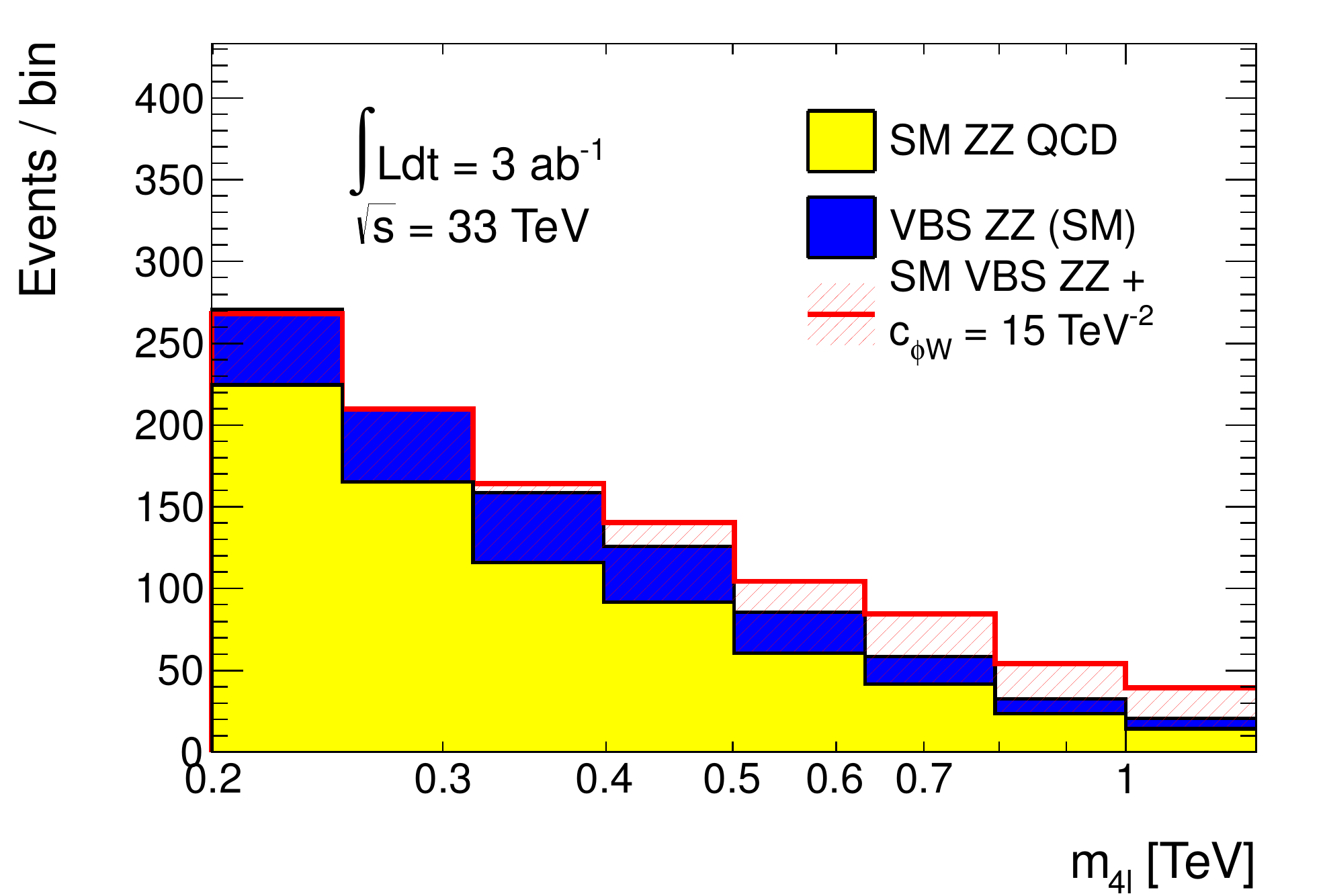}
  \caption{\label{fig:zz_ft8_ft9_cphiw} 
 In the $pp \to ZZ + 2j \to 4 \ell + 2j$ process, 
 the reconstructed 4-lepton mass ($m_{4 \ell}$)
 spectrum comparisons between Standard Model and dimension-8 operator coefficient $f_{T8}/\Lambda^{4} = 1.5$~TeV$^{-4}$ (top), $f_{T9}/\Lambda^{4} = 3$~TeV$^{-4}$ (middle) and dimension-6 operator coefficient $c_{\phi W}/\Lambda^2 = 15$~TeV$^{-2}$ (bottom) are shown after requiring $m_{jj} > 1$~TeV at $\sqrt{s}=14$ TeV (left) and 33 TeV (right). The overflow and underflow bins are included in the  plots. The UV bound is applied. }
\end{figure}

 The unitarity violation (UV) bounds are  calculated using the form factor tool avaliable with VBFNLO~\cite{unitarityCalculator}. The bounds, which are shown in the left plots of Figure~\ref{fig:zz_cphiw_curve} and Figure~\ref{fig:zz_ft89_curve}, vary as a function of the coefficients
 of the higher-dimension operators.  One method of imposing unitarity preservation is to apply the UV bound as an upper bound
 on the invariant mass of the final-state bosons at truth-level. We use this method through this study. 
 As expected, when the coefficients approach zero the bound goes to infinity, since the SM amplitude respects
 unitarity at all energies.

 Application of the bound lowers the sensitivity of the search, especially when the coefficient is large. We present the $5 \sigma$-significance discovery values and 95\% CL limits with and without
  applying the UV bound in Table~\ref{tab:zzSignificance_noUVCutOff}. 
 The reconstructed 4-lepton invariant mass distributions are shown in Figure~\ref{fig:zz_ft8_ft9_cphiw_noUVCutOff} and Figure~\ref{fig:zz_ft8_ft9_cphiw} without and with the
 UV bound, respectively. The right plots of Figure~\ref{fig:zz_cphiw_curve} and Figure~\ref{fig:zz_ft89_curve} show the signal significance as a function of $f_{T8}/\Lambda^4$, $f_{T9}/\Lambda^4$ 
 and $c_{\phi W}/\Lambda^2$ without the UV bound applied. 

\begin{table}[h]
\centering
\begin{tabular}{c|c|c|c|c|c}
\hline\hline
\multirow{2}{*}{Parameter}                           & Luminosity & \multicolumn{2}{|c|}{14 TeV} & \multicolumn{2}{|c}{33 TeV}  \\
\cline{3-6}
                                                     & [fb$^{-1}$]             & $5 \sigma$ & 95\% CL              & $5 \sigma$ & 95\% CL                \\
\hline
\multirow{2}{*}{$c_{\phi W}/\Lambda^2$ [TeV$^{-2}$]} & 3000                                   & 16.2 (16.2) &  9.7 (9.7)  & 13.2 (13.2)  & 8.2 (8.2) \\
\cline{3-6}
                                                     & 300                                    & 31.3 (31.5) & 18.2 (18.3)  & 23.8 (23.8)  & 14.7 (14.7) \\
\hline
\multirow{2}{*}{$f_{T8}/\Lambda^{4}$ [TeV$^{-4}$]}   & 3000                                   & 2.9 (4.7)  & 1.7 (2.4)   & 1.6 (1.7)   & 1.0 (1.3) \\
\cline{3-6}
                                                     & 300                                    & 5.5 (8.4)  & 3.2 (5.3)   & 2.8 (2.3)   & 1.8 (1.8) \\
\hline
\multirow{2}{*}{$f_{T9}/\Lambda^{4}$ [TeV$^{-4}$]}   & 3000                                   & 5.7 (6.3)  & 3.9 (4.6)   & 3.8 (6.6)   & 2.5 (3.5) \\
\cline{3-6}
                                                     & 300                                    & 8.7 (9.0)  & 6.2 (6.7)   & 6.3 (10.1)   & 4.2 (8.2) \\
\hline\hline
\end{tabular}
\caption{In $pp \to ZZ + 2j \to 4 \ell + 2j$ processes, $5 \sigma$-significance discovery values and 95\% CL limits are shown
 for coefficients of high-dimension operators with 300 fb$^{-1}$/3000 fb$^{-1}$ of integrated luminosity.
 To show the impact of the UV bound, the corresponding results are shown in parentheses.}
\label{tab:zzSignificance_noUVCutOff}
\end{table}

\section{VBS $WZ \to \ell \nu \ell \ell$ }

We parameterize new physics in this channel using the dimension-8 operator
\begin{equation}
{\cal L}_{T,1} = \frac{f_{T1}}{\Lambda^4} {\rm Tr} [\hat{W}_{\alpha \nu} \hat{W}^{\mu \beta}] \times {\rm Tr} [\hat{W}_{\mu \beta}\hat{W}^{\alpha \nu}]
\label{eqn_FT1_lagrangian}
\end{equation}

and dimension-6 operator
\begin{equation}
{\cal L}_{\phi d} = \frac{c_{\phi d}}{\Lambda^2} \partial_{\mu} (\phi^\dagger \phi) \partial^{\mu} (\phi^\dagger \phi) \; \; .
\end{equation}
As shown in Table~\ref{tab:WZ_xsec_ILC}, the ${\cal L}_{T,1}$ operator produces the largest cross section enhancement for the VBS $WZ$ final state. The ${\cal L}_{\phi d} $ operator is chosen because
 its coefficient can be easily translated into a modification of the Higgs boson couplings to SM particles. Therefore this operator provides a way to tune the $\phi \to WW$ and $\phi \to ZZ$ couplings 
 and check the impact on unitarity violation in the VBS process if the Higgs couplings deviated from the SM.

The fully leptonic VBS $WZ \rightarrow \ell\nu\ell\ell $ channel can be reconstructed  by solving for the neutrino $p_z$ using the $W$ boson mass constraint. Its cross section is larger 
 than that of VBS $ZZ \to 4 \ell $ and it can probe some operators better than the latter process. The electron and muon decay channels are used in this study. 
 Mis-identification backgrounds are small in this channel, as shown in~\cite{atlas:WZconf} and therefore neglected in this sensitivity study.  
 The lepton from the $W$ boson decay must be identified in order to use the $W$ mass constraint. The procedure described in~\cite{ATLAS-Collaboration:1496527, ATLAS-Collaboration:1558703} is also used here.

Non-VBS $WZ$ production in association with  radiation of
two jets (SM WZ QCD) was simulated using {\sc madgraph}~\cite{Alwall:2011uj}.
{\sc madgraph} 5.1.5.10 was used to generate SM and non-SM VBS $WZ$  events.

\subsection{Event Selection}
After {\sc pythia} 6.4~\cite{pythia6} parton showering, additional detector effects are applied using {\sc delphes} 3.0.9~\cite{deFavereau:2013fsa} with the Snowmass parameterization~\cite{delphes1,
  delphes2,   delphes3}.
Events are considered VBS $WZ$ candidates provided they meet the following criteria:

\begin{itemize}
    \item Exactly three selected leptons (each with $p_T > 25$~GeV) which can be
        separated into an opposite sign, same flavor pair and an
        additional single lepton
    \item At least two selected jets with $p_T > 50$~GeV
    \item $m_{jj} > 1$ TeV, where $m_{jj}$ is the invariant mass of the
        two highest-$p_T$ selected jets
\end{itemize}

\subsection{Statistical Analysis}

\begin{figure}[h]
  \centering
  \includegraphics[width=0.49\textwidth]{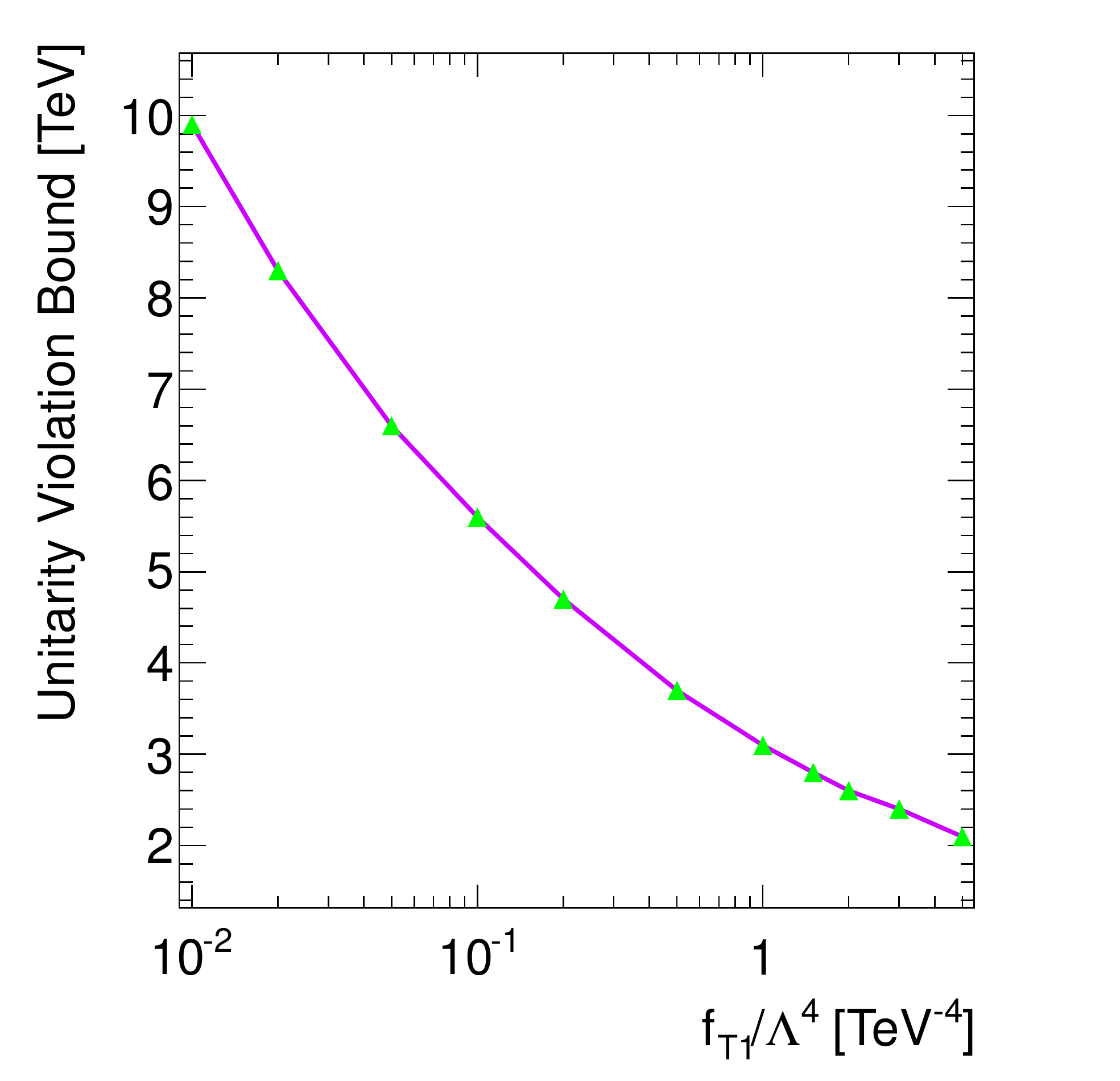}
  \includegraphics[width=0.49\textwidth]{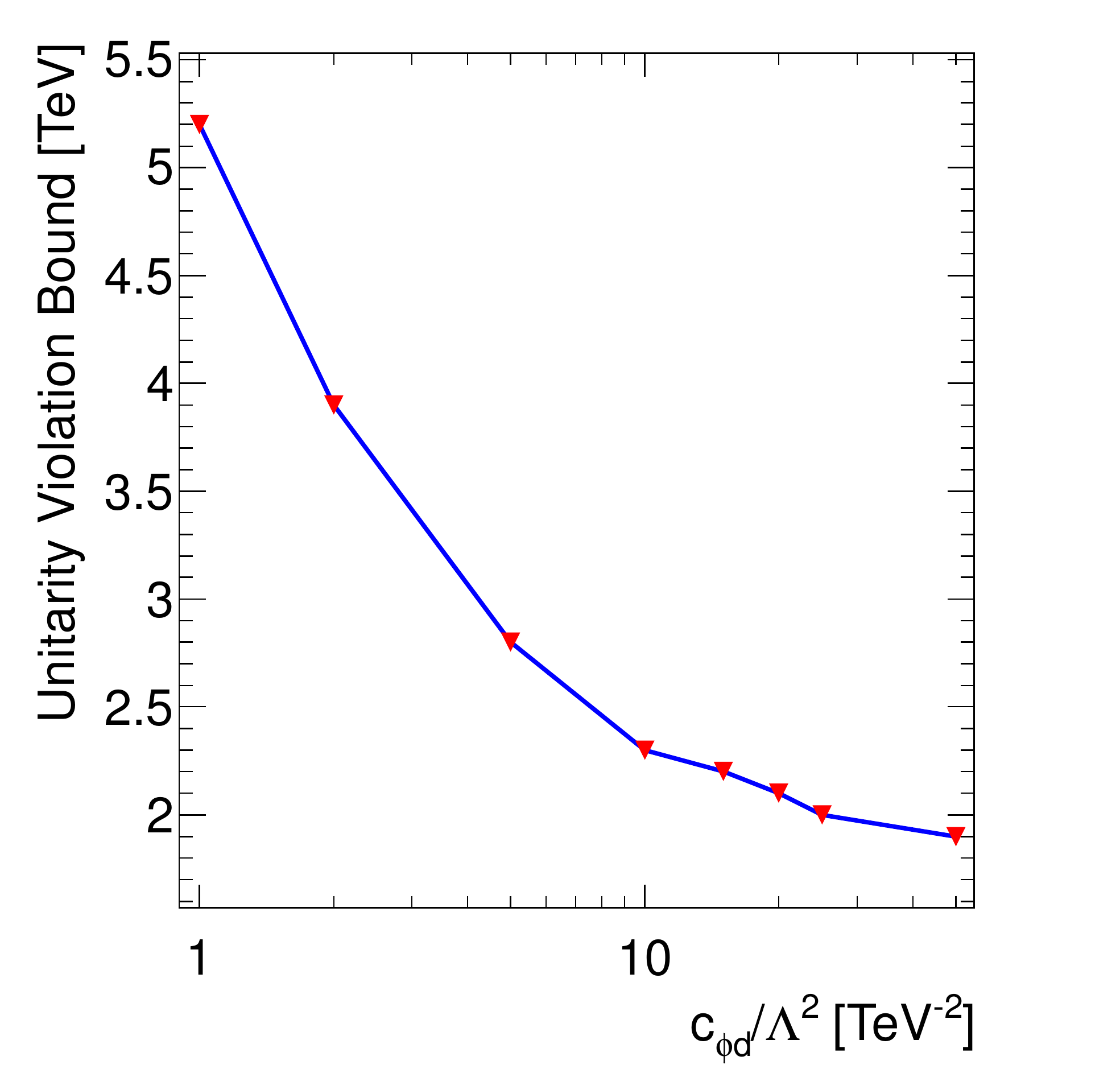}
  \caption{
The unitarity violation bounds for the  dimension-8 operator ${\cal L}_{T,1}$ (left) and the dimension-6 operator ${\cal L}_{\phi d}$ (right) are shown 
 as functions of operator coefficient values in $pp \to WZ + 2j \to \ell \nu \ell \ell + 2j$ processes.}
\label{fig:wz_UV_boundary}
\end{figure}

\begin{figure}[h]
  \centering
  \includegraphics[width=0.49\textwidth]{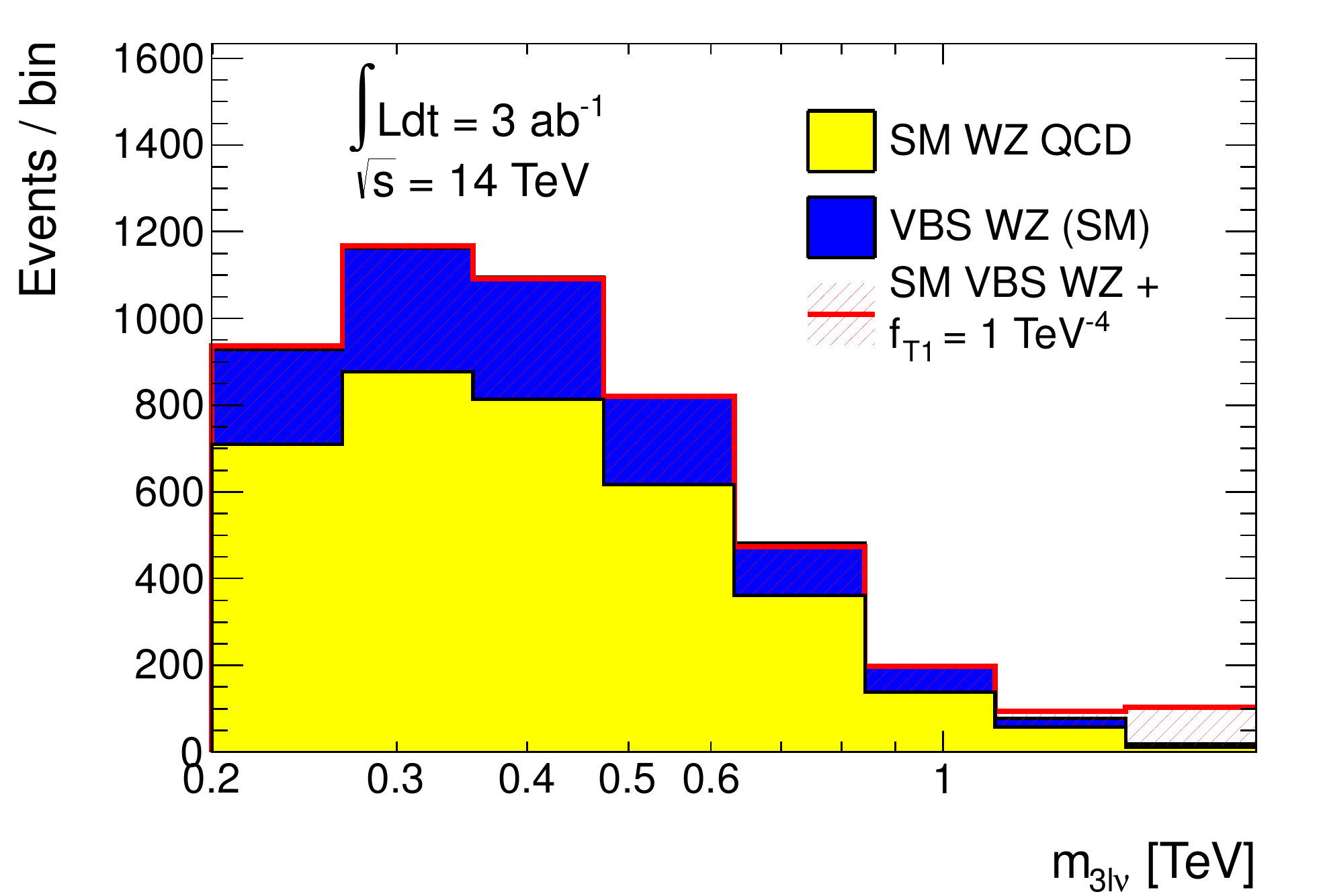}
  \includegraphics[width=0.49\textwidth]{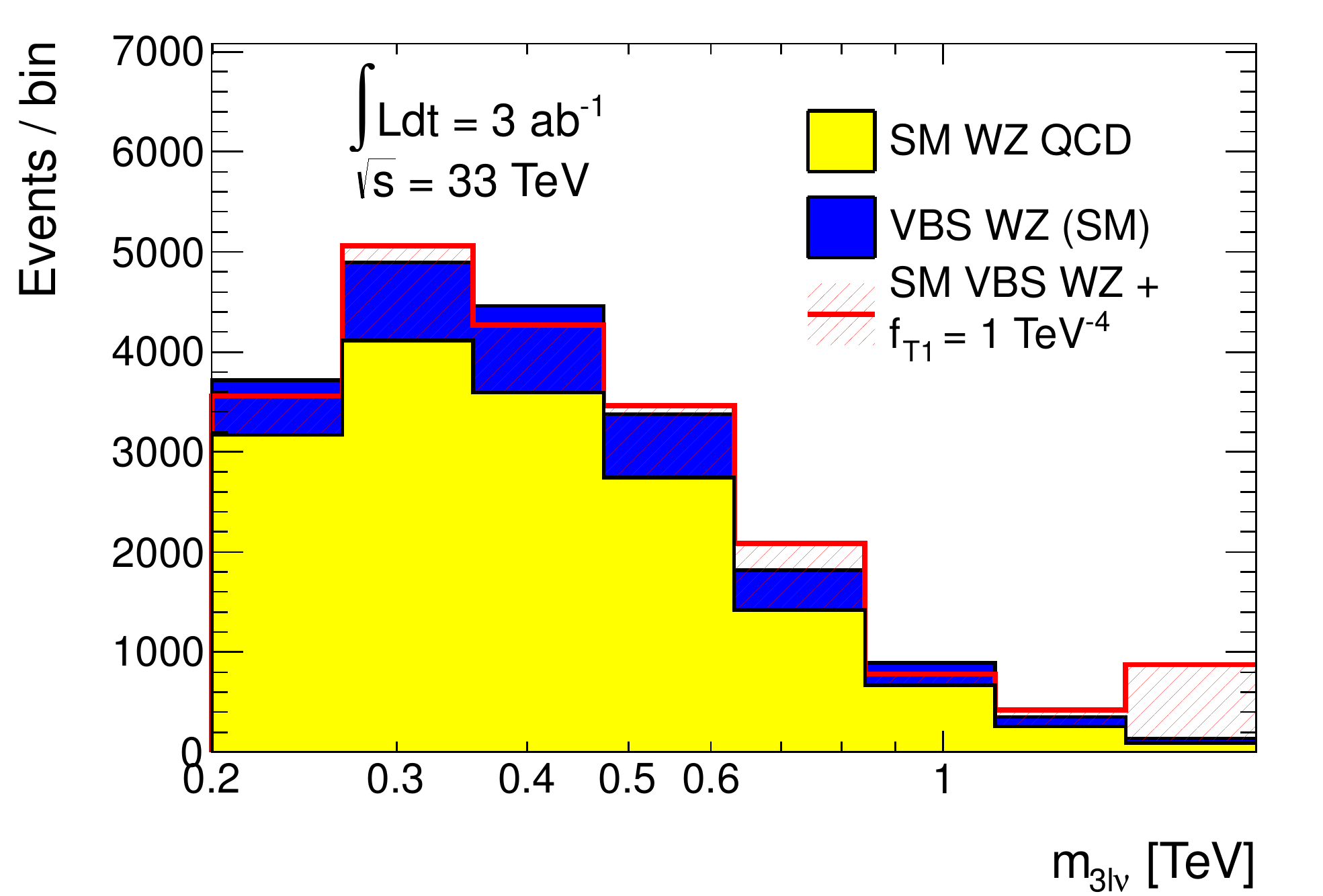} \\
  \includegraphics[width=0.49\textwidth]{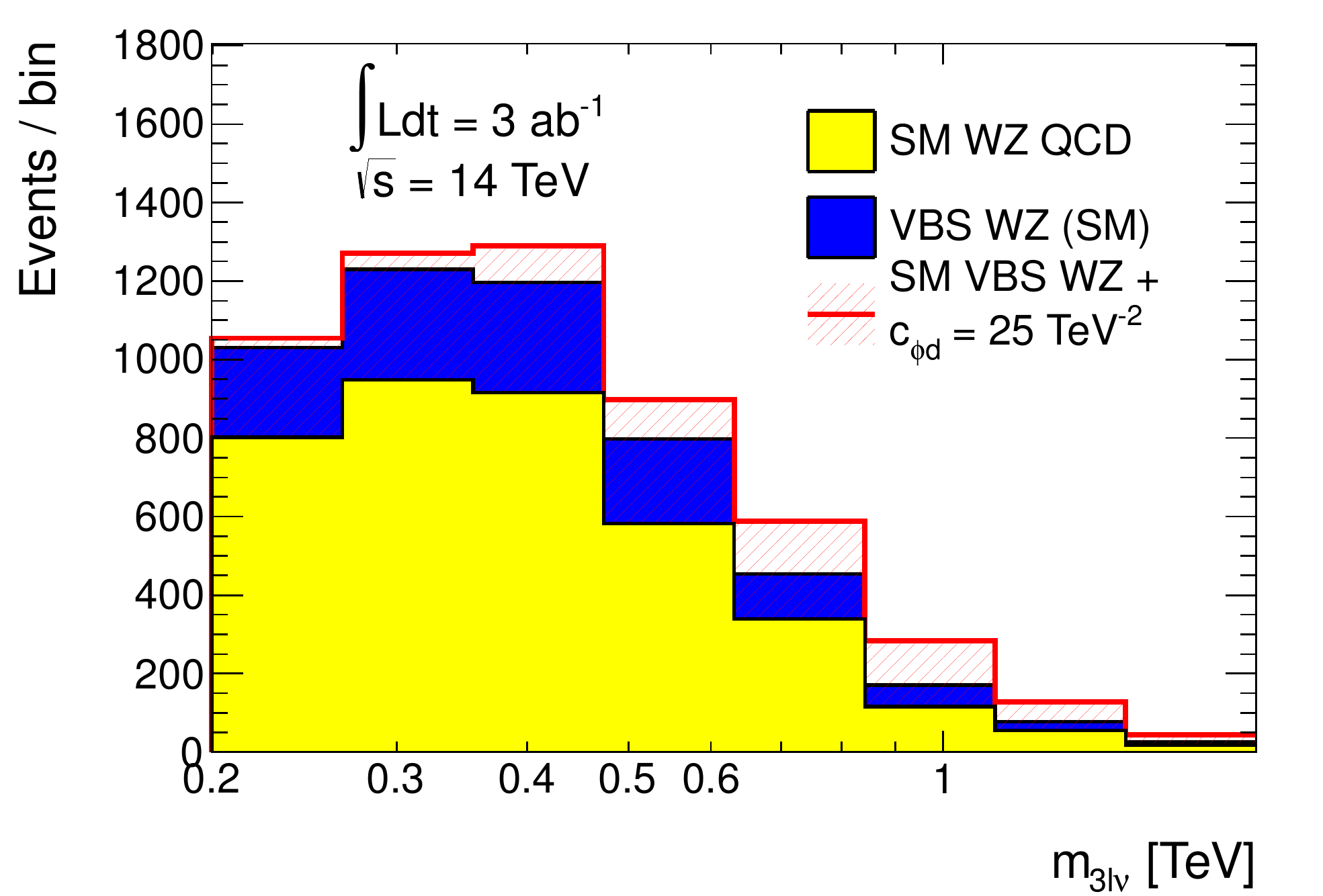}
  \includegraphics[width=0.49\textwidth]{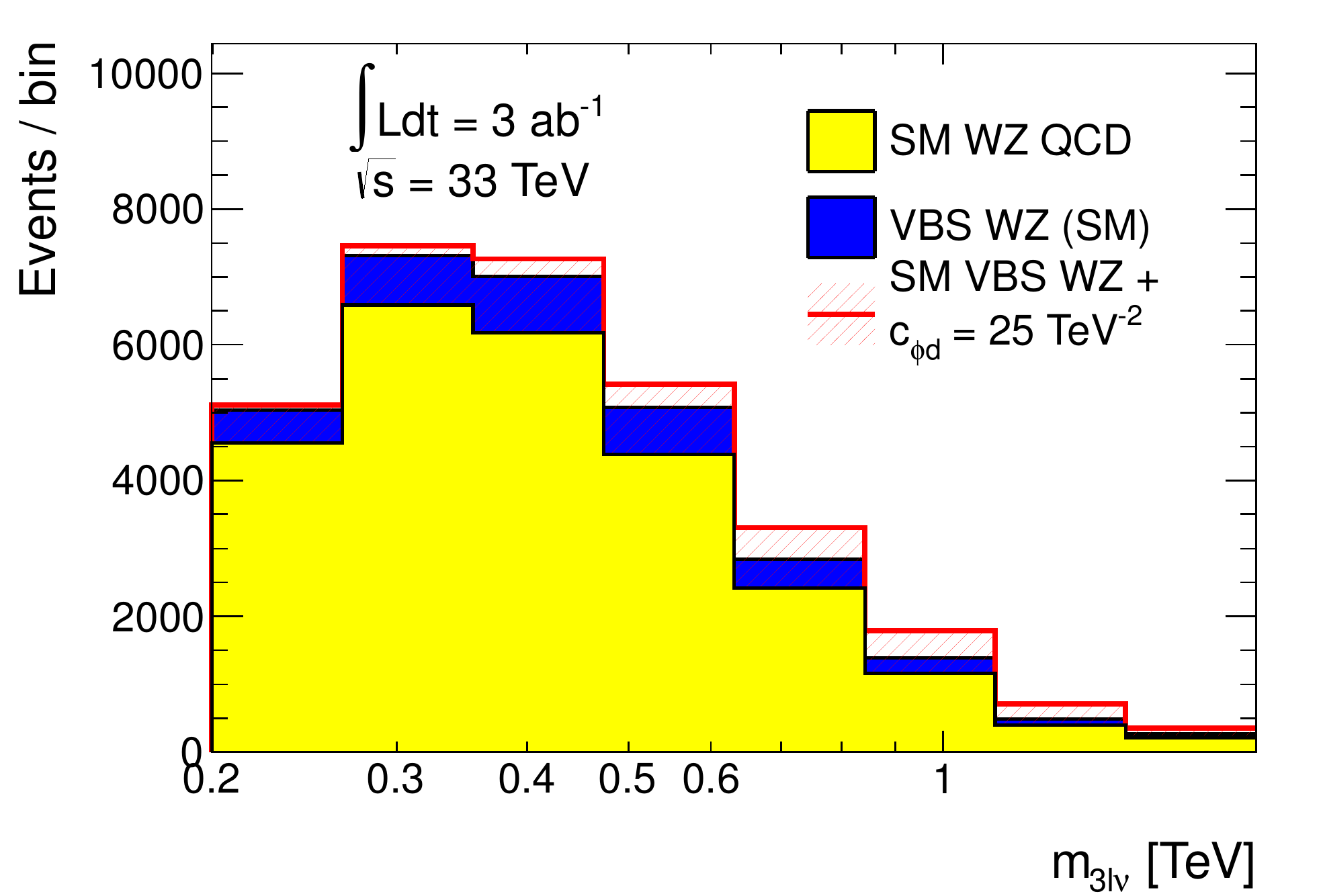}
  \caption{
In the $pp \to WZ + 2j \to \ell \nu \ell \ell + 2j$ channel, the reconstructed $WZ$ mass spectrum comparisons between Standard Model 
and dimension-8 operator coefficient $f_{T1}/\Lambda^4 = 1$ TeV$^{-4}$ (top) and dimension-6 operator coefficient $c_{\phi d}/\Lambda^2 = 25 $~TeV$^{-2}$ (bottom) 
are shown using the charged leptons and the neutrino solution after requiring $m_{jj} > 1 $~TeV at $\sqrt{s}=14$ TeV (left) and 33 TeV (right). 
The overflow and underflow bins are included in the  plots. The UV bounds have not been applied. }
\label{fig:wz_ft1_cphid_noUVCutOff}
\end{figure}

\begin{figure}[h]
  \centering
  \includegraphics[width=0.49\textwidth]{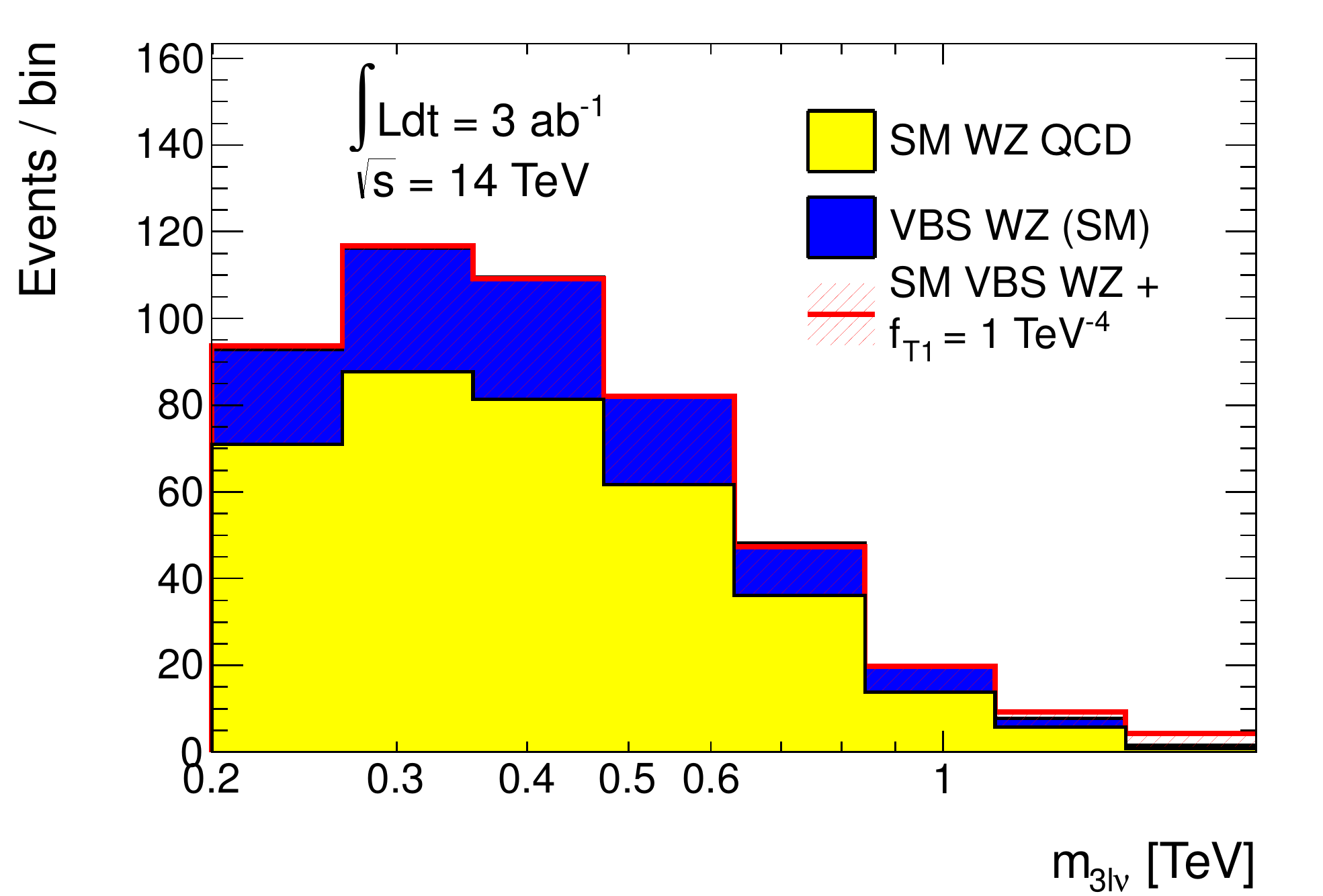} 
  \includegraphics[width=0.49\textwidth]{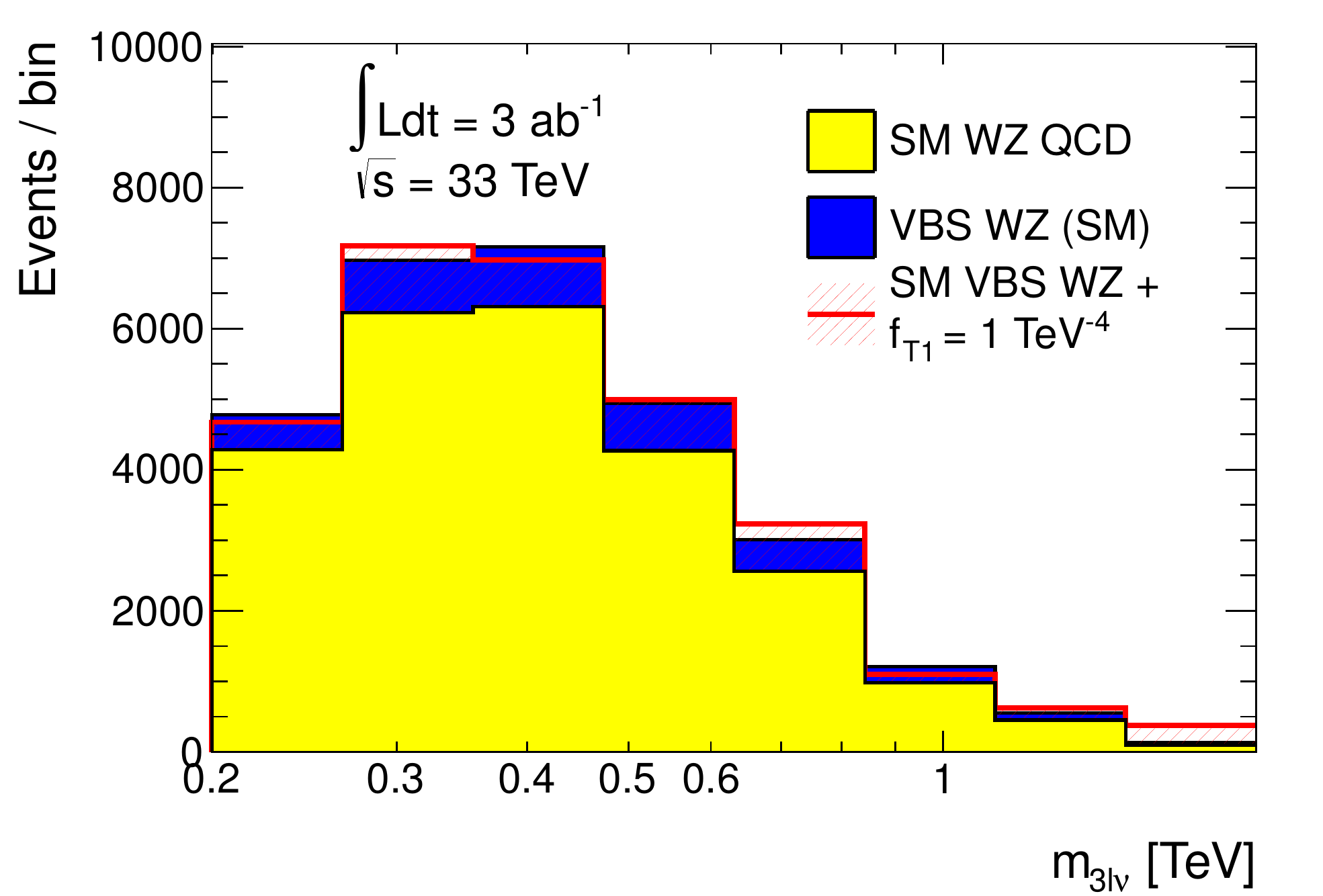} \\
  \includegraphics[width=0.49\textwidth]{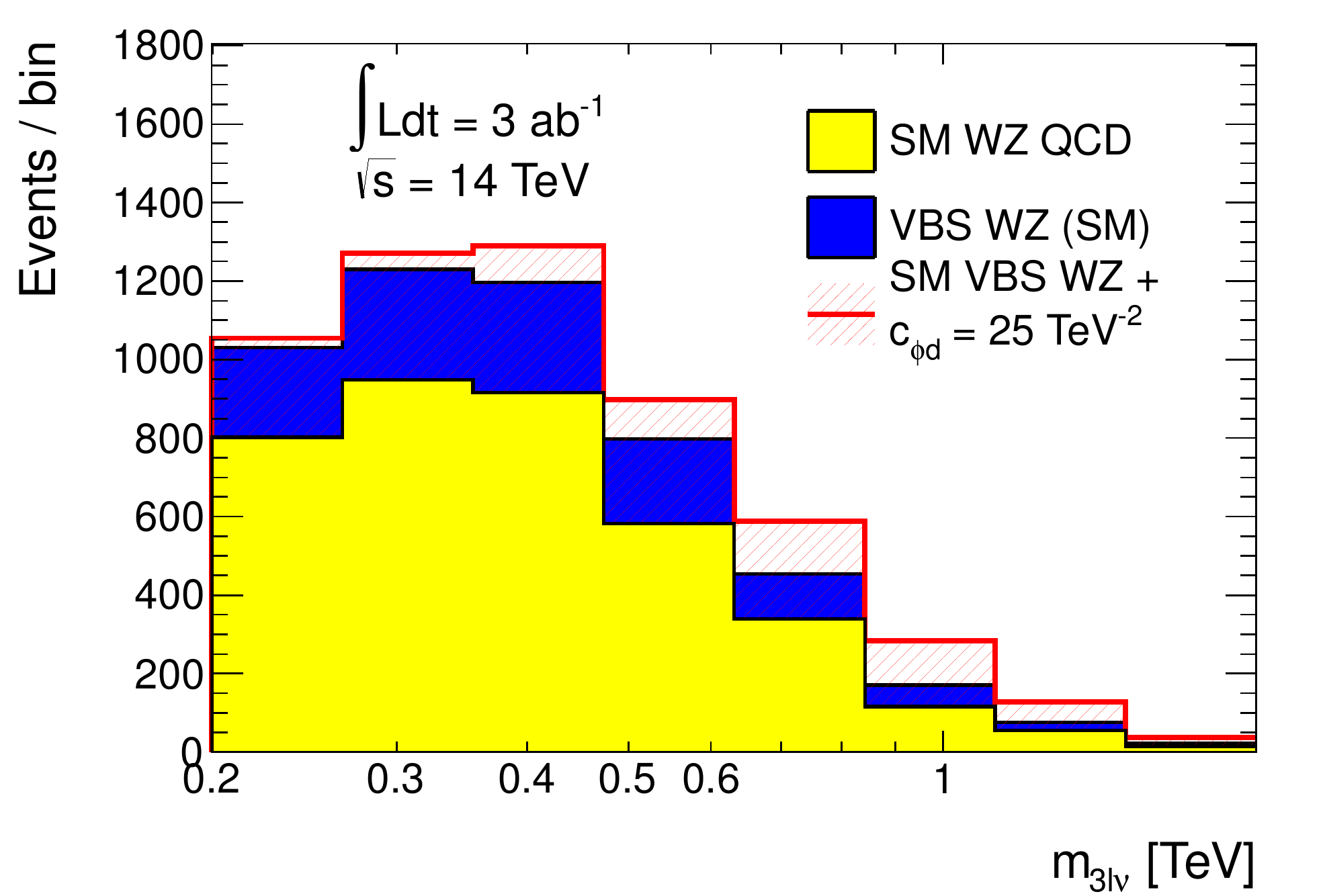}
  \includegraphics[width=0.49\textwidth]{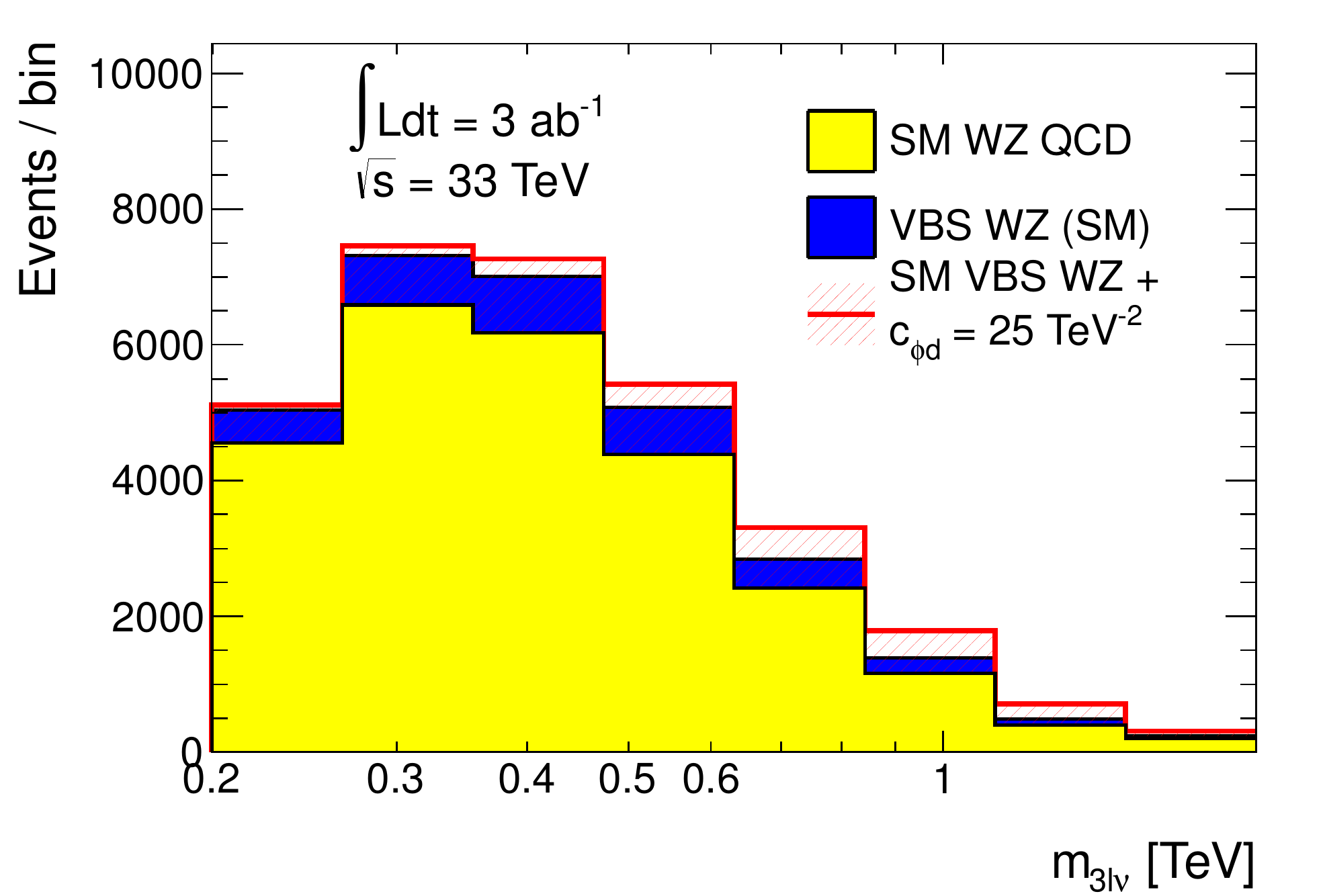}
  \caption{
In the $pp \to WZ + 2j \to \ell \nu \ell \ell + 2j$ channel, the reconstructed $WZ$ mass spectrum comparisons between dimension-8 operator coefficient 
$f_{T1}/\Lambda^4 = 1$~TeV$^{-4}$ (top), dimension-6 operator coefficient $c_{\phi d}/\Lambda^2 = 25$~TeV$^{-2}$ (bottom) and Standard Model are shown 
using the charged leptons and the neutrino solution after requiring $m_{jj} > 1 $~TeV at $\sqrt{s}=14$ TeV (left) and 33 TeV (right). 
The overflow and underflow bins are included in the plots. The UV bounds have been applied. }
\label{fig:wz_ft1_cphid} 
\end{figure}

\begin{figure}[h]
  \centering
  \includegraphics[width=0.49\textwidth]{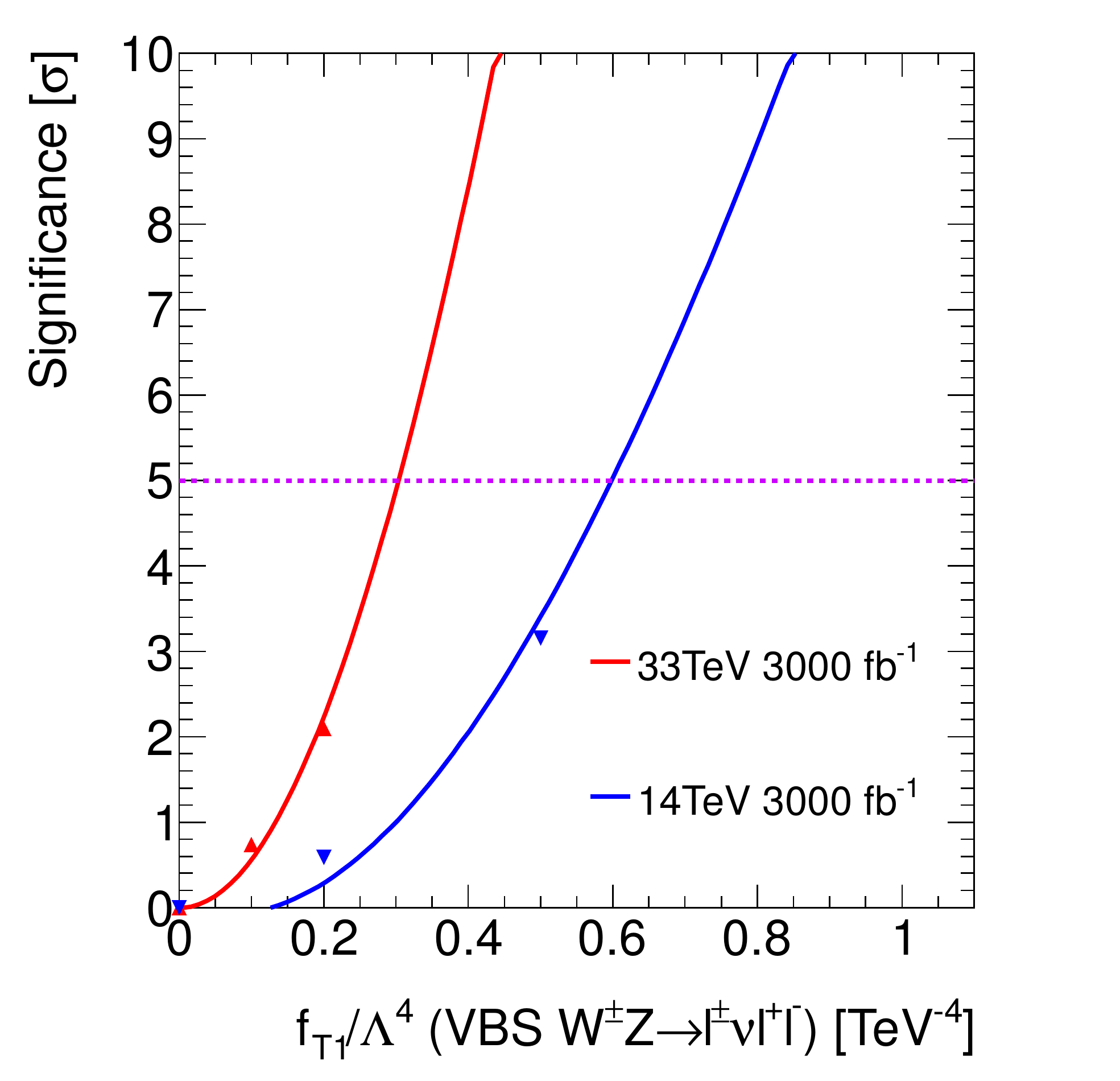}
  \includegraphics[width=0.49\textwidth]{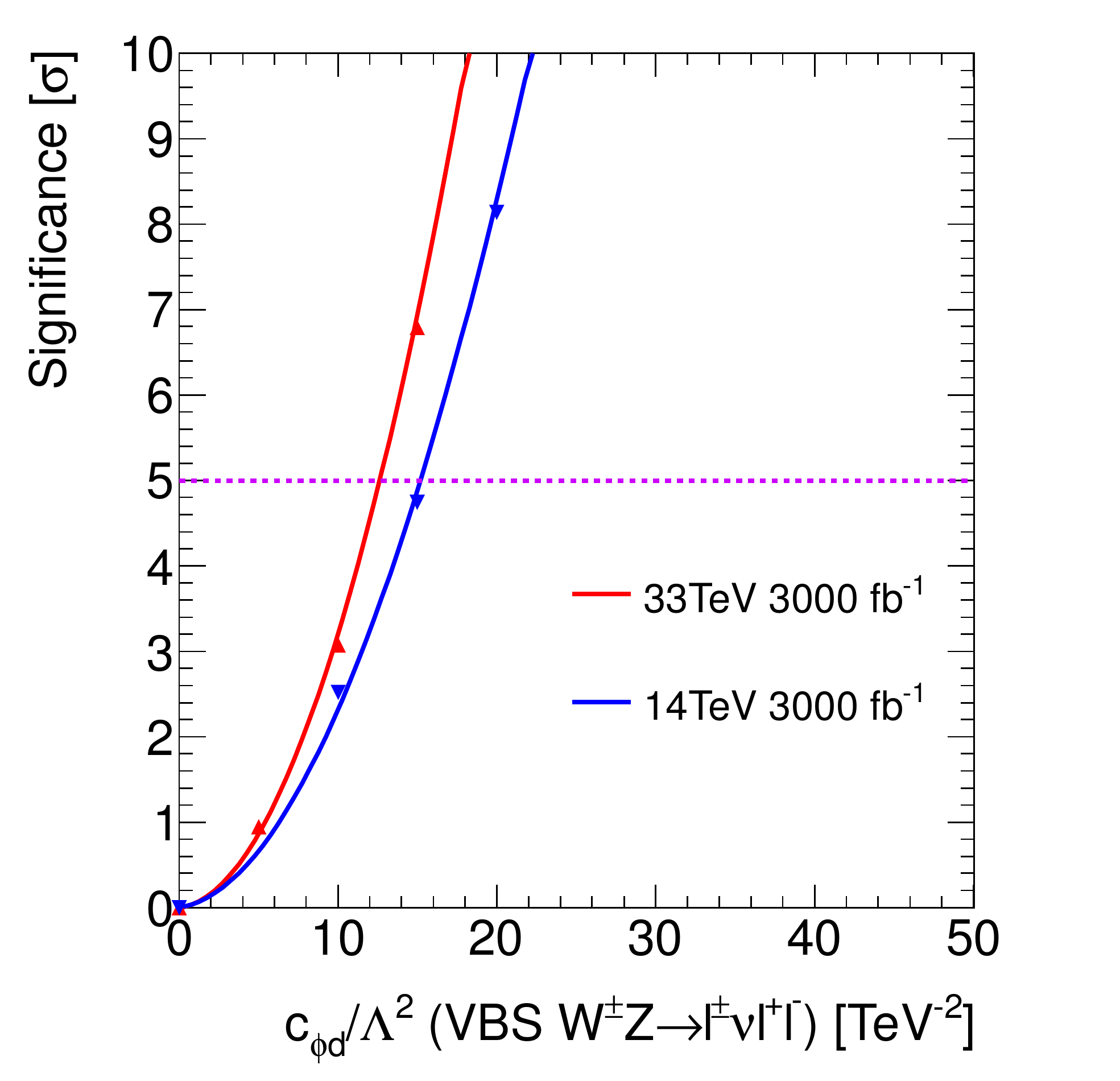}
  \caption{
$pp \to WZ + 2j \to \ell \nu \ell \ell + 2j$ signal significance as a function of $f_{T1}/\Lambda^4$ (left) and $c_{\phi d}/\Lambda^2$ (right) calculated from reconstructed $WZ$ mass spectra at $\sqrt{s}=14$ TeV and 33 TeV.
 The UV bounds have not been applied. }
\label{fig:wz_sig_noUVCutOff}
\end{figure}

The statistical analysis is identical to that employed in Sec.~\ref{zzjjStats}. 
As with the VBS $ZZ$ channel, we present sensitivity studies with and without the UV bound applied, to show the impact of the UV bound. The UV bounds for different operators are shown in Figure~\ref{fig:wz_UV_boundary}. 
  We present the $5 \sigma$-significance discovery values and 95\% CL limits  in Table~\ref{tab:wzSignificance_noUVCutOff}. Figure~\ref{fig:wz_sig_noUVCutOff} shows the signal significance as a function of $f_{T1}/\Lambda^4$ and $c_{\phi d}/\Lambda^2$ without the UV bound and the corresponding reconstructed 4-lepton invariant mass distributions are shown in Figure~\ref{fig:wz_ft1_cphid_noUVCutOff}. The same distributions with the UV bounds are shown in Figure~\ref{fig:wz_ft1_cphid}.

\begin{table}[h]
\centering
\begin{tabular}{c|c|c|c|c|c}
\hline\hline
\multirow{2}{*}{Parameter}                           & Luminosity  & \multicolumn{2}{|c|}{14 TeV} & \multicolumn{2}{|c}{33 TeV}  \\
\cline{3-6}
                                                     & [fb$^{-1}$]   & $5 \sigma$      & 95\% CL          & $5 \sigma$       & 95\% CL                \\
\hline
\multirow{2}{*}{$c_{\phi d}/\Lambda^2$ [TeV$^{-2}$]} & 3000                                   & 15.2 (15.2) &  9.1 (9.1)  & 12.6 (12.7)  & 7.7 (7.7) \\
\cline{2-6}
                                                     & 300                                    & 28.5 (28.7) &  17.1 (17.1) & 23.1 (23.3)  & 14.1 (14.2) \\
\hline
\multirow{2}{*}{$f_{T1}/\Lambda^{4}$ [TeV$^{-4}$]}   & 3000                                   & 0.6 (0.9)  & 0.4 (0.5)   & 0.3 (0.6)   & 0.2 (0.3) \\
\cline{2-6}
                                                     & 300                                    & 1.1 (1.6)  & 0.7 (1.0)   & 0.6 (0.9)   & 0.3 (0.6) \\
\hline\hline
\end{tabular}
\caption{In $pp \to WZ + 2j \to \ell \nu \ell \ell + 2j$ processes, $5 \sigma$-significance discovery values and 95\% CL limits are shown 
 for coefficients of higher-dimension operators with 300 fb$^{-1}$/3000 fb$^{-1}$ of integrated luminosity at
 $\sqrt{s} = 14$ TeV and $\sqrt{s} = 33$ TeV. The results obtained after applying the UV bounds are shown in parentheses.}
\label{tab:wzSignificance_noUVCutOff}
\end{table}

\section{VBS $W^{\pm}W^{\pm} \to \ell \nu \ell \nu$ }

The sensitivity to new physics was examined in the $pp \to W^{\pm} W^{\pm} + 2j \to \ell^{\pm} \nu \ell^{\pm} \nu + 2j$ (ssWW) channel where $\ell$ is an electron or muon. The dimension-8 operator
  ${\cal L}_{T,1}$ as shown in 
 Eqn.~\ref{eqn_FT1_lagrangian}, was used to probe deviations from SM predictions. This operator causes  the strongest enhancement of the ssWW VBS cross section
 relative to the SM value. Table~\ref{tab:ssWW_operators} shows the relative cross section yield due to each operator relevant to ssWW production.

\begin{table}[h]
\centering
\begin{tabular}{c|c}
\hline
operator & cross section ratio \\
\hline
 ${\cal L}_{S,0}$ &  1.1  \\
 ${\cal L}_{S,1}$ &  1.0  \\
 ${\cal L}_{M,0}$ &  1.3  \\
 ${\cal L}_{M,1}$ &  1.1  \\
 ${\cal L}_{T,0}$ &  33  \\
 ${\cal L}_{T,1}$ &  150  \\
 ${\cal L}_{T,2}$ &  17  \\
\hline
\end{tabular}
\caption{Ratio of cross sections due to each dimension-8 operator with respect to the Standard Model value,  for operator coefficients $f/\Lambda^4 =10$~TeV$^{-4}$ for VBS ssWW production
 at the LHC with $\sqrt{s} = 14$~TeV.
  The SM cross section is 9 fb.}
\label{tab:ssWW_operators}
\end{table}

A range of coefficient values with sensitivities near the  5$\sigma$ significance level were studied for 14 TeV and 100 TeV $pp$ machine scenarios for various pileup conditions. The effect of a UV bound was also considered.

\subsection{Monte Carlo Predictions}

The samples were generated using {\sc madgraph} version 5.1.5.11~\cite{Alwall:2011uj} 
 for backgrounds, SM VBS, and new physics. The main backgrounds in ssWW VBS production are ssWW QCD diagrams, $WZ$, $W \gamma$, and mis-identification backgrounds including
 charge mis-identification. Following~\cite{ATLAS-Collaboration:1558703}, 
 the $W \gamma$ and mis-identification backgrounds were assumed to have the same shape as the $WZ$ background and therefore, the $WZ$ background was scaled appropriately (by a factor of 2)
  to account for these other backgrounds as well. 

\subsection{Event Selection}
After {\sc pythia} 6.4~\cite{pythia6} parton showering, additional detector effects are applied using {\sc delphes} 3.0.9~\cite{deFavereau:2013fsa} with the Snowmass
  parameterization~\cite{delphes1, delphes2, delphes3}.
The analysis selection of the VBS ssWW candidates were given by the following criteria:
\begin{itemize}
    \item Exactly two leptons of same sign each with $p_T > 25$~GeV 
    \item At least two jets with $p_T > 50$~GeV
    \item $m_{jj} > 1$ TeV, where $m_{jj}$ is the invariant mass of the two highest-$p_T$ jets
\end{itemize}
The high jet $p_T$ cut helps protect against pileup jets while the $m_{jj}$ cut strongly selects for the VBS signal region. 

\begin{figure}[h]
  \centering
 \includegraphics[width=0.6\textwidth]{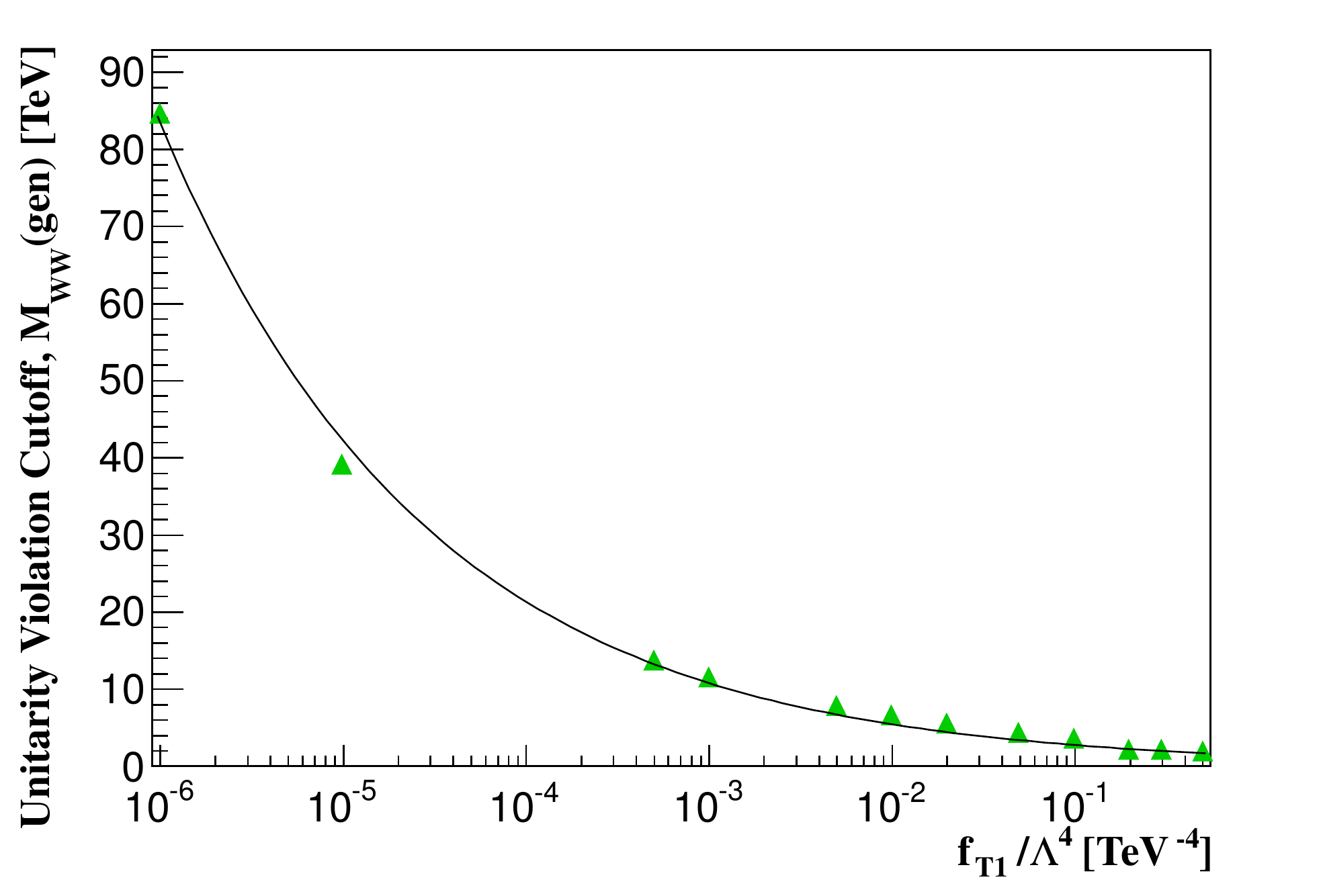} 
  \caption{The Unitarity Violation (UV) cut-off values for a given $f_{T1}/\Lambda^{4}$ value for the VBS  ssWW
 final state.}
  \label{fig:ssWW_UV_limits} 
\end{figure}

 The UV bound for each $f_{T1}/\Lambda^{4}$ value used for this channel
  is shown in Fig.~\ref{fig:ssWW_UV_limits}. 

\subsection{Statistical Analysis}
Following~\cite{ATLAS-Collaboration:1496527,ATLAS-Collaboration:1558703}, 
 the 4-body invariant mass of the two leading jets and the two leptons, $m_{jjll}$, was used to discriminate new physics from the SM. The statistical analysis approach is identical to that in Sec.~\ref{zzjjStats}. 

\begin{figure}[h]
  \centering
  \includegraphics[width=0.49\textwidth]{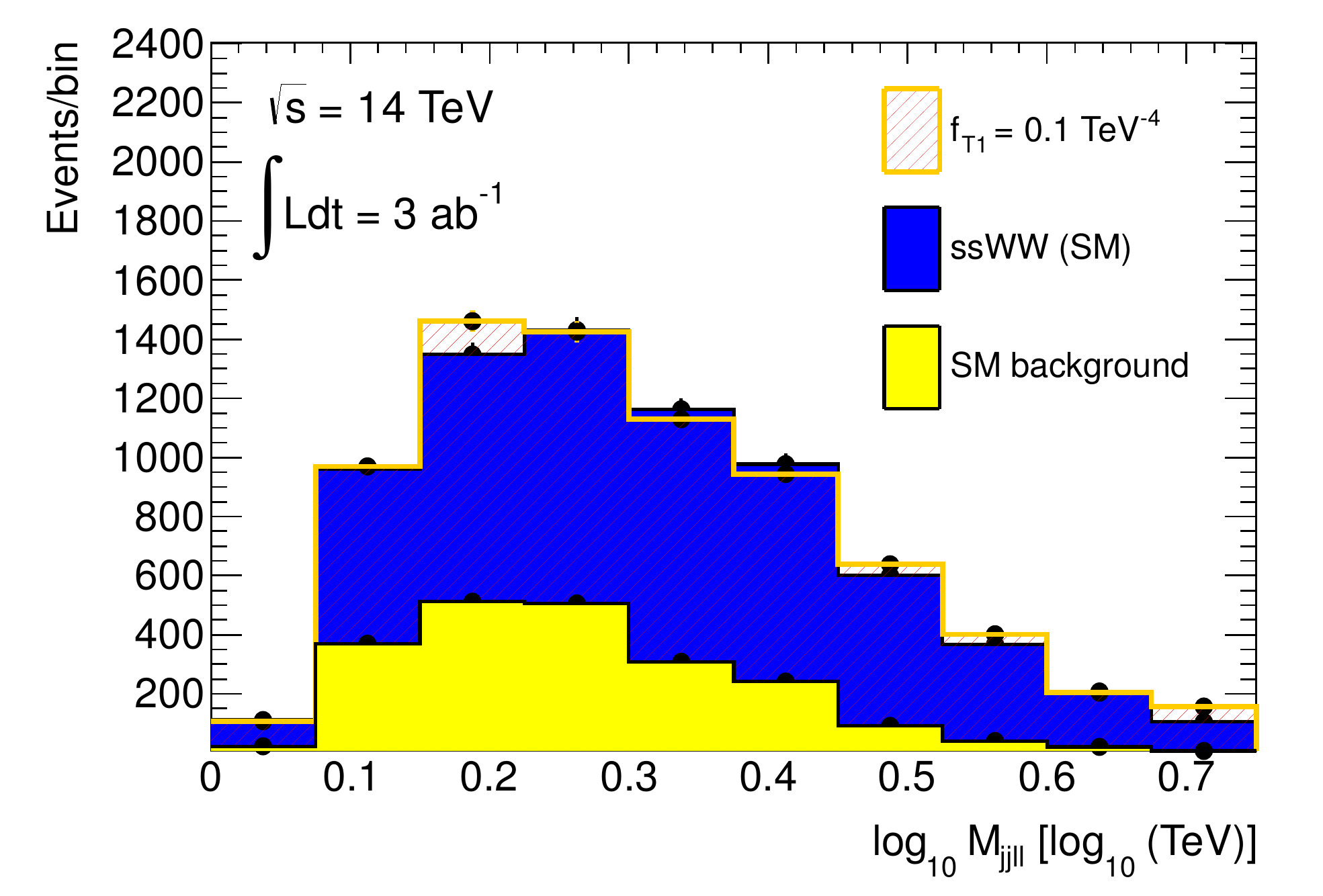} 
  \includegraphics[width=0.49\textwidth]{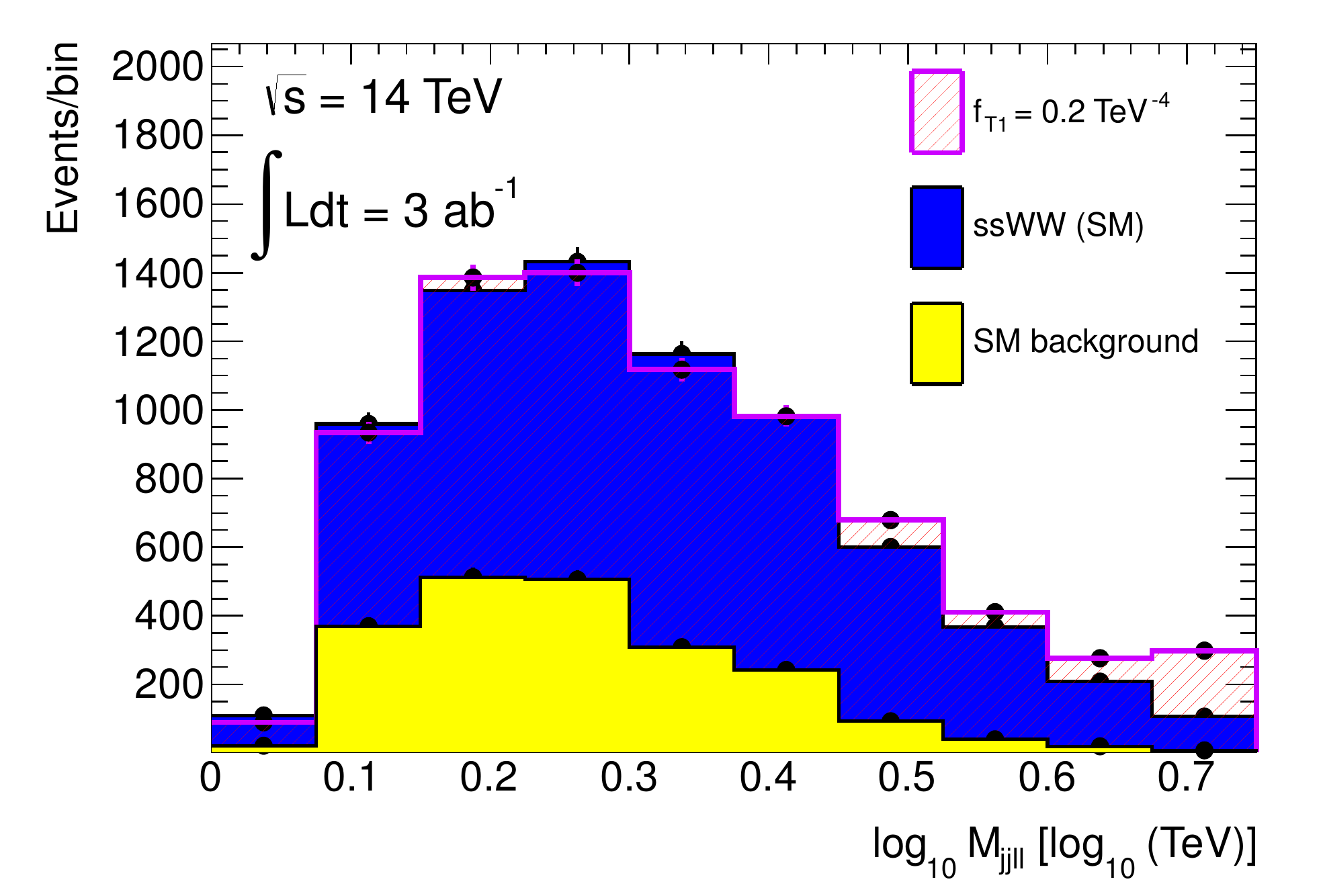} 
  \caption{The invariant mass of the 4-body $m_{jjll}$ system in ssWW events is shown for $f_{T1}/\Lambda^{4}$ equal to 0.1 TeV$^{-4}$ corresponding to a significance of 4.2$\sigma$ (left) and $f_{T1}/\Lambda^{4}$ = 0.2 TeV$^{-4}$ with 17$\sigma$ significance  (right) for $\sqrt{s} = 14$~TeV, 140 pileup events per crossing, without the UV cut-off applied, and 3000 fb$^{-1}$ scenario.}
  \label{fig:ssWW_Mjjll_14TeV} 
\end{figure}

\begin{figure}[h]
  \centering
  \includegraphics[width=0.49\textwidth]{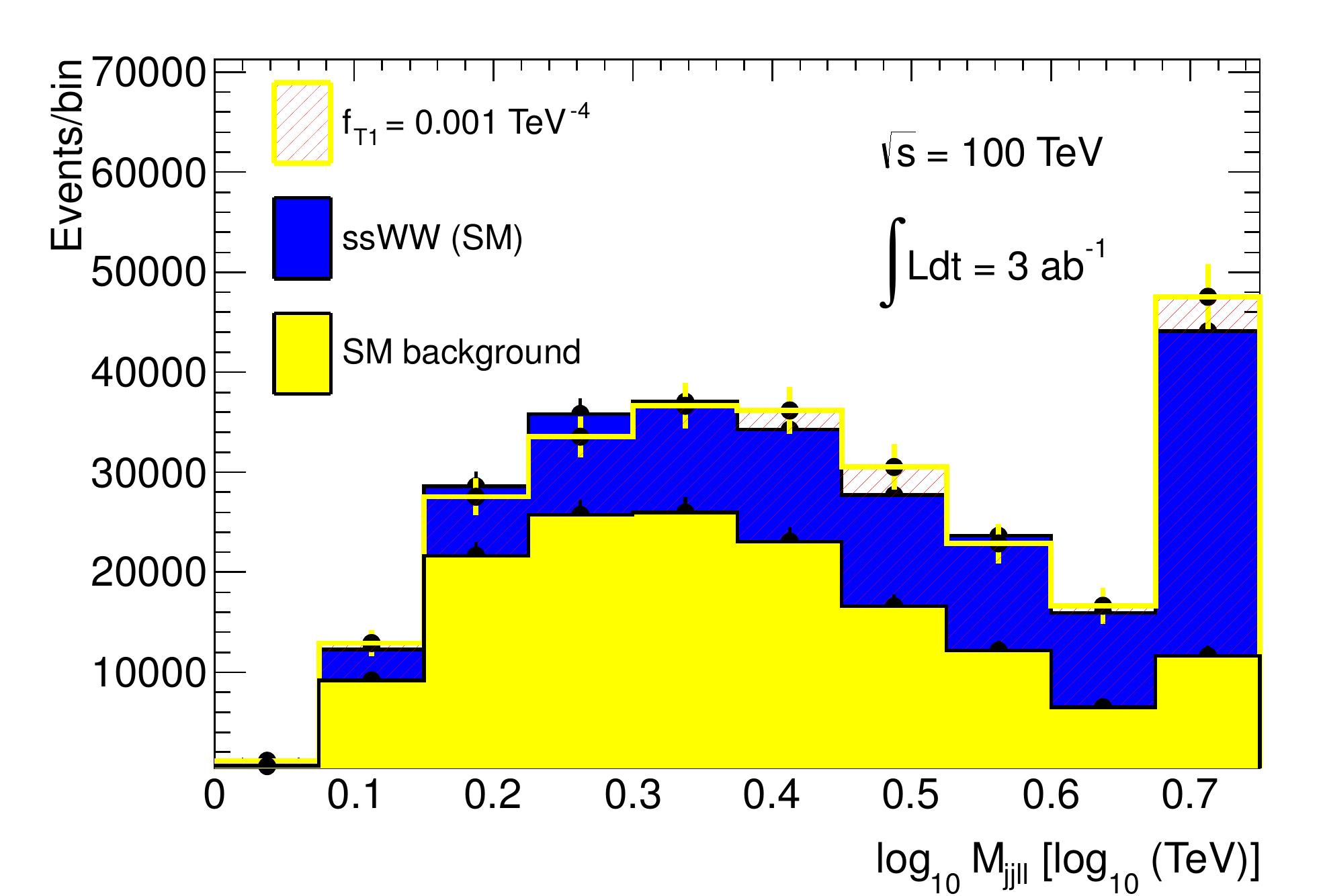} 
  \caption{In ssWW events, $m_{jjll}$ for $f_{T1}/\Lambda^{4}$ =  0.001 corresponding to a 4$\sigma$ significance for the case of a  $\sqrt{s} = 100$~TeV $pp$ machine with 263 pileup, without the UV cut-off applied, at 3000 fb$^{-1}$ is shown.}
  \label{fig:ssWW_Mjjll_100TeV} 
\end{figure}

Figure~\ref{fig:ssWW_Mjjll_14TeV} compares the shape of the $m_{jjll}$ distributions for the SM  and two values of $f_{T1}/\Lambda^{4}$ (0.1~TeV$^{-4}$ and 0.2~TeV$^{-4}$, respectively) 
 at 14 TeV. Increasing the anomalous quartic coupling increases the event rate
 at the high end of the $m_{jjll}$ spectrum. 
A similar plot for 100 TeV is shown in Figure~\ref{fig:ssWW_Mjjll_100TeV} for $f_{T1}/\Lambda^{4} =  0.001$~TeV$^{-4}$.

\begin{figure}[h]
  \centering
  \includegraphics[width=0.49\textwidth]{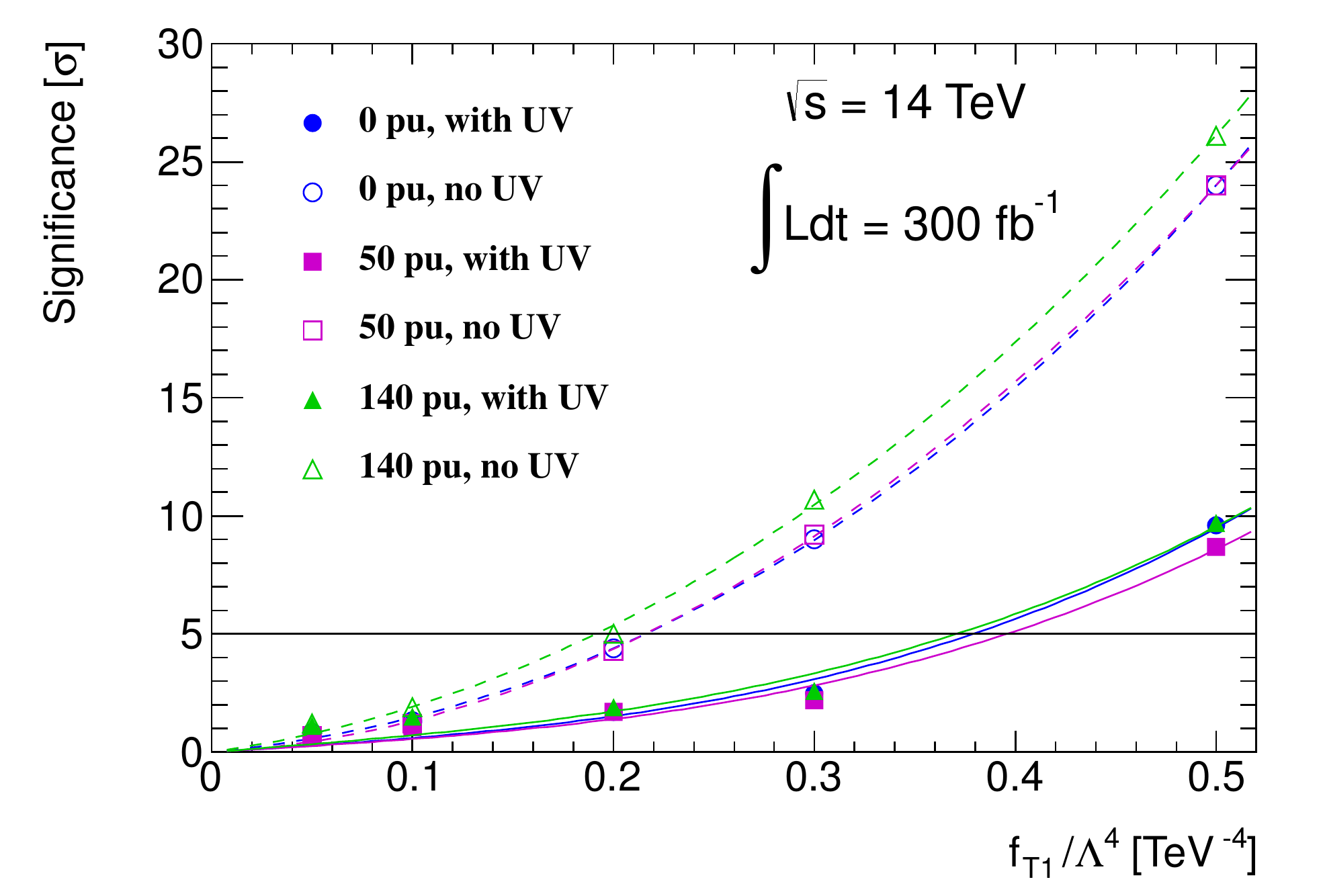} 
  \includegraphics[width=0.49\textwidth]{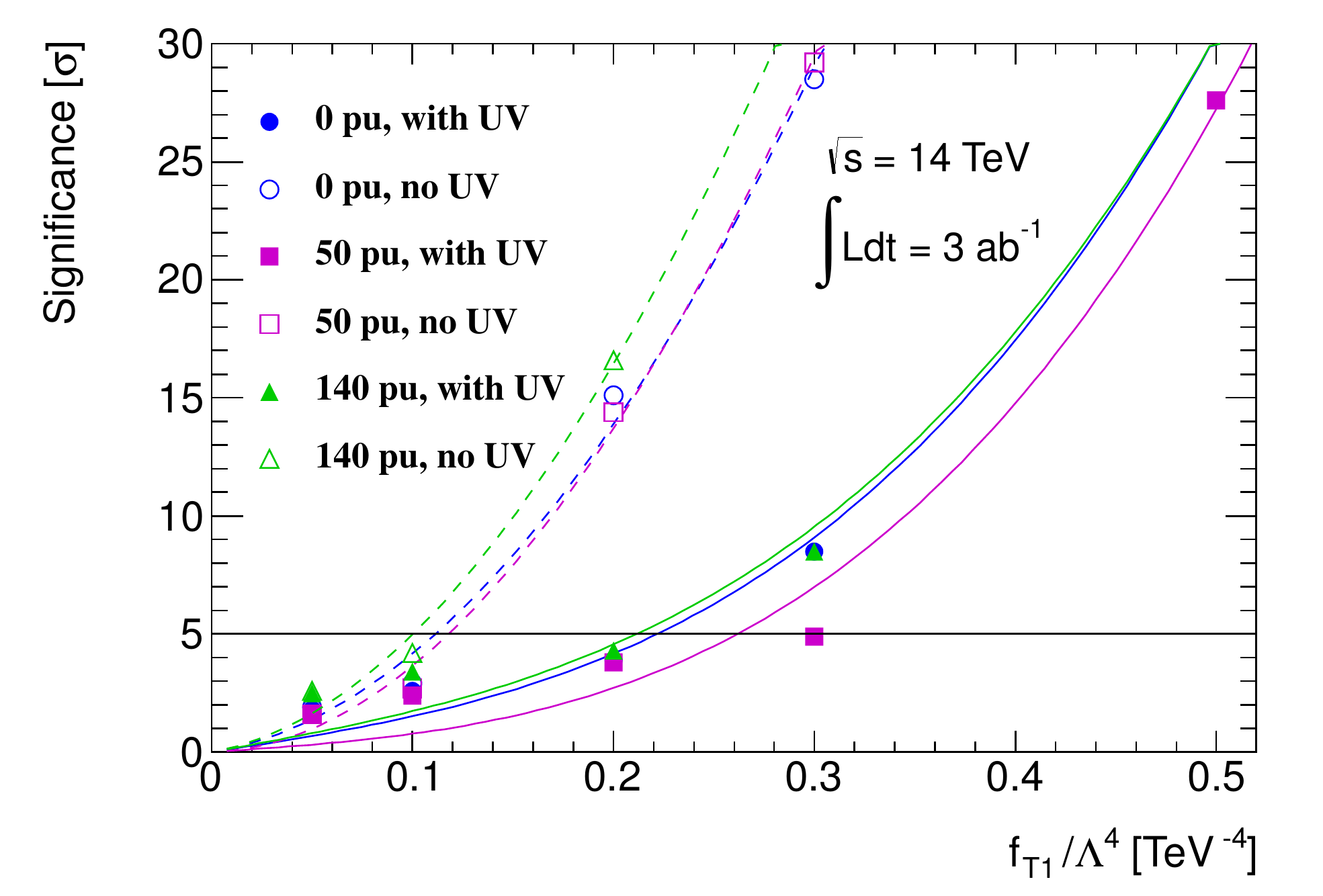} 
  \includegraphics[width=0.49\textwidth]{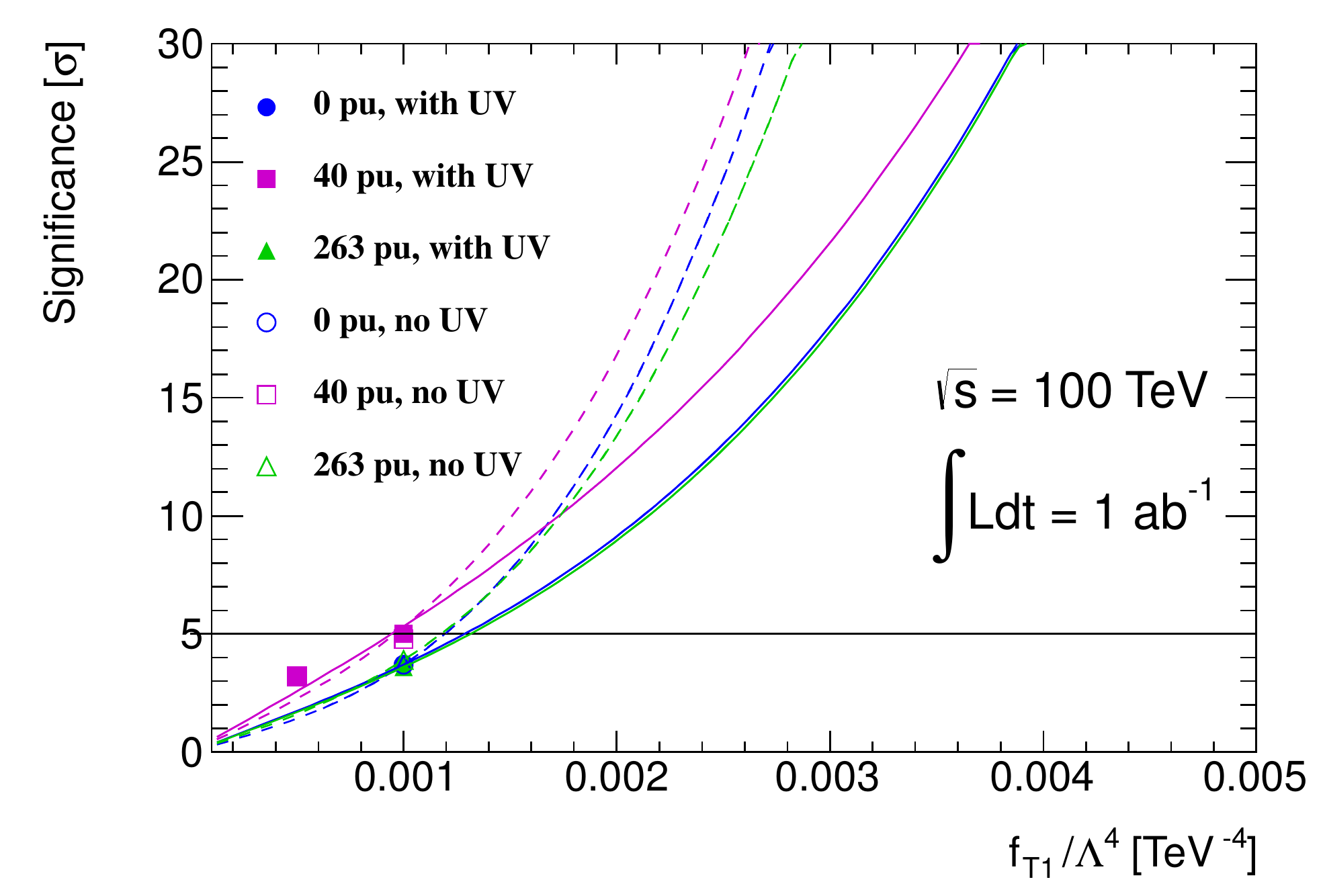} 
  \includegraphics[width=0.49\textwidth]{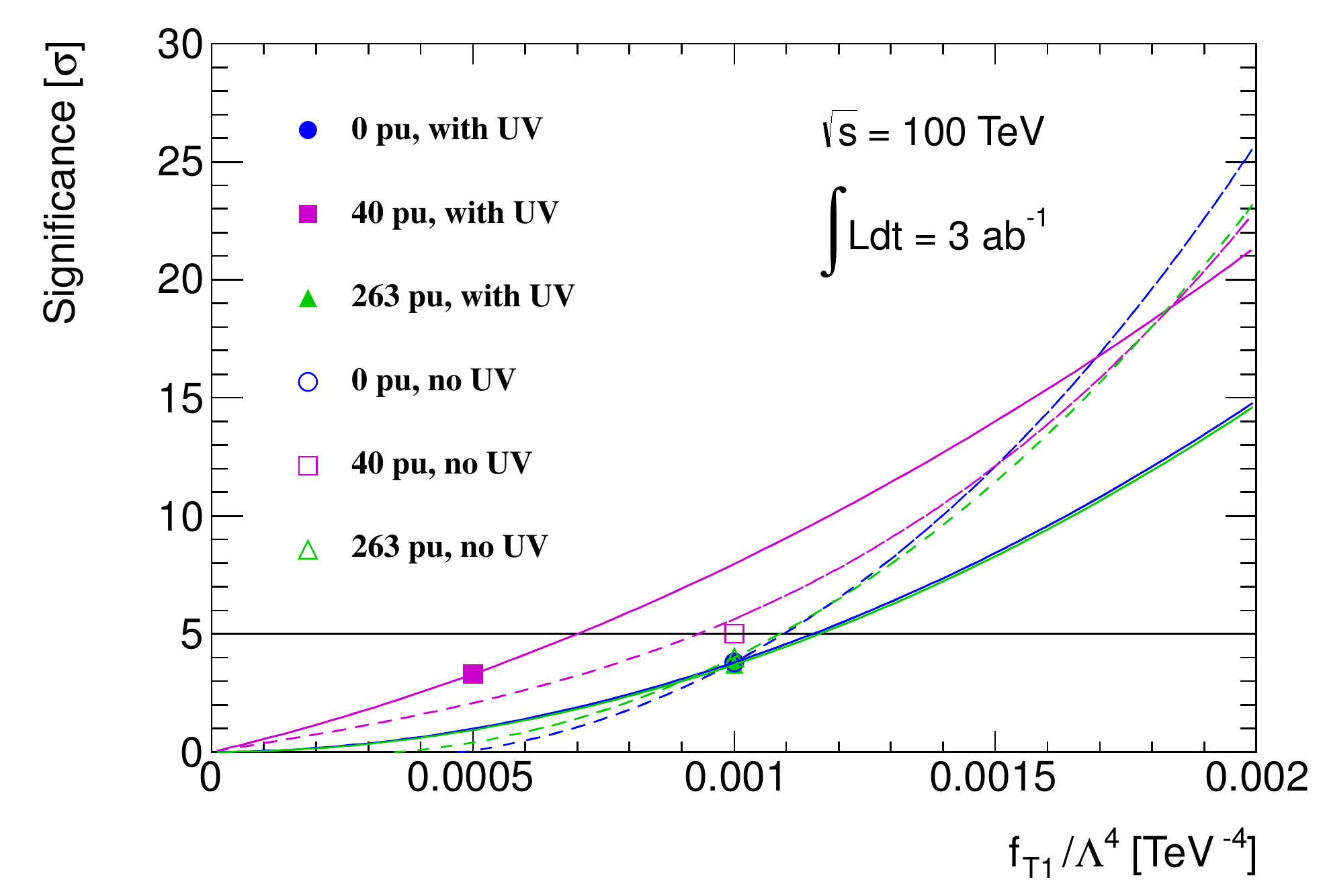} 
  \caption{The significance trends for each beam energy and luminosity are shown for the various pileup scenarios. The fits are done with a 3rd order polynomial (curves). For the each set of machine conditions, each pileup scenario was considered with (solid points/solid curves) and without (open points/dashed curves) the UV bound. The top row displays the $\sqrt{s} = 14$~TeV cases at 300 fb$^{-1}$ (left) and 3000 fb$^{-1}$ (right). The bottom row shows 
 the $\sqrt{s} = 100$~TeV cases at 1000 fb$^{-1}$ (left) and 3000 fb$^{-1}$ (right). }
  \label{fig:ssWW_Significance} 
\end{figure}

The different pileup scenarios as well as the effect of the UV bound are shown in Figure~\ref{fig:ssWW_Significance} for different $pp$ collider energies. The pileup has a small, but non-negligible effect whereas the UV bound has a larger effect. The significance values reported in Table~\ref{tab:ssWWSignificance_scenarios} are without the UV bound applied and with the UV cut-off in parentheses. 

A summary of the 5$\sigma$ significance and the 95\% Confidence Level (CL) for each machine scenario is listed in Table~\ref{tab:ssWWSignificance_scenarios} along with a direct comparison of 14 TeV and 100 TeV machines at 3000 fb$^{-1}$ for zero pileup. At 14 TeV, by accumulating 3000 fb$^{-1}$ compared to 300 fb$^{-1}$, the sensitivity to $f_{T1}/\Lambda^{4}$ is improved by a factor of two. Comparing  a 14 TeV machine to a 100 TeV machine, 
  we obtain at least a factor of 100 gain in sensitivity to the operator coefficent $f_{T1}/\Lambda^{4}$ for a 5$\sigma$ discovery value.

\begin{table}[h]
\centering
\begin{tabular}{c|c|c|c|c|c}
\hline
\hline
{Parameter} & $\sqrt{s}$ & Luminosity & pileup & $5 \sigma$  & {95\% CL }  \\
                      & [TeV] & [fb$^{-1}$]&  & [TeV$^{-4}$] & [TeV$^{-4}$]  \\
\hline
        $f_{T1}/\Lambda^{4}$ & 14  & 300  &  50 &  0.2 (0.4)  & 0.1 (0.2)    \\
        $f_{T1}/\Lambda^{4}$ & 14  & 3000  &  140 &  0.1 (0.2)  & 0.06 (0.1)    \\
        $f_{T1}/\Lambda^{4}$ & 14  & 3000  &  0 &  0.1 (0.2)  & 0.06 (0.1)    \\
        $f_{T1}/\Lambda^{4}$ & 100  & 1000  &  40 &  0.001 (0.001) & 0.0004 (0.0004) \\
        $f_{T1}/\Lambda^{4}$ & 100  & 3000 &  263 & 0.001 (0.001) & 0.0008 (0.0008) \\
        $f_{T1}/\Lambda^{4}$ & 100  & 3000 &  0 &   0.001 (0.001) & 0.0008 (0.0008) \\
\hline
\hline
\end{tabular}
\caption{In $pp \to W^{\pm} W^{\pm} + 2j \to \ell \nu \ell \nu + 2j$ processes, $5 \sigma$-significance discovery values and 95\% CL limits are shown for coefficients of the higher-dimension operator, $f_{T1}/\Lambda^{4}$, for different machine scenarios without the UV cut and with the UV cut in parentheses.
 Pileup refers to the number of $pp$ interactions per crossing. }
\label{tab:ssWWSignificance_scenarios}
\end{table}

%

\section{$WWW\rightarrow \ell \nu  \ell \nu  \ell \nu$}
\label{WWW}
In the Standard Model (SM), the only allowed quartic coupling terms in the Lagrangian are $WWWW$, $WWZZ$, $WWZ  \gamma$ and $WW \gamma \gamma$,  and they are completely specified. Measuring these couplings will provide stringent tests on the SM and guide searches on physics beyond the SM. These couplings can be  measured using triboson production as well as vector boson scattering. The triboson $WWW$ production probes  the $WWWW$ coupling, 
 while $WWZ$ and $WW\gamma$ production probe the $WWZZ, WWZ \gamma$ and $WW\gamma \gamma$ couplings, respectively~\cite{Green:2003mn, LHC:QGC1998}. This section describes a cross section scan of  $WWW, WWZ, WZZ$ and $ZZZ$ production for different
 anomalous couplings induced by higher-dimension operators, followed by  a case study of $WWW$ production for both dimension-8 and dimension-6 operators. 

The dimension-8 operator studied most extensively was the $\cal L$$_{T,0}$ operator, given below
\begin{equation}
{\cal L}_{T,0} = \frac{f_{T0}}{\Lambda^4} {\rm Tr} [\hat{W}_{\mu \nu} \hat{W}^{\mu \nu}] \times {\rm Tr} [\hat{W}_{\alpha \beta}\hat{W}^{\alpha \beta}]
\label{eq:t0dim8}
\end{equation}

The dimension-6 operator used is $\cal L$$_{WWW}$, given below

\begin{equation}
{\cal L}_{WWW} = \frac{c_{WWW}}{\Lambda^2} {\rm Tr}[W_{\mu \nu}W^{\nu \rho}W^{\mu}_{\rho}] \; \; .
\label{eq:cwwwdim6}
\end{equation}

\subsection{Monte Carlo Predictions}
{\sc madgraph} 5.1.5.11~\cite{Alwall:2011uj} was used to generate all $WWW$ signal and SM samples and the background samples. In all cases $W$ and $Z$ bosons were required to decay leptonically to electrons or muons.  Default generator-level  cuts on the leptons were  $p_{T} >  10$~GeV and   $|\eta| < 2.5$. 

We performed a cross section scan to compare the SM cross sections to anomalous coupling cross sections for various higher-dimension operators (Tables~\ref{tab:dim8scan} 
and~\ref{tab:dim6scan}). From the study, the $\cal {L}$$_{T,0}$ 
 operator is found to have the largest effect on triboson production from  the dimension-8 operators, and  $\cal{L}$$_{WWW}$ from the dimension-6 operators, particularly in the $WWW$ channel.  For this reason, we focus on this channel
  and these operators for the remainder of this section.

\begin{table}[htdp]
\begin{center}
\begin{tabular}{c|c|c|c|c}
\hline
\hline
operator  & $WWW$ & $WWZ$ & $WZZ$ & $ZZZ$ \\
SM cross section [ab] & 603 & 124 & 9.63 & 0.972 \\
\hline
$\cal{L}$$_{S,0}$/SM & 1.0 & 1.0 & 1.0 & 1.0 \\
\hline
$\cal{L}$$_{S,1}$/SM & 1.0 & 1.0 & 1.0 & 1.0 \\
\hline 
$\cal{L}$$_{M,0}$/SM & 1.46 & 1.09 & 1.05 & 1.02 \\
\hline
$\cal{L}$$_{M,1}$/SM & 1.17 & 1.02 & 1.04 & 1.03 \\
\hline
$\cal{L}$$_{M,2}$/SM & 1.0 & 1.05 & 1.0 & 1.02 \\
\hline
$\cal{L}$$_{M,3}$/SM & 1.0 & 1.01 & 1.00 & 1.01 \\
\hline
$\cal{L}$$_{T,0}$/SM & 18.31 & 3.96 & 3.38 & 2.90\\
\hline
$\cal{L}$$_{T,1}$/SM & 15.15 & 2.10 & 2.83 & 2.90 \\
\hline
$\cal{L}$$_{T,2}$/SM & 4.48 & 1.32 & 1.35 & 1.54 \\
\hline
$\cal{L}$$_{T,8}$/SM & 1.0 & 1.0 & 1.0 & 1.31 \\
\hline
$\cal{L}$$_{T,9}$/SM & 1.0 & 1.0 & 1.0 & 1.08 \\
\hline 
\hline
\end{tabular}
\end{center}
\caption{The ratios of cross sections for various dimension-8 operators  to SM values are shown
 for a $\sqrt{s} = 14$~TeV $pp$ collider. In each case, the coefficient of the dimension-8 operator was set to 10 TeV$^{-4}$. All channels are fully leptonic 
 decays. }
\label{tab:dim8scan}
\end{table}

\begin{table}[htdp]
\begin{center}
\begin{tabular}{c|c|c|c|c}
\hline
\hline
operator & $WWW$ & $WWZ$ & $WZZ$ & $ZZZ$ \\
SM cross section [ab] & 603 & 124 & 9.63 & 0.972 \\
\hline
$\cal{L}$$_{WWW}$/SM & 1.4 & 1.3 & 1.4 & 1.0 \\
\hline
$\cal{L}$$_{W}$/SM & 1.1 & 1.1 & 1.2 & 1.1 \\
\hline 
$\cal{L}$$_{b}$/SM & 1.0 & 1.0 & 1.0 & 1.0 \\
\hline
\hline
\end{tabular}
\end{center}
\caption{The ratios of cross sections for various dimension-6 operators  to SM values are shown 
 for a $\sqrt{s} = 14$~TeV $pp$ collider. In each case, the coefficient of the dimension-6 operator was set to 5 TeV$^{-2}$. All channels are fully
 leptonic decays. }
\label{tab:dim6scan}
\end{table}

\subsection{Event Selection}
\label{wwwsel}
After {\sc pythia} 6.4~\cite{pythia6} parton showering, additional detector effects are applied using {\sc delphes} 3.0.9~\cite{deFavereau:2013fsa} with the Snowmass parameterization~\cite{delphes1, delphes2, delphes3}.  This parameterization includes effects from pile-up events.  We have studied these effects and found them to be negligible for this analysis, which focuses on events with high invariant mass
 of the triboson system.  Thus, all results presented in this section are extracted using Monte Carlo events generated without pile-up in the interest of reducing computational time.

Events are considered to be part of the $WWW$ signal if they meet the following criteria, where $p_T(\ell)$ is the transverse momentum of the lepton, $M($all lep$)$ is the invariant mass of all the 
 charged leptons with $p_T(\ell) > 25$~GeV and \met~ is the missing transverse energy of the event:

\begin{itemize}
 \item At least three leptons, where leptons must have $p_T(\ell) > 25$~GeV
 \item No two leptons may have the same flavor and opposite charge (to suppress diboson $WZ$ background) 
 \item $M($all lep$) > 400$~GeV
 \item $\met > 150$~GeV
 \end{itemize}
 
These selections were specifically chosen to optimize the signal and reduce backgrounds.  The backgrounds considered include $Z$+jets, $W$+2~jets, \ttbar, diboson ($WW$, $WZ$, $ZZ$), and $Z+\gamma$.  The first selection reduces particularly the $W$+jets, $Z$+jets and $WW$ backgrounds, and the second reduces backgrounds with $Z$ bosons.  The selection on lepton number also helps to reduce the \ttbar\ background, but as only one jet has to fake a lepton and the cross-section for this process is much higher than the signal process, we still have a comparatively large contribution from \ttbar\ production.  The last two selections help to reduce this remaining \ttbar\ contanimation.  After these selections, the results are not significantly affected by the remaining background events and these background processes are neglected in the final analysis.  

\begin{figure}[htbp]
\begin{center}
\includegraphics[width=0.49\textwidth]{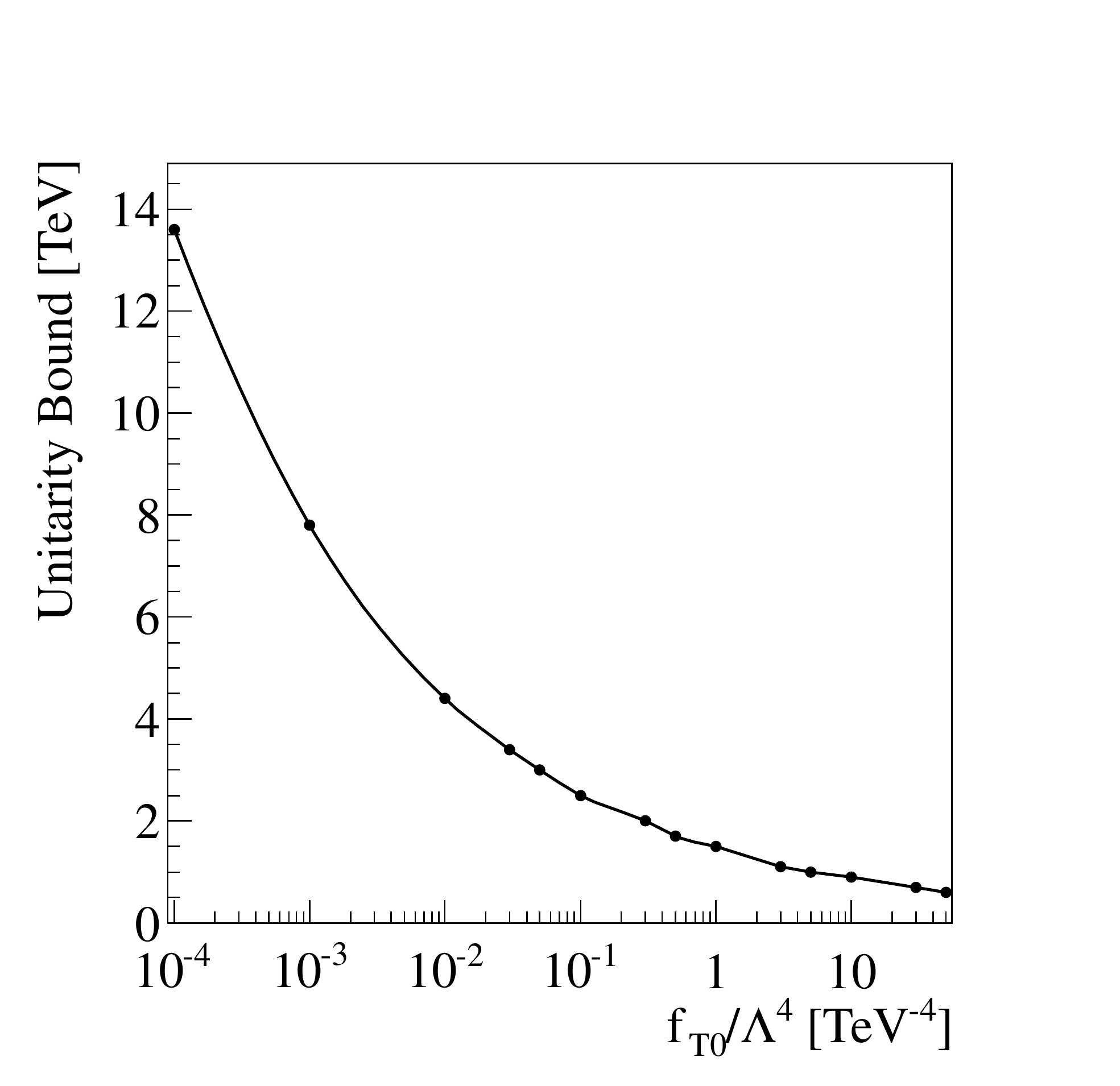}
\includegraphics[width=0.49\textwidth]{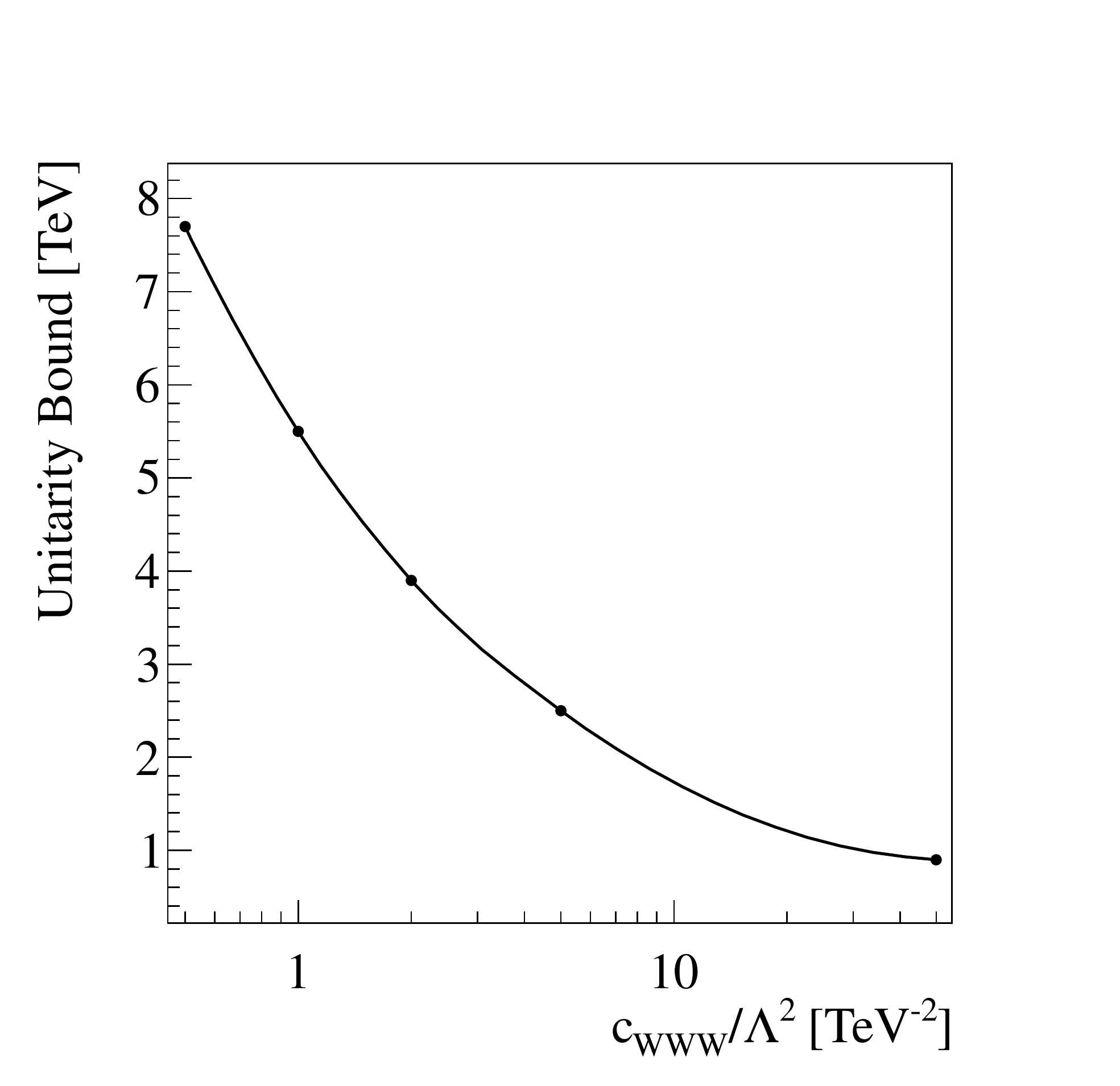}
\caption{Unitarity bounds for the ${\cal L}_{T,0}$ operator (left) and for the ${\cal L}_{WWW}$ operator (right) are shown in various scenarios.}
\label{fig:unitarity}
\end{center}
\end{figure}

There is an additional selection at the parton (truth) level to remove events in the kinematic region where the anomalous coupling 
 amplitude would violate unitarity.  Events are removed if the invariant mass of the three $W$ bosons is larger than the 
 UV bound.  These bounds are estimated using the form factor tool available with VBFNLO~\cite{unitarityCalculator} and are presented for various values of $f_{T0}/\Lambda^4$ and $c_{WWW}/ \Lambda^2$ in Figure~\ref{fig:unitarity}.  The bound rises rapidly for lower values of these coefficients, leading to a reduced impact of this bound.  Conversely, for higher values of the coefficients, where the cross-section is higher, the impact of this bound is stronger.

In the next section, for both $\cal{L}$$_{T,0}$ and $\cal{L}$$_{WWW}$ operators, we present results without the UV bound applied.  We also comment on expectations with the application of the  UV bound in Sec.~\ref{ubcut}.

\subsection{Statistical Analysis}
\begin{figure}[htbp]
\begin{center}
\includegraphics[width=0.49\textwidth]{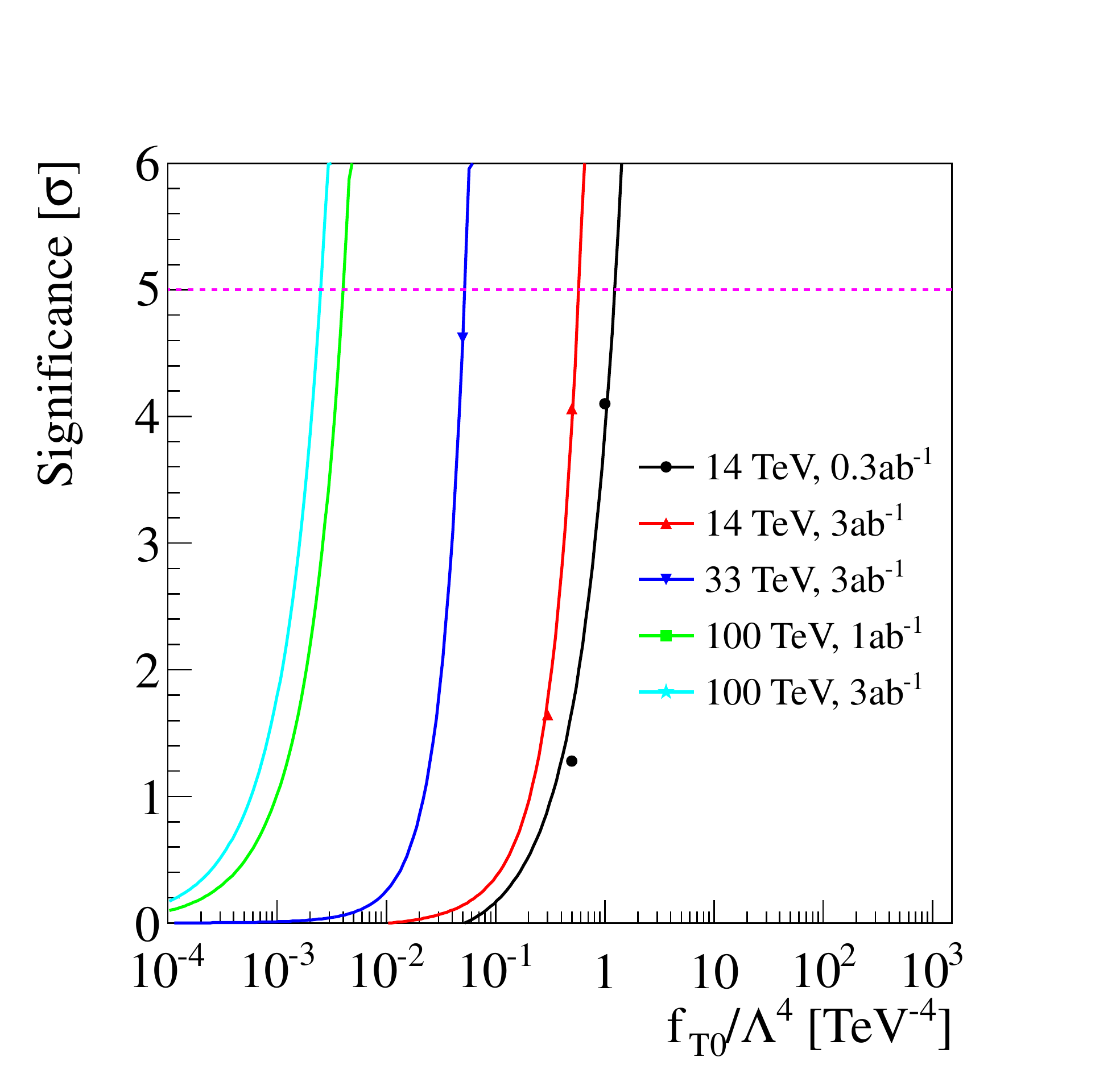}
\includegraphics[width=0.49\textwidth]{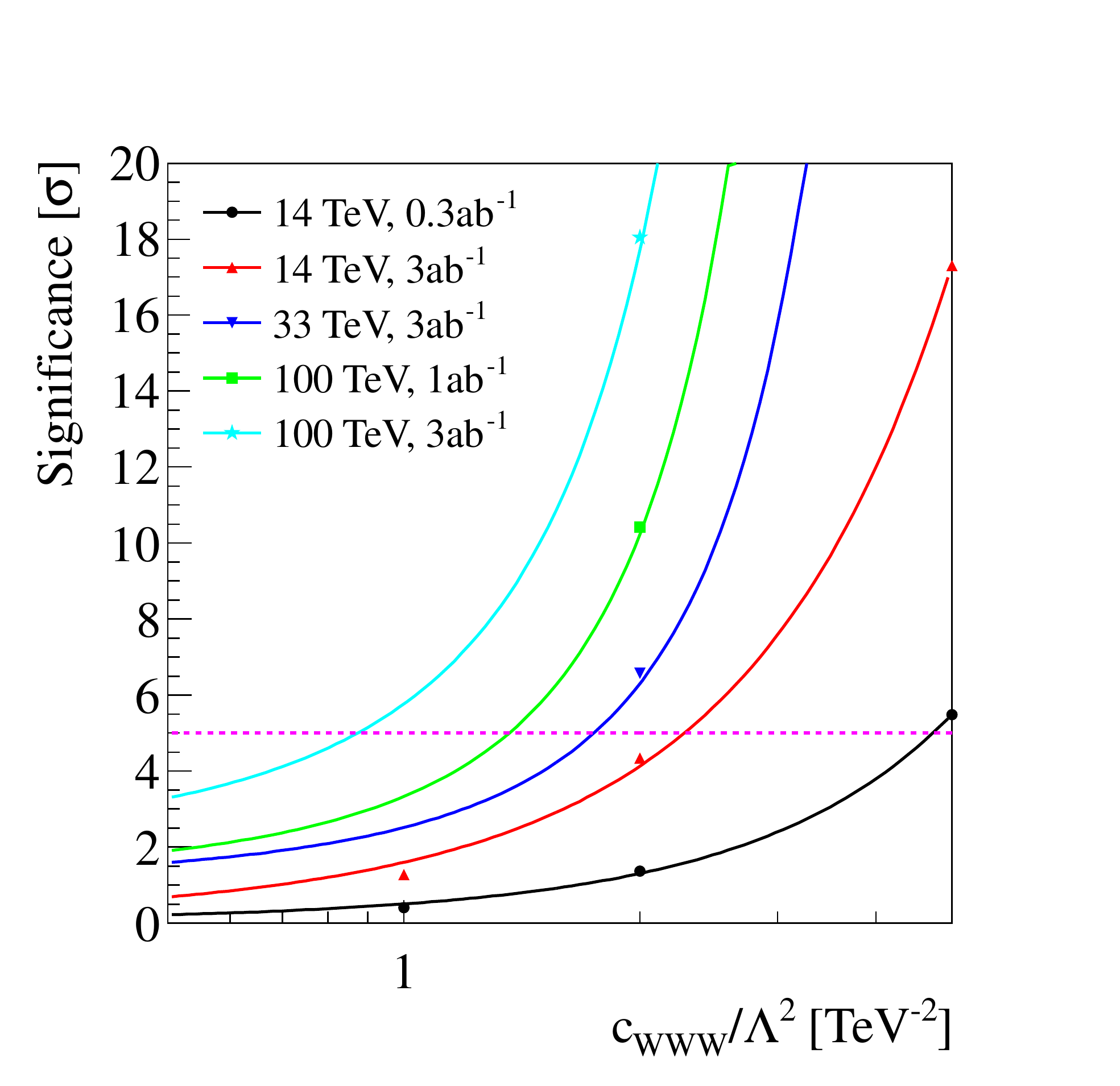}
\caption{Significance values without the application of the UV bound as a function of $f_{T0} / \Lambda^4$  (left) and $c_{WWW} / \Lambda^2$ (right) are shown in various scenarios.}
\label{fig:unitaritysigNOUB}
\end{center}
\end{figure}

The distribution of $M($all lep$)$ is used for hypothesis testing.  We compare the standard model prediction with the prediction for a non-zero value of a given higher-dimension operator.  The statistical analysis is identical to that employed in Sec.~\ref{zzjjStats}.  Figure~\ref{fig:t0www33NOUB} shows the $WWW$ templates used for the $\sqrt{s} = 33$~TeV $pp$ collider, before the lepton invariant mass selection.  The significance estimate uses distributions with this cut applied and different binning depending on the collider energy.   The fits of the significance curves use quadratic functions except for the 100~TeV and 33~TeV $c_{WWW}$ curves without (with) the application of the UV bound, which are fit better by an exponential function (a third-order polynomial).  Figure~\ref{fig:unitaritysigNOUB} shows the significance estimates for various $f_{T0}$ and $c_{WWW}$ values for different hadron collider machines being studied for Snowmass. As the machine energies and integrated luminosities increase we are able to put tighter constraints on these operators, and of course we are also able to discover and probe new physics at increasingly higher mass scales and/or smaller couplings.

\begin{table}[h]
\centering
\begin{tabular}{c|c|c|c|c|c}
\hline\hline
Parameter & dim. & Luminosity [fb$^{-1}$] & 14 TeV & 33 TeV & 100 TeV \\
\hline 
\multirow{3}{*}{$c_{WWW}/\Lambda^2$  [TeV$^{-2}$] }& \multirow{3}{*}{6} & 300 & 4.8 (8)  & - & - \\
\cline{3-6}
 &   & 1000 & - & - & 1.3 (1.5)  \\
 \cline{3-6}
  & &  3000 & 2.3 (2.5)  & 1.7 (2.0) & 0.9 (1.0) \\
  \hline
\multirow{3}{*}{  $f_{T0} / \Lambda^4$ [TeV$^{-4}$] }&\multirow{3}{*}{ 8} & 300 & 1.2  & - & - \\
\cline{3-6}
   &  & 1000 & - & - & 0.004 \\
   \cline{3-6}
   & & 3000 & 0.6  & 0.05 & 0.002 \\
   
  \hline\hline
  \end{tabular}
  \caption{In the $pp \to WWW \to 3\ell + 3\nu$ process, the $5\sigma$-significance discovery values are shown for the coefficients of higher-dimension operators. The values in parentheses are obtained with the UV bound applied. 
 $pp$ colliders at $\sqrt{s} = 14, \; 33$ and 100~TeV are studied.  }
  \label{tab:WWW5sigma}
  \end{table}

\begin{figure}[htbp]
\begin{center}
\includegraphics[width=0.49\textwidth]{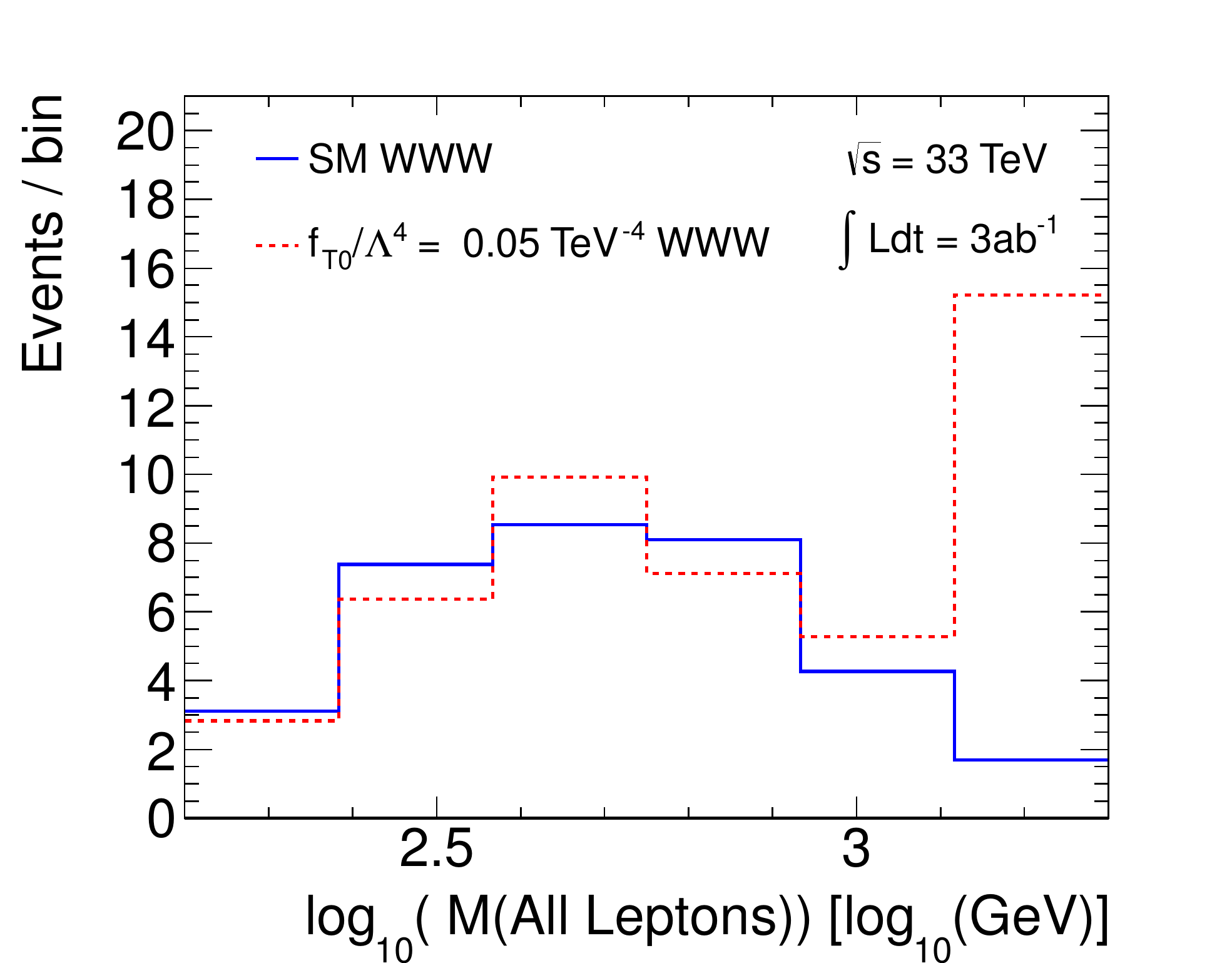}
\includegraphics[width=0.49\textwidth]{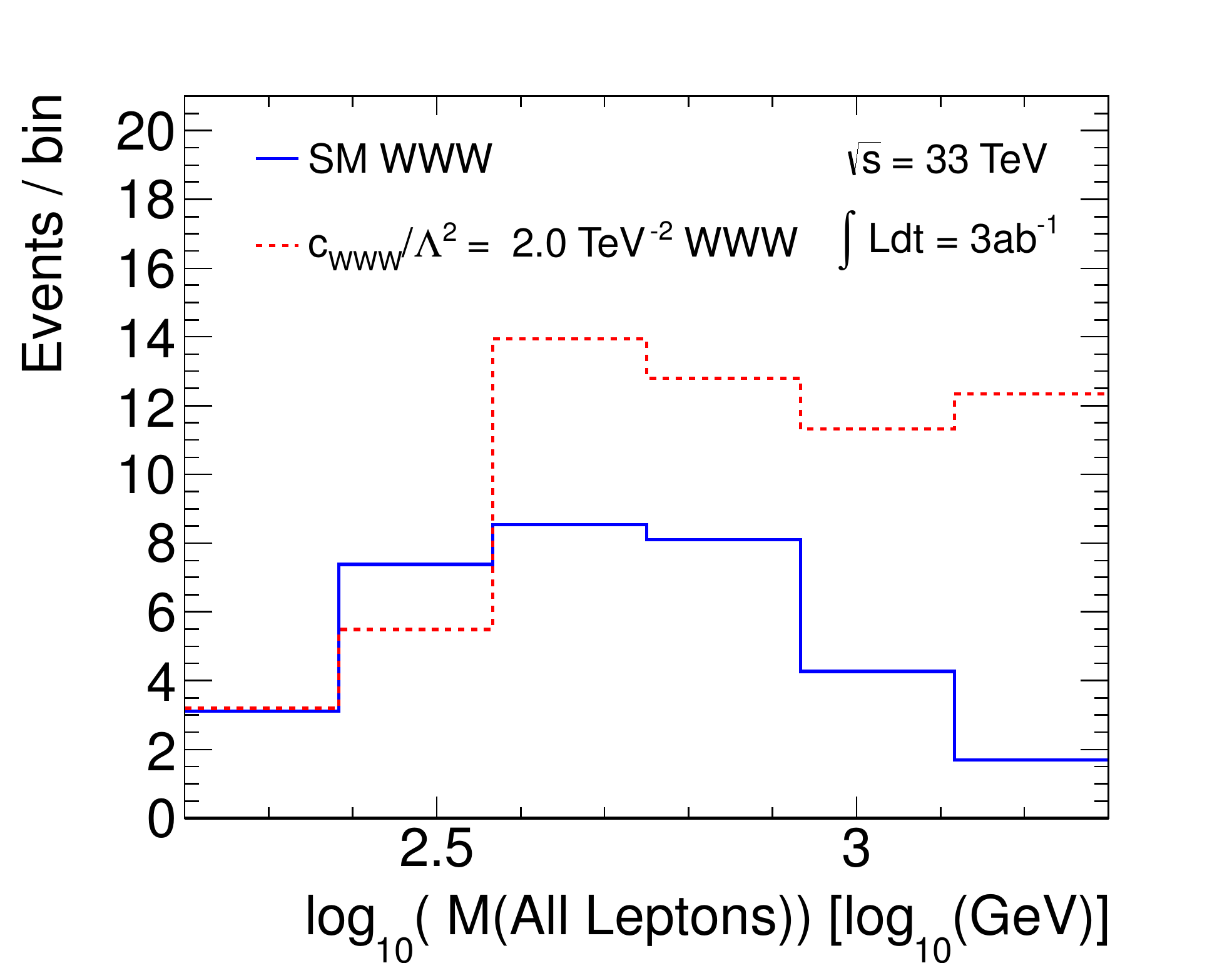}
\caption{The invariant mass of all leptons  is shown without applying the UV bound, for the SM and with $f_{T0} / \Lambda^4 = 0.05$~TeV$^{-4}$ (left) and $c_{WWW}/ \Lambda^2 = 2$~TeV$^{-2}$ (right) for $\sqrt{s} = 33$~TeV.
   These distributions were made without the lepton invariant mass selection.}
\label{fig:t0www33NOUB}
\end{center}
\end{figure}


\subsubsection{Impact of the Unitarity Violation Selection}
\label{ubcut}



As mentioned previously, we want to consider what happens when a UV bound is applied, which can be done in different ways.  Here, we apply a selection on the generator-level $WWW$ mass using the values discussed in Section~\ref{wwwsel} as an upper bound.   Figure~\ref{fig:t0www33NOUB} shows the $WWW$ templates used for the 33 TeV $pp$ collider, before the lepton invariant mass selection was applied, with and without the UV bound applied for a given
 value of the ${\cal L}_{T,0}$ coefficient.  The application of the UV bound lowers the expected number of signal events in the highest-mass bin.  The UV bound used in this study has a large impact on the ${\cal L}_{T,0}$ results with this simple event removal, increasing the values from Table~\ref{tab:WWW5sigma} of $f_{T0}/\Lambda^4$ at which we might expect a $5 \sigma$-significance discovery by more than a factor of 20.
 The large impact of the UV bound on the significance indicates that the breakdown of the EFT is imminent, and this channel is sensitive to the direct production of new particles which have been integrated out in the EFT for
 this dimension-8 operator. We expect this conclusion to be generically true for triboson production induced by dimension-8 operators, where the growth of the cross section with energy is more rapid  than with dimension-6 operators.  

\begin{figure}[htbp]
\begin{center}
\includegraphics[width=0.49\textwidth]{WWW_33tevtemplate_dim6_NOUB-eps-converted-to.pdf}
\includegraphics[width=0.49\textwidth]{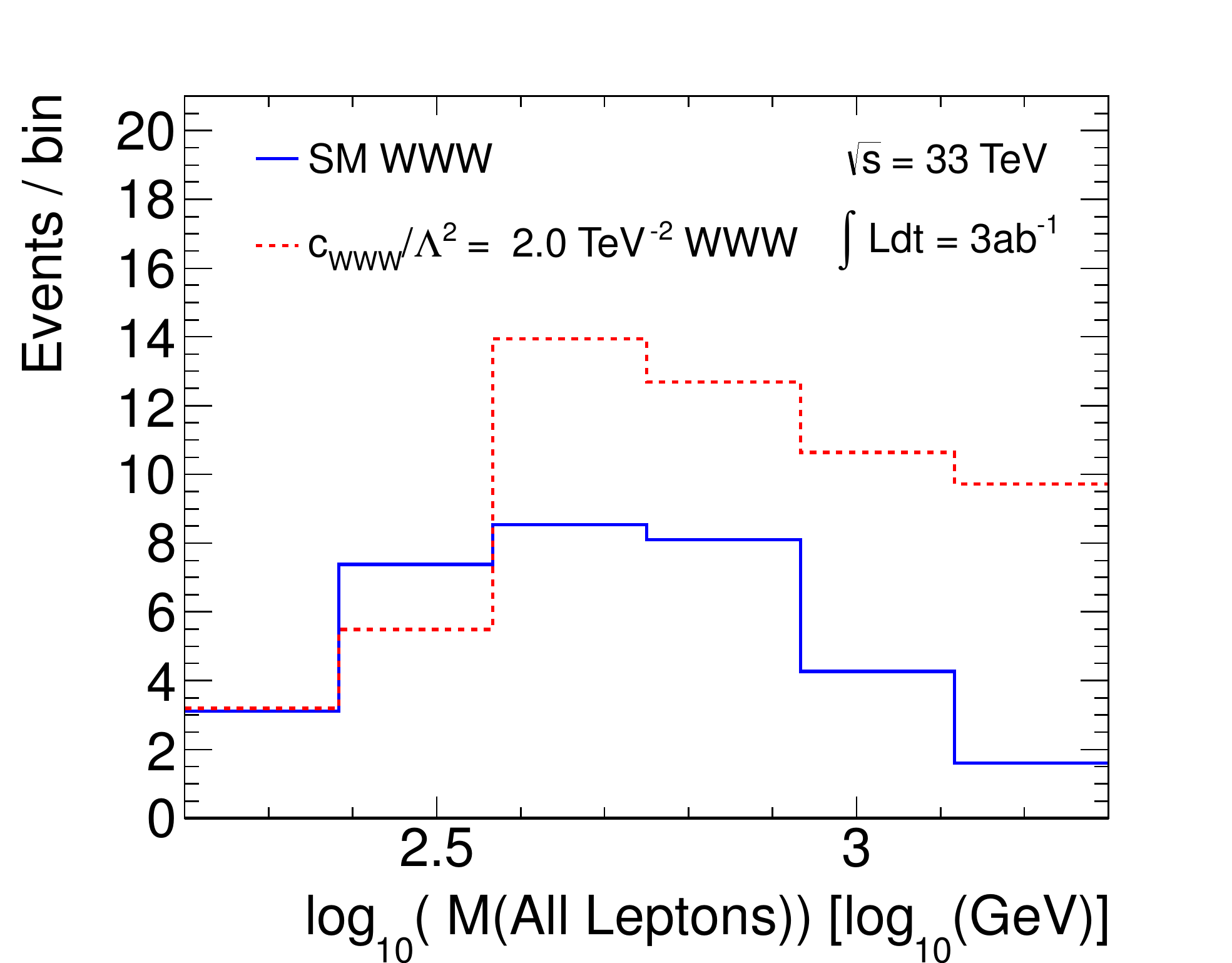}
\caption{The invariant mass of all leptons is shown for the SM $WWW$ process and with $c_{WWW} / \Lambda^2 = 2$~TeV$^{-2}$ for $\sqrt{s} = 33$ TeV without the UV bound (left) and with this bound applied (right).  This distribution was made without the lepton invariant mass selection.}
\label{fig:t0www33_dim6}
\end{center}
\end{figure}

\begin{figure}[htbp]
\begin{center}
\includegraphics[width=0.49\textwidth]{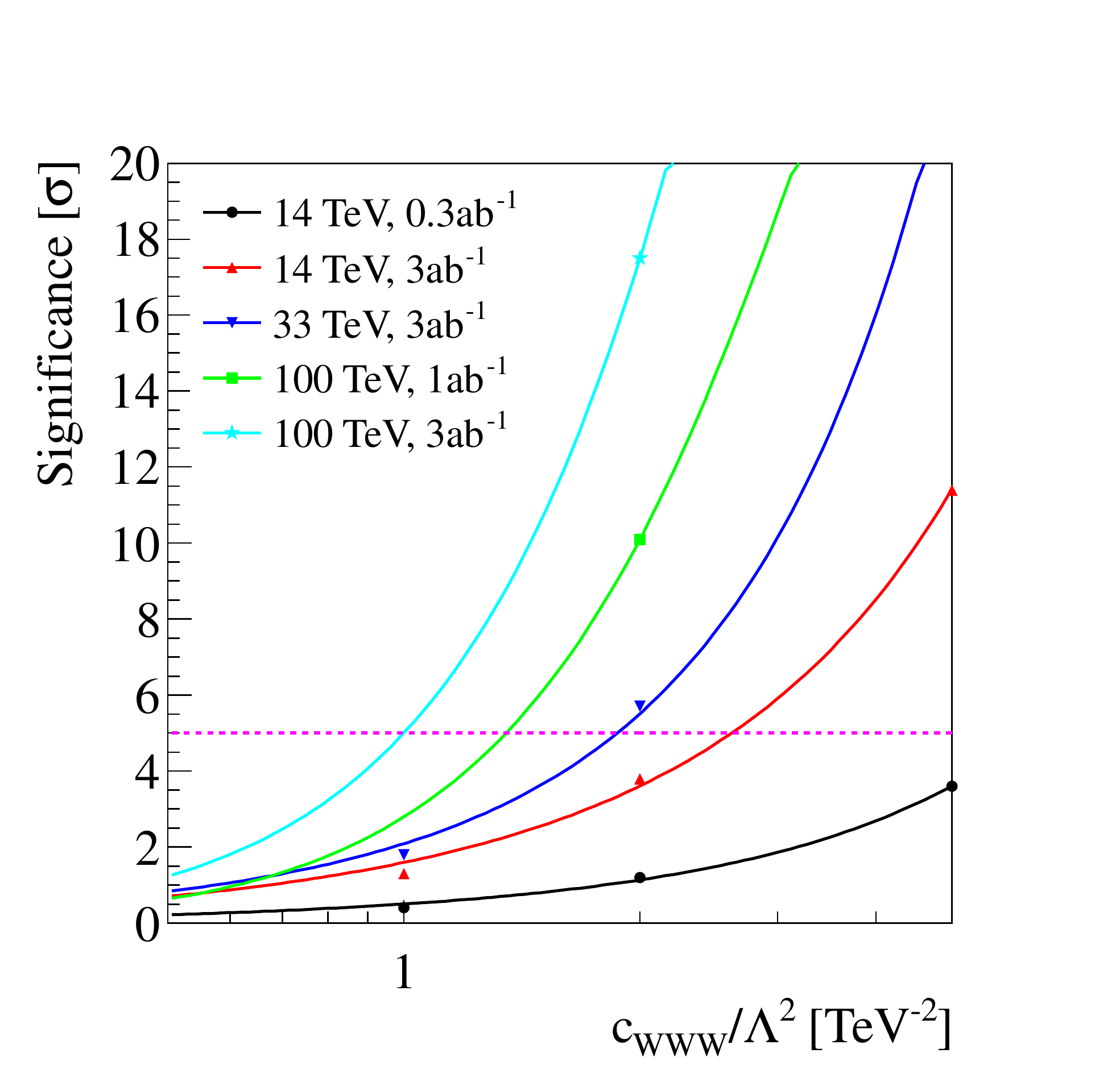}
\caption{Significance values for the ${\cal L}_{WWW}$ operator coefficient are shown in various scenarios. The UV bound has been applied for these results. }
\label{fig:unitaritysigdim6}
\end{center}
\end{figure}

The impact of applying the UV bound is less severe for the $\cal L$$_{WWW}$ operator.  Figure~\ref{fig:t0www33_dim6} shows the $WWW$ templates used for the 33 TeV machine, before the lepton invariant mass selection for the same operator and coefficient, with and without the UV bound applied.  There are fewer events in the last bins of Figure~\ref{fig:t0www33_dim6} due to the application of the UV bound, but overall the distributions are relatively similar, much more so than for ${\cal L}_{T,0}$ operator (see Figure~\ref{fig:t0www33NOUB}).  Figure~\ref{fig:unitaritysigdim6} shows the significance estimates for various $c_{WWW}/\Lambda^2$ values for the different hadron collider machines being studied for Snowmass, with the UV bound applied. These values are generally 10-20$\%$ higher than the values of $c_{WWW}/\Lambda^2$ at which we might expect a $5 \sigma$-significance discovery without the removal of the high-mass events. As mentioned in the previous paragraph, the dimension-6 operator amplitude does not violate unitarity as rapidly as a dimension-8 operator, when the energy is raised.



\begin{figure}[htbp]
\begin{center}
\includegraphics[width=0.49\textwidth]{WWW_33tevtemplate_514-eps-converted-to.pdf}
\includegraphics[width=0.49\textwidth]{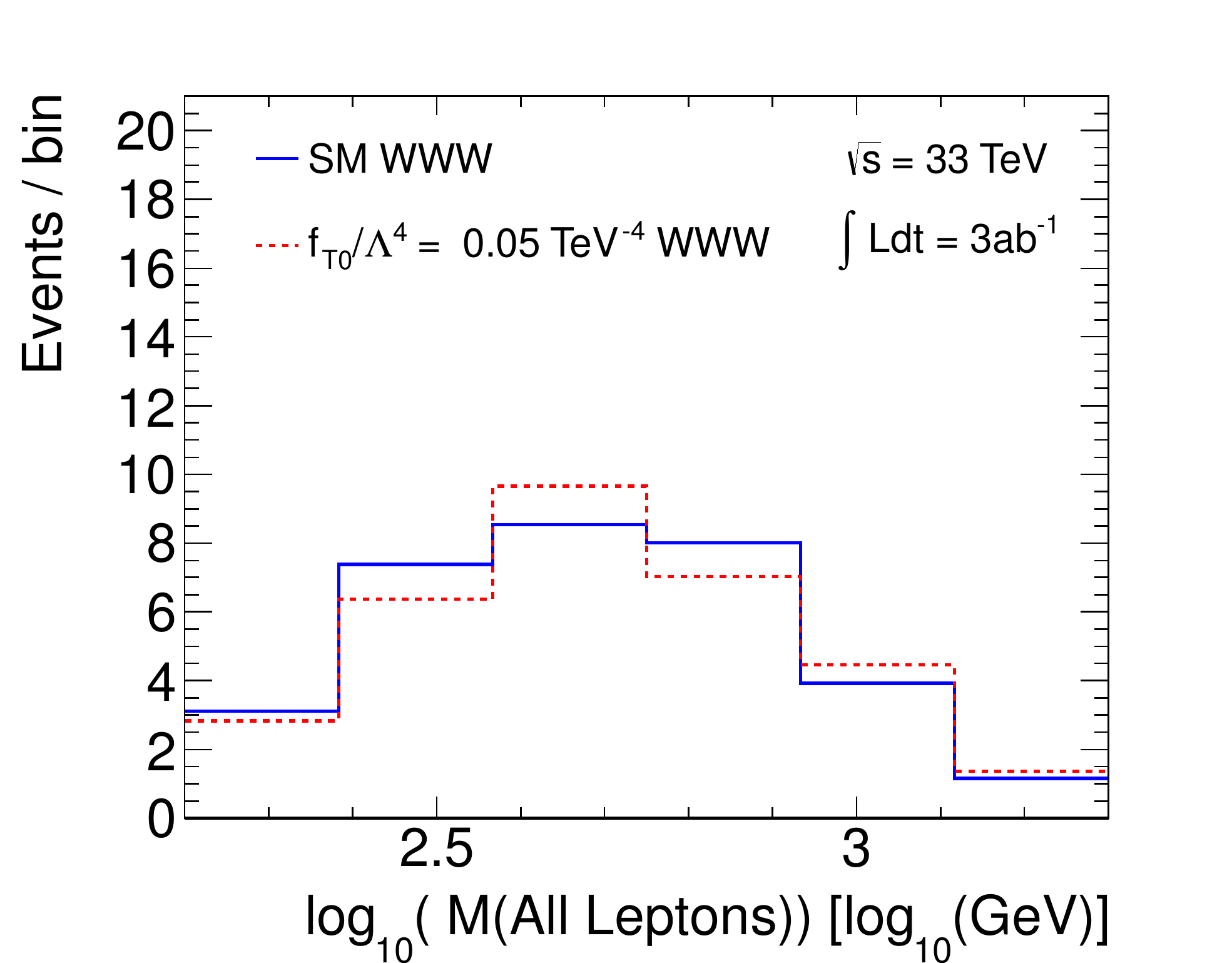}
\caption{The invariant mass of all leptons is shown for the SM $WWW$ process and with $f_{T0}/\Lambda^{-4} = 0.05$~TeV$^{-4}$ for $\sqrt{s} = 33$ TeV without the UV bound (left) and with this bound
  applied (right).  This distribution was made without the lepton invariant mass selection.}
\label{fig:t0www33NOUB}
\end{center}
\end{figure}


\section{$Z\gamma\gamma \to l^{+} l^{-}\gamma\gamma$}

The $Z\gamma\gamma$ mass spectrum at high mass is sensitive to BSM triboson contributions.  The lepton-photon channel allows  full reconstruction  of the final state  and the 
 calculation of 
the $Z\gamma\gamma$ invariant mass $m_{Z \gamma \gamma}$.
We parameterize the BSM physics using the following operators
\begin{eqnarray}
{\cal L}_{M,0} & = & \frac{f_{M0}}{\Lambda^4} Tr[\hat{W}_{\mu \nu}\hat{W}^{\mu \nu}] \times [(D_{\beta}\phi)^\dagger D^{\beta}\phi] \\
{\cal L}_{M,1} & = & \frac{f_{M1}}{\Lambda^4} Tr[\hat{W}_{\mu \nu}\hat{W}^{\nu \beta}] \times [(D_{\beta}\phi)^\dagger D^{\mu}\phi] \\
{\cal L}_{M,2} & = & \frac{f_{M2}}{\Lambda^4} [B_{\mu\nu}B^{\mu\nu}] \times [(D_{\beta}\phi)^\dagger D^{\beta}\phi] \\
{\cal L}_{M,3} & = & \frac{f_{M3}}{\Lambda^4} [B_{\mu\nu}B^{\nu\beta}] \times [(D_{\beta}\phi)^\dagger D^{\mu}\phi]. 
\label{ZAAeqns}
\end{eqnarray}

They are selected because of the possibility to be converted to $a_{0}$ and $a_{C}$, the non-linear parametrization of anomalous couplings adopted by the LEP results.

{\sc madgraph} 5.1.5.10~\cite{Alwall:2011uj} was used to generate all $Z\gamma\gamma$ samples and background samples, $Z\gamma j$ and $Zjj$.
In all  cases $Z$ bosons were required to decay to electron or muon pairs.

\subsection{Event Selection}
After {\sc pythia} 6.4~\cite{pythia6} parton showering,
the reconstruction effects of  resolution and identification efficiency are applied  using {\sc delphes}~\cite{Ovyn:2009tx,delphes1,delphes2,delphes3}.
A constant jet-to-photon fake rate of $10^{-3}$  is applied to each jet 
 in the $Z\gamma j$ and $Zjj$ samples to construct smooth background templates.
Events are considered $Z\gamma\gamma$ candidates provided
they meet the following criteria:

\begin{itemize}
    \item $p_T(\ell) > 25$~\GeV,  $|\eta(\ell)|<2.0$
    \item $p_T(\gamma) > 25$~GeV,  $|\eta(\gamma)|<2.0$
    \item leading $\gamma$  $p_T > 160$~GeV
    \item $| m_{\ell \ell} - 91 \GeV | < 10$~ GeV
    \item $\Delta R (\gamma,\gamma) > 0.4$; $\Delta R (\ell,\gamma) > 0.4$; $\Delta R (\ell,\ell) > 0.4$.
\end{itemize}

The $p_T > 160$~GeV requirement on the photon suppresses fake background.
The 10~GeV invariant mass window cut around the $Z$ boson mass peak suppresses the $\gamma*$ contribution to the dilepton system.
The large angular separation ($\Delta R$) between the photon and the lepton and the high transverse-momentum requirement of the photon reduces the final-state photon radiation  contribution.
This leads to the phase space which is uniquely sensitive to the quartic gauge coupling (QGC).

\subsection{Statistical Analysis}

\begin{figure}[h]
  \centering
  \includegraphics[width=0.60\textwidth]{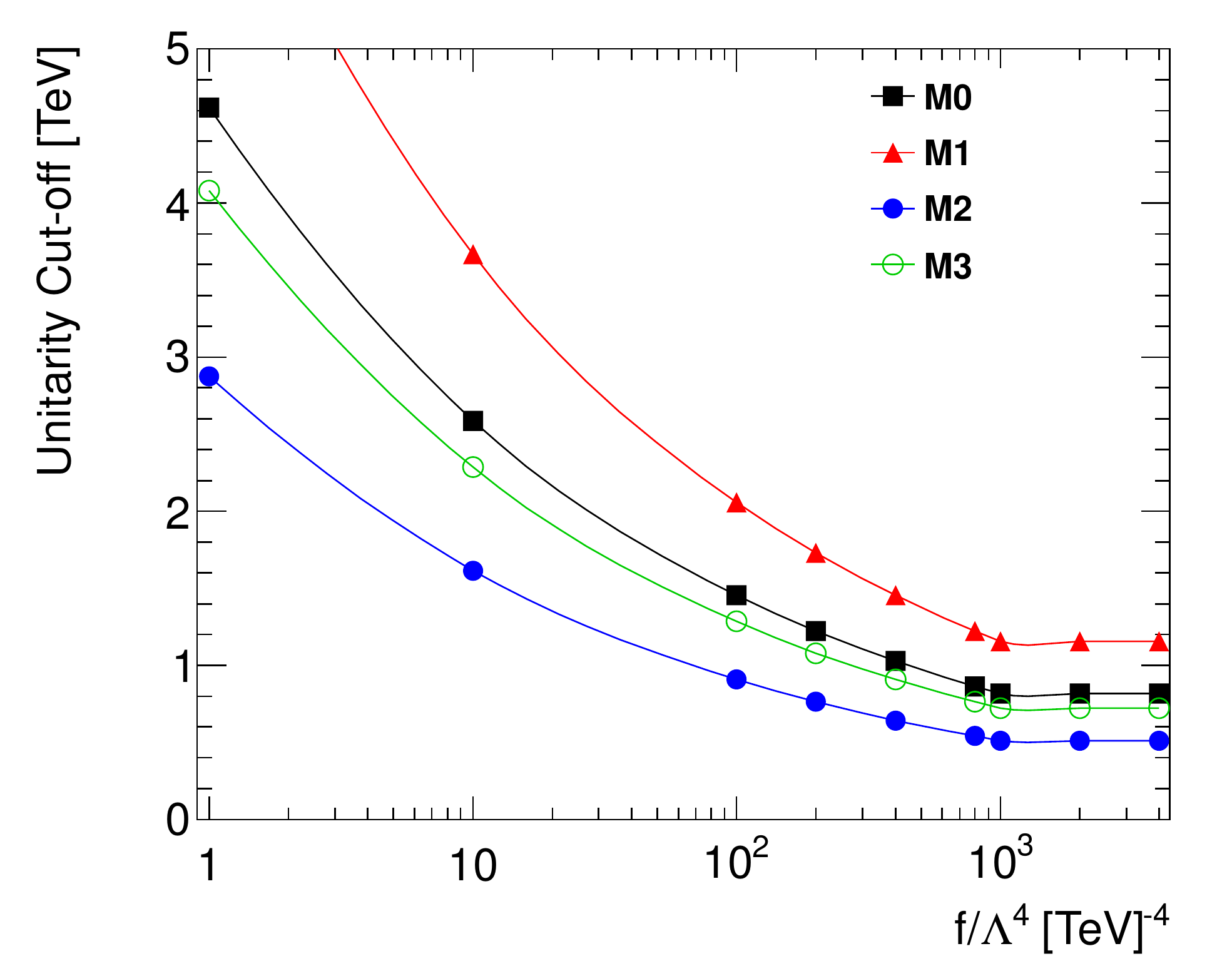}
  \caption{
The unitarity violation bounds for ${\cal L}_{M,i}, i=0...3$  operators are shown as functions of their  coefficient values in the $pp \to Z\gamma\gamma$ process.
}
\label{fig:ZAA_UV_boundary}
\end{figure}

The statistical analysis is identical to that employed in Sec.~\ref{zzjjStats}. 
The distribution of $m_{Z \gamma \gamma}$ is used for hypothesis testing.
The dominant process in the QGC-sensitive kinematic phase space is $Z\gamma\gamma$ production
while the fake backgrounds $Z\gamma j$ and $Zjj$ are sub-dominant.
We present sensitivity studies with and without the UV bound applied. The bounds for different operators are shown in Figure~\ref{fig:ZAA_UV_boundary}. 

\begin{figure}[h]
  \centering
  \includegraphics[width=0.47\textwidth]{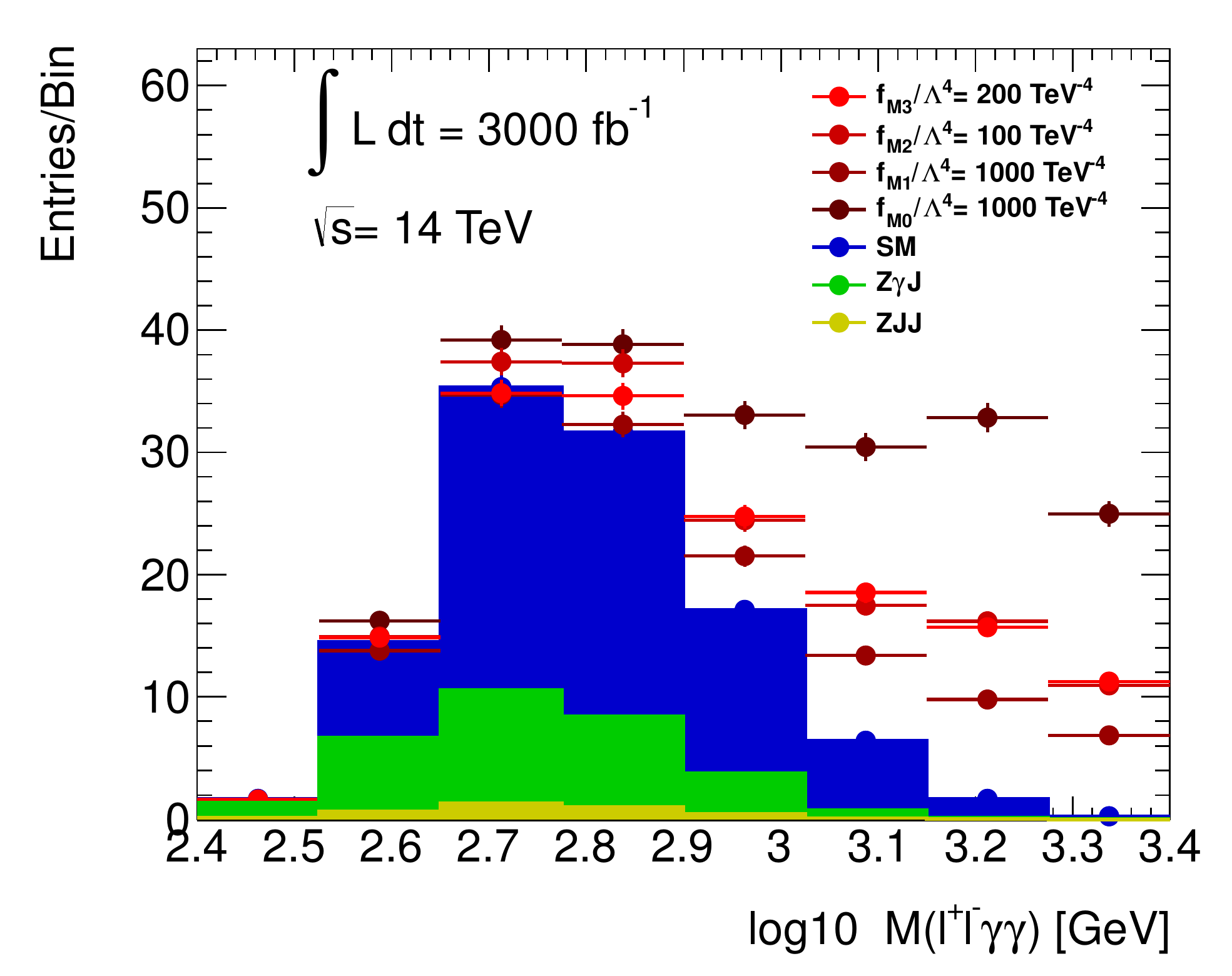}
  \includegraphics[width=0.47\textwidth]{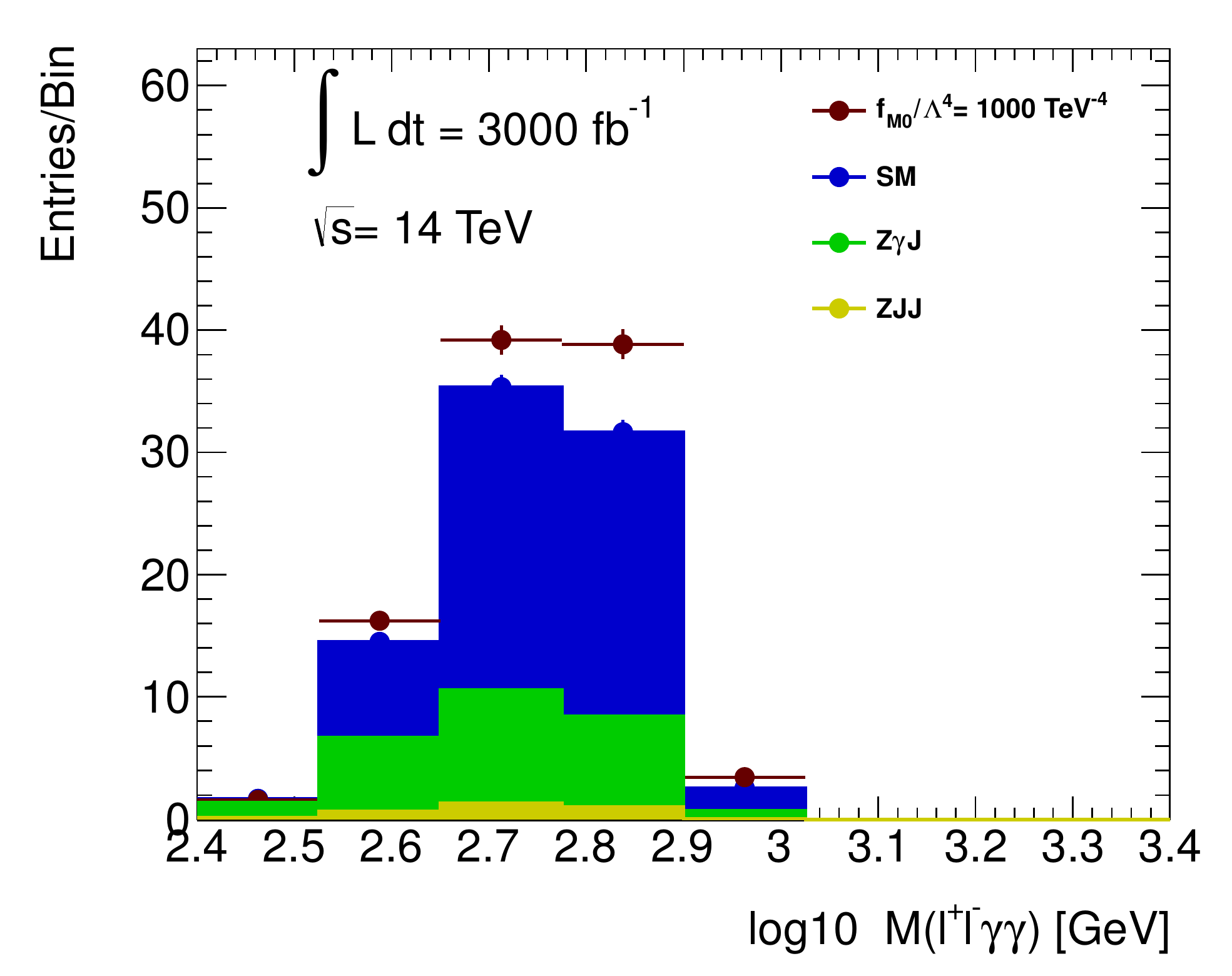}
  \includegraphics[width=0.47\textwidth]{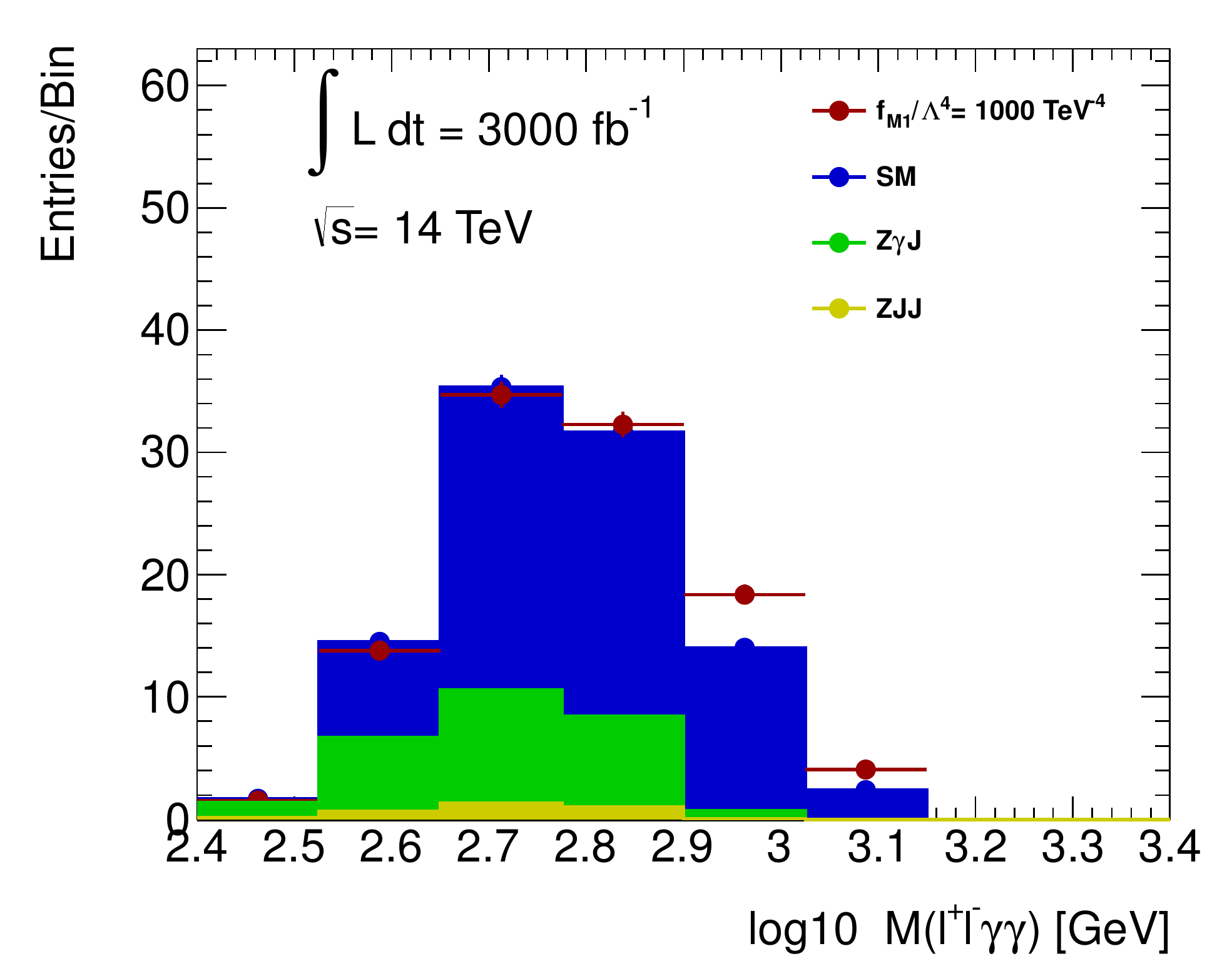}
  \includegraphics[width=0.47\textwidth]{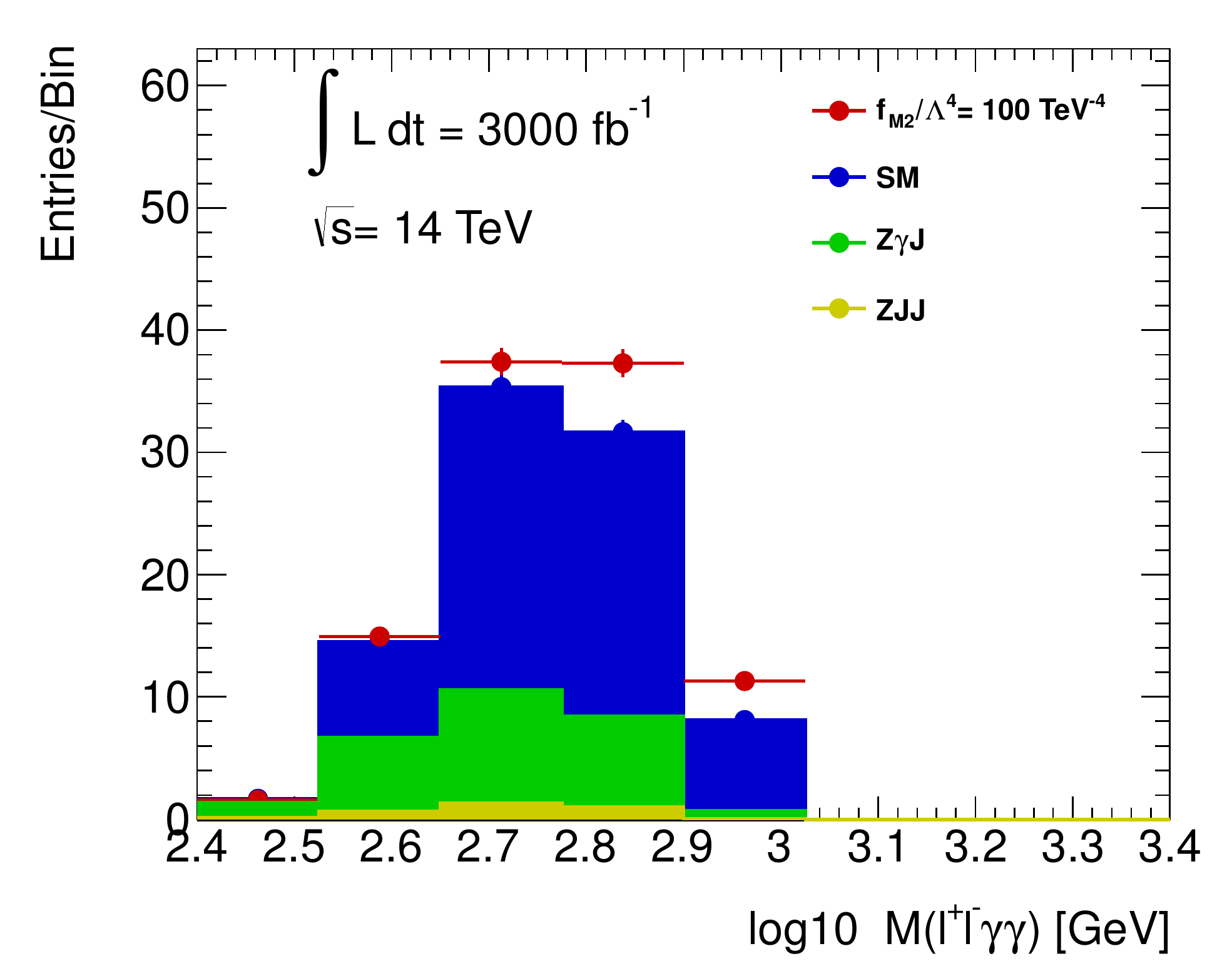}
  \caption{
Reconstructed  mass spectra using the charged leptons and  photons after event selection are shown without the UV bound (top left) and with the UV bound (top right and bottom). The coefficient is selected to be near
 the 95\% CL limit for each operator after applying the UV bound.
}
\label{fig:mZAA}
\end{figure}

Figure~\ref{fig:mZAA} shows  the reconstructed 4-body invariant mass
distribution for this channel at the $\sqrt {s} = 14$~TeV $pp$ collider. The left upper figure shows the distributions without the UV bound. The other three plots show the mass distribution due to each
 anomalous QGC operator with the  corresponding UV bound applied. 
The bound is applied as an additional event selection requirement on the 4-body mass.

\begin{figure}[h]
  \centering
  \includegraphics[width=0.49\textwidth]{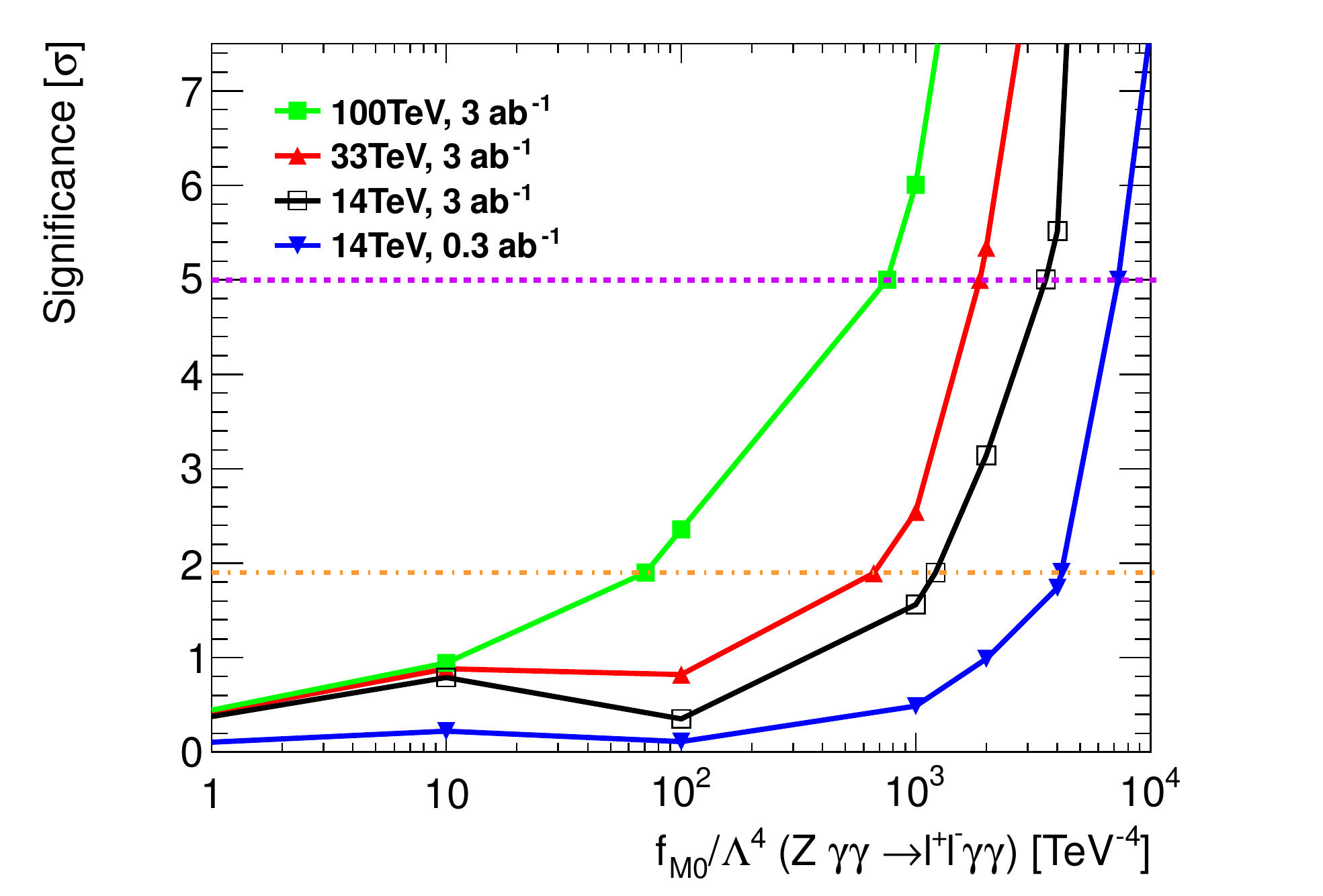}             
  \includegraphics[width=0.49\textwidth]{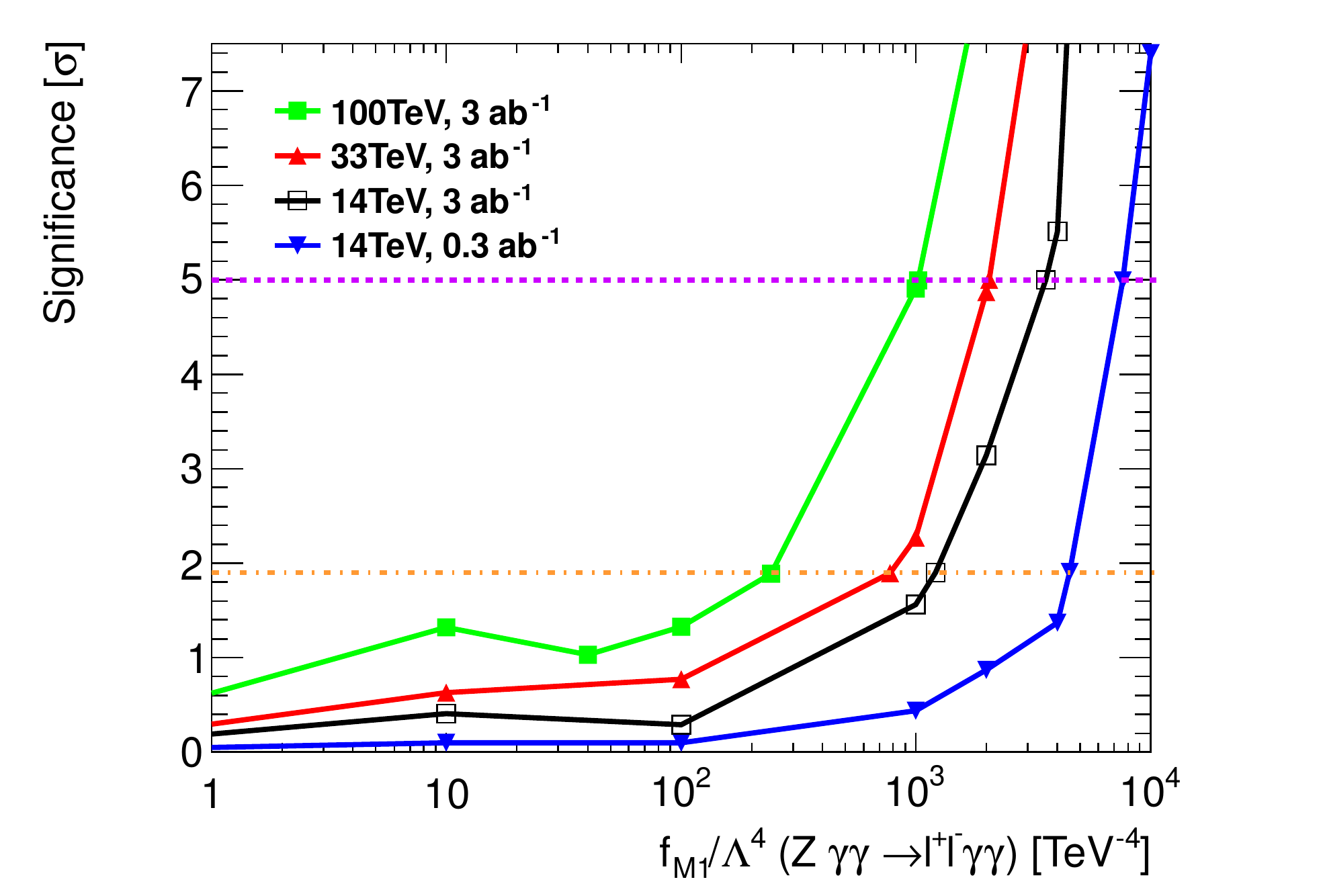}
  \includegraphics[width=0.49\textwidth]{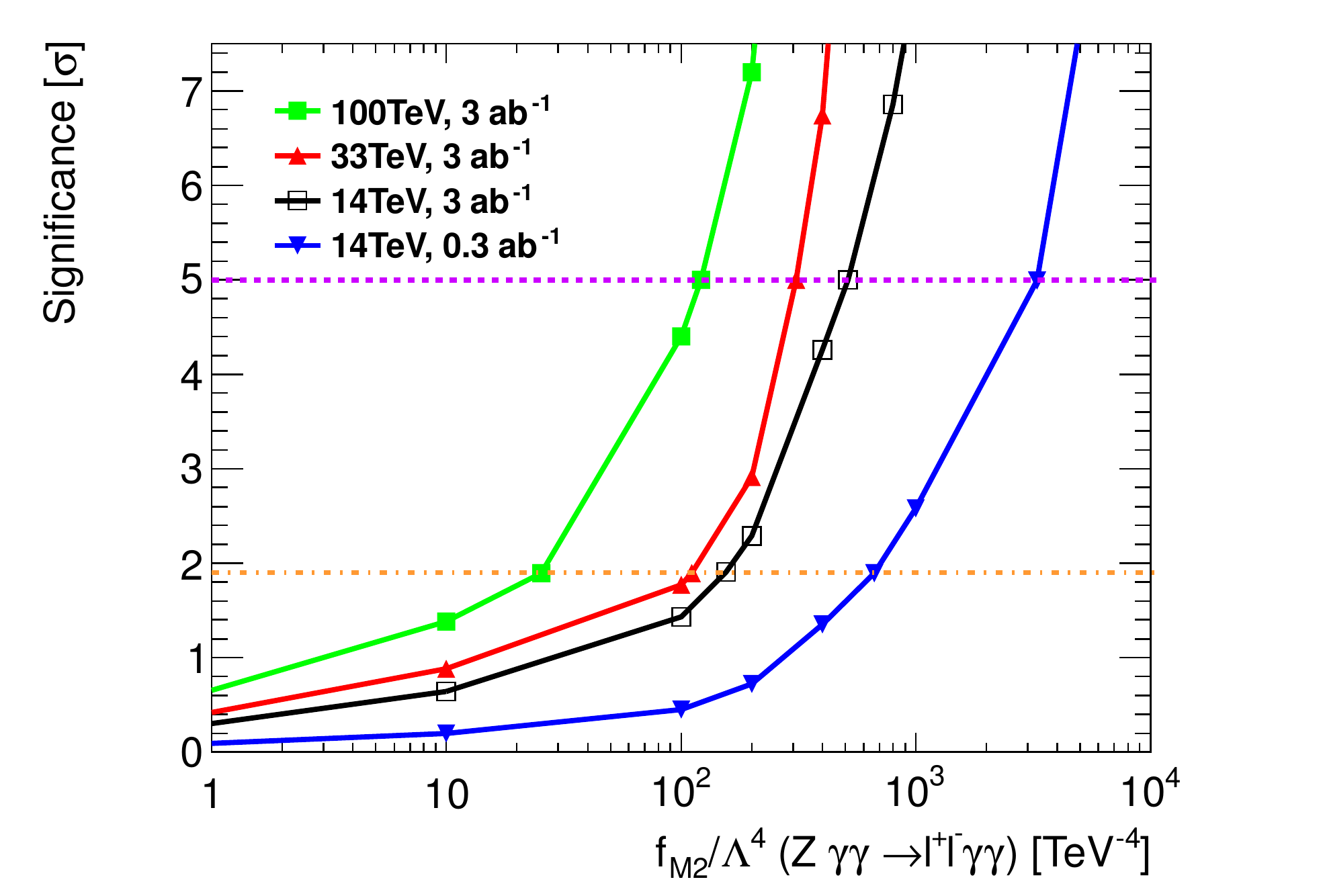}
  \includegraphics[width=0.49\textwidth]{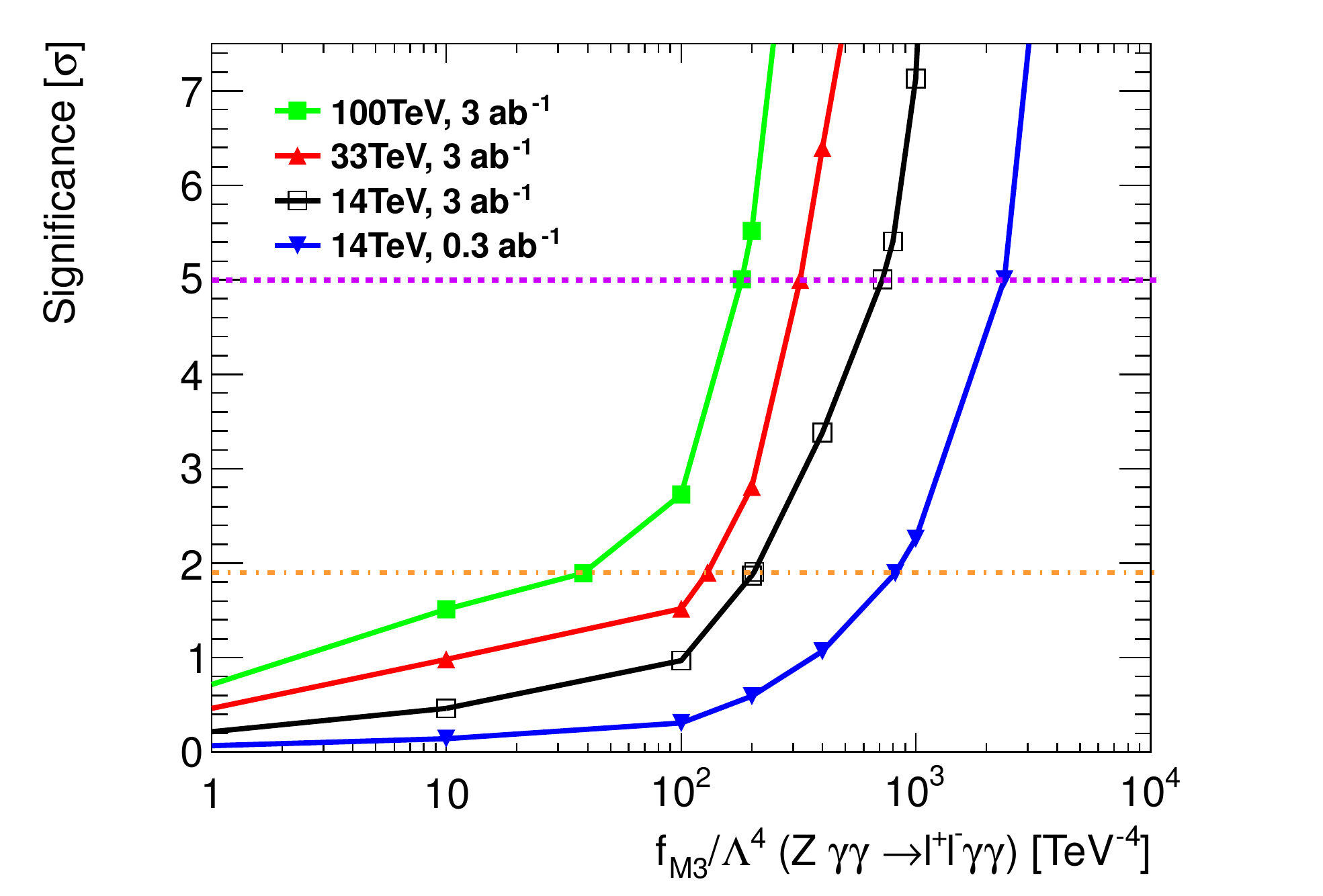}
  \caption{
The signal significance as a function of $f_{Mi}/\Lambda^4$, i = 0,1,2,3 with the UV bound applied, is shown for various collider options. The dashed lines indicate the 5$\sigma$ and 95\% CL limit values.
}
\label{fig:zaa_Sig}
\end{figure}

To show the impact of the UV bound,
  we present the $5 \sigma$-significance discovery values and 95\% CL limits both without (in parentheses) and with  the UV bounds applied, in Table~\ref{tab:zaaSignificance}.
Figure~\ref{fig:zaa_Sig} shows the significance as function of coefficient values with the UV bound.
The effect of UV cutoff is significant and as large as a factor of $5-10$ for large coefficient values. 
The smaller the coefficient value, the smaller the UV cutoff effect.

\begin{table}[h]
\centering
\begin{tabular}{c|c|c|c|c|c}
\hline\hline
\multirow{2}{*}{Parameter}     & $\sqrt{s}$           & 14 TeV & 14 TeV & 33 TeV & 100 TeV \\
\cline{2-6}
                               & Lum.                 & 300 fb$^{-1}$ & \multicolumn{3}{c}{3000 fb$^{-1}$}  \\
\hline
\multirow{2}{*}{$f_{M0}/\Lambda^4$ [TeV$^{-4}$]} & 5$\sigma$  & 7300 (830) & 3600 (310) & 1900 (190) & 750 (120) \\
\cline{2-6}
                                                 & 95\% CL  & 4200 (360) & 1200 (160) & 660 (120)  & 71 (59) \\
\hline
\multirow{2}{*}{$f_{M1}/\Lambda^4$ [TeV$^{-4}$]} & 5$\sigma$  &  7600 (1600) & 3600 (680)  & 2100 (340) & 1000 (220) \\
\cline{2-6}
                                                 & 95\% CL  &  4500 (800)  & 1200 (290) & 770 (160) & 240 (126)  \\

\hline
\multirow{2}{*}{$f_{M2}/\Lambda^4$ [TeV$^{-4}$]} & 5$\sigma$  & 3300 (130) & 510 (48) & 310 (26) & 120 (16) \\
\cline{2-6}
                                                 & 95\% CL  & 670 (56) & 160 (21) & 110 (13) & 25 (10) \\

\hline
\multirow{2}{*}{$f_{M3}/\Lambda^4$ [TeV$^{-4}$]} & 5$\sigma$  & 2400 (250) & 720 (120)  & 320 (66) & 180 (34) \\
\cline{2-6}
                                                 & 95\% CL  & 820 (133) & 210 (52) & 130 (23) & 38 (15)\\
\hline\hline
\end{tabular}
\caption{In $pp \to Z\gamma\gamma \to l^{+}l^{-}\gamma\gamma$ processes, $5 \sigma$-significance discovery values and 95\% CL limits are shown 
 for coefficients of dimension-8 operators with integrated luminosity of 300 fb$^{-1}$ at $\sqrt{s} = 14$~TeV and 3000 fb$^{-1}$  at $\sqrt{s} = 14$~TeV, 33 TeV and 100 TeV, respectively.  
 To show the impact without the UV bound, the corresponding results are shown in parentheses.}
\label{tab:zaaSignificance}
\end{table}

\section{Conclusions}
\label{conclusions}
We have presented sensitivity studies for gauge-invariant
 dimension-6 and dimension-8 operators which can parameterize new
 physics involving Higgs and gauge boson fields. We have explored anomalous vector boson scattering and 
 triboson production due to such operators at $pp$ colliders with $\sqrt{s} = 14, \; 33$ and 100~TeV
 center-of-mass energy. 

 Our conclusions are as follows:
\begin{itemize}
\item The VBS $ZZ$ final state, when used to probe the ${\cal L}_{\phi W}$ dimension-6 operator, increases in
 sensitivity to the operator coefficient by a factor of $\approx 1.9$ when the luminosity is increased by a 
 factor of 10 from 300 fb$^{-1}$ to 3000 fb$^{-1}$, and by a factor of $\approx 1.2$ when the collider 
 energy 
 is increased from 14~TeV to 33~TeV. When considering the dimension-8 operators ${\cal L}_{T,8}$ (${\cal L}_{T,9}$),
 the sensitivity increases by a factor of 1.9 (1.5) due to the same luminosity increase and by a factor of
 $\approx 1.8$ (1.5) due to the energy increase. The sensitivity to the dimension-6 operator is not affected by
 imposing a UV bound, while the sensitivity to the dimension-8 operator is reduced by a factor of about 1.8
 when the bound is applied. 

\item The VBS $WZ$ final state, when used to probe the ${\cal L}_{\phi d}$ dimension-6 operator, increases in
 sensitivity to the operator coefficient by a factor of $\approx 1.9$ when the luminosity is increased 
 from 300 fb$^{-1}$ to 3000 fb$^{-1}$, and by a factor of $\approx 1.2$ when the collider 
 energy 
 is increased from 14~TeV to 33~TeV. When considering the dimension-8 operator ${\cal L}_{T,1}$,
 the sensitivity increases by a factor of $\approx 1.8$ due to the same luminosity increase and by a factor of
 $\approx 2$ due to the energy increase. The sensitivity to the dimension-6 operator is not affected by
 imposing a UV bound, while the sensitivity to the dimension-8 operator is reduced by a factor of about 1.8
 when the bound is applied. 

\item The VBS ss$WW$ final state, when used to probe the ${\cal L}_{T,1}$ dimension-8 operator, increases in
 sensitivity to the operator coefficient by a factor of $\approx 2$ when the luminosity is increased 
 from 300 fb$^{-1}$ to 3000 fb$^{-1}$ at $\sqrt{s} = 14$~TeV. An increase in  collider  energy 
 from 14 TeV to 100~TeV increases the sensitivity by a factor of 100. 
 The sensitivity is not affected at $\sqrt{s} = 100$~TeV by
 imposing a UV bound because the bound is very high for the value of the coefficient probed. 
 The sensitivity at $\sqrt{s} = 14$~TeV is reduced by a factor of about 2 
 when the bound is applied. 

\item The triboson $WWW$ final state, when used to probe the ${\cal L}_{WWW}$ dimension-6
  operator, increases in
 sensitivity to the operator coefficient by a factor of $\approx 2$ when the luminosity
 is increased 
 from 300 fb$^{-1}$ to 3000 fb$^{-1}$ at $\sqrt{s} = 14$~TeV. An increase in  collider 
  energy 
 from 14 TeV to 33~TeV (100~TeV) increases the sensitivity by a factor of 1.3 (2.5). 
 These results are affected at the 10\% level by the application of the UV bound. 
 When probing the dimension-8 operator ${\cal L}_{T,0}$, the sensitivity to the operator
 coefficient increases by a factor of $\approx 2$ when the luminosity is increased
 from 300 fb$^{-1}$ to 3000 fb$^{-1}$ at $\sqrt{s} = 14$~TeV. An increase 
 in  collider  energy 
 from 14 TeV to 33~TeV (100~TeV) increases the sensitivity by a factor of 12 (300).
 This dramatic increase is tamed by the UV bound; we take this as an indication that
 $WWW$ triboson production is a sensitive channel for direct production of new particles
 as the collider energy is raised. 

\item The triboson $Z \gamma \gamma$ final state, when used to probe the ${\cal L}_{M,i}$ dimension-8
  operators, increases in
 sensitivity to the operator coefficient by a factor of $2-6$ (depending on the operator considered) when the luminosity
 is increased 
 from 300 fb$^{-1}$ to 3000 fb$^{-1}$ at $\sqrt{s} = 14$~TeV. An increase in  collider 
  energy 
 from 14 TeV to 33~TeV (100~TeV) increases the sensitivity by a factor of $\approx 2$ (4 to 5). 
 These results are strongly affected by the application of the UV bound. 
 \end{itemize}

 The sensitivity to pileup effects has been studied and it is found that, for the leptonic decay modes presented here, the
 results are not sensitive to pileup effects. 

 It is important to note that the sensitivities for the 33~TeV and 100~TeV colliders
 are based on 
 analyses that have not been re-optimized for higher energy colliders; the analyses were
  optimized for
  14 TeV only. Optimization of the analyses for higher collider energies is important and should be revisited in the future as it will 
 lead to further improvements of the
  sensitivity to new physics at  those
 machines.  Furthermore, the sensitivity can be improved using multivariate techniques.

\newpage
\bibliographystyle{atlasBibStyleWithTitle}
\bibliographystyle{atlasBibStyleWithTitle}


\newpage
\appendix
\section{Comparisons of $W^{\pm}Z$ and $ZZ$ VBS cross sections at ILC and LHC}
The following cross sections were computed with {\sc madgraph} at leading order for the VBS $WZ$ and $ZZ$ final states, inclusive of all decay channels. 
\begin{table}[h]
\small
\centering
\begin{tabular}{c|c|c}
\hline\hline
coefficient (TeV$^{-4}$)                   & Cross Section at LHC [fb] & Cross Section at ILC1000 [fb]  \\
\hline
$f_{T0}/\Lambda^{4}$ = 1                             & 612                           & 13.6			\\
$f_{T1}/\Lambda^{4}$ = 1                             & 744                           & 13.6				\\
$f_{T2}/\Lambda^{4}$ = 1                             & 553                           & 13.6			\\
$f_{M0}/\Lambda^{4}$ = 1                             & 539                           & 13.5			\\
$f_{M1}/\Lambda^{4}$ = 1                             & 536                           & 13.6			\\
$f_{M2}/\Lambda^{4}$ = 1                             & 537                           & 13.6			\\
$f_{M3}/\Lambda^{4}$ = 1                             & 539                           & 13.5			\\
$f_{S0}/\Lambda^{4}$ = 1                             & 537                           & 13.5			\\
$f_{S1}/\Lambda^{4}$ = 1                             & 534                           & 13.5			\\
\hline                                                                  
Standard Model                      		     & 537                           & 13.6			\\
\hline\hline
\end{tabular}
\caption{$pp \to WZ + 2j$ (LHC at $\sqrt{s} = 14$ TeV) and $e^{+}e^{-} \to WZ + 2j$ (ILC at $\sqrt{s} = 1$ TeV) Vector Boson Scattering cross section comparisons with dimension-8 operator coefficients.}
\label{tab:WZ_xsec_ILC}
\end{table}

\begin{table}[h]
\small
\centering
\begin{tabular}{c|c|c}
\hline\hline
coefficient (TeV$^{-4}$)                   & Cross Section at LHC [fb] & Cross Section at ILC1000 [fb]  \\
\hline
$f_{T0}/\Lambda^{4}$ = 1                             & 327                     & 0.808                       \\
$f_{T1}/\Lambda^{4}$ = 1                             & 238                     & 0.808                                \\
$f_{T2}/\Lambda^{4}$ = 1                             & 159                     & 0.784                       \\
$f_{T8}/\Lambda^{4}$ = 1                             & 194                     & 0.898                       \\
$f_{T9}/\Lambda^{4}$ = 1                             & 144                     & 0.824                       \\
$f_{M0}/\Lambda^{4}$ = 1                             & 138                     & 0.760                       \\
$f_{M1}/\Lambda^{4}$ = 1                             & 134                     & 0.768                       \\
$f_{M2}/\Lambda^{4}$ = 1                             & 136                     & 0.735                       \\
$f_{M3}/\Lambda^{4}$ = 1                             & 133                     & 0.790                       \\
$f_{S0}/\Lambda^{4}$ = 1                             & 132                     & 0.763                       \\
$f_{S1}/\Lambda^{4}$ = 1                             & 133                     & 0.763                       \\
\hline                                                                  
Standard Model                                       & 133                     & 0.765                       \\
\hline\hline
\end{tabular}
\caption{$pp \to ZZ + 2j$ (LHC at $\sqrt{s} = 14$ TeV) and $e^{+}e^{-} \to ZZ + 2j$ (ILC at $\sqrt{s} = 1$ TeV) Vector Boson Scattering cross section comparisons with dimension-8 operator coefficients.}
\label{tab:ZZ_xsec_ILC}
\end{table}

\end{document}